\documentclass[iop]{emulateapj}

\usepackage{graphicx}
\usepackage[space]{grffile}
\usepackage{latexsym}
\usepackage{amsfonts,amsmath,amssymb}
\usepackage{url}
\usepackage[utf8]{inputenc}
\usepackage{fancyref}
\usepackage{hyperref}
\usepackage{textcomp}
\usepackage{longtable}
\usepackage{multirow,booktabs}
\usepackage{subfigure}
\usepackage{natbib}
\usepackage{longtable}

\newcommand{\Ks}{K$_{\rm s}$}
\newcommand{\Halpha}{H$\alpha$}
\newcommand{\Hbeta}{H$\beta$}

\newcommand{\msol}{M$_\odot$}

\newcommand{\SII}{[\hbox{{\rm S}\kern 0.1em{\sc ii}}]}
\newcommand{\AlIII}{\hbox{{\rm Al}\kern 0.1em{\sc iii}}}
\newcommand{\NII}{[\hbox{{\rm N}\kern 0.1em{\sc ii}}]}
\newcommand{\OII}{[\hbox{{\rm O}\kern 0.1em{\sc ii}}]}
\newcommand{\OIII}{[\hbox{{\rm O}\kern 0.1em{\sc iii}}]}
\newcommand{\MgII}{\hbox{{\rm Mg}\kern 0.1em{\sc ii}}}
\newcommand{\MgI}{\hbox{{\rm Mg}\kern 0.1em{\sc i}}}
\newcommand{\FeII}{\hbox{{\rm Fe}\kern 0.1em{\sc ii}}}
\newcommand{\CIII}{\hbox{{\rm C}\kern 0.1em{\sc iii}}}
\newcommand{\CIV}{\hbox{{\rm C}\kern 0.1em{\sc iv}}}

\newcommand{\CII}{\hbox{{\rm C}\kern 0.1em{\sc ii}}}
\newcommand{\OI}{\hbox{{\rm O}\kern 0.1em{\sc i}}}
\newcommand{\NeIII}{[\hbox{{\rm Ne}\kern 0.1em{\sc iii}}] }
\newcommand{\NeII}{[\hbox{{\rm Ne}\kern 0.1em{\sc ii}}] }
\newcommand{\NaI}{[\hbox{{\rm Na}\kern 0.1em{\sc i}}] }
\newcommand{\around}{$\sim$}

\newcommand{\NMAD}{$\sigma_{\mathrm{NMAD}}$}
\newcommand{\zspec}{$z_{\mathrm{spec}}$}
\newcommand{\zgrism}{$z_{\mathrm{grism}}$}
\newcommand{\zphoto}{$z_{\mathrm{photo}}$}
\newcommand{\mass}{M$_*$/M$_\odot$}

\slugcomment{To appear in The Astrophysical Journal Supplements (ApJS)}

\shorttitle{The ZFIRE\ Survey}
\shortauthors{Nanayakkara et al.} 

\begin{document}


\title{ZFIRE: A KECK/MOSFIRE Spectroscopic Survey of Galaxies in Rich
Environments at $z\sim2$} 
% spectroscopy of galaxies in rich environments at $z\sim2$:
% Catalogue Release I and a comparison of  spectroscopic and photometric derived properties of galaxies}


\author{Themiya Nanayakkara\altaffilmark{1,*} }
\author{Karl Glazebrook\altaffilmark{1}}
\author{Glenn G. Kacprzak\altaffilmark{1}}
\author{Tiantian Yuan\altaffilmark{2}}
\author{Kim-Vy Tran\altaffilmark{3}}
\author{Lee Spitler\altaffilmark{5,6}}
\author{Lisa Kewley\altaffilmark{2}}
\author{Caroline Straatman\altaffilmark{4}}
\author{Michael Cowley\altaffilmark{5,6}}
\author{David Fisher\altaffilmark{1}}
\author{Ivo Labbe\altaffilmark{4}}
\author{Adam Tomczak\altaffilmark{3}}
\author{Rebecca Allen\altaffilmark{1,6}}
\author{Leo Alcorn\altaffilmark{3}}

\altaffiltext{1}{Centre for Astrophysics and Supercomputing, Swinburne University of Technology, Hawthorn, Victoria 3122, Australia.}
\altaffiltext{*}{tnanayak@astro.swin.edu.au}
\altaffiltext{2}{Research School of Astronomy and Astrophysics, The Australian National University, Cotter Road, Weston Creek, ACT 2611, Australia.}
\altaffiltext{3}{George P. and Cynthia W. Mitchell Institute for Fundamental Physics and Astronomy, Department of Physics and Astronomy, Texas A \& M University, College Station, TX 77843.}
\altaffiltext{4}{Leiden Observatory, Leiden University, PO Box 9513, 2300
RA Leiden, Netherlands.}
\altaffiltext{5}{Department of Physics \& Astronomy, Macquarie University,
Sydney, NSW 2109, Australia.}
\altaffiltext{6}{Australian Astronomical Observatory, PO Box 915, North
Ryde, NSW 1670, Australia.}

\begin{abstract}
We present an overview and the first data release of
ZFIRE, a spectroscopic
redshift survey of star-forming galaxies that utilizes the MOSFIRE
instrument on Keck-I to study galaxy properties in rich environments
at $1.5<z<2.5$.  ZFIRE measures accurate spectroscopic redshifts and
basic galaxy properties derived from multiple emission lines.  The 
galaxies are selected from a stellar mass limited sample based on deep
near infrared imaging ($\mathrm{K_{AB}<25}$) and precise photometric
redshifts from the ZFOURGE and UKIDSS surveys as well as grism
redshifts from 3DHST.  Between 2013 and 2015 ZFIRE has observed the COSMOS and UDS
legacy fields over 13 nights and has obtained 211 galaxy redshifts over
$1.57<z<2.66$ from a combination of nebular emission lines (such as
\Halpha, \NII, \Hbeta, \OII, \OIII, \SII) observed at 1--2\micron.
Based on our medium-band near infra-red photometry, we are able to
spectrophotometrically flux calibrate our spectra to
\around10\% accuracy.  ZFIRE reaches $5\sigma$ emission line flux
limits of \around$\mathrm{3\times10^{-18}~erg/s/cm^2}$ with a
resolving power of $R=3500$ and reaches masses down to
\around10$^{9}$\msol.   
We confirm that the primary input survey,
ZFOURGE, has produced photometric redshifts for star-forming galaxies
(including highly attenuated ones) accurate to $\Delta
z/(1+z\mathrm{_{spec})}=0.015$ with $0.7\%$ outliers.  We measure a
slight redshift bias of $<0.001$, and we note that the redshift bias tends to
be larger at higher masses.  We also examine the
role of redshift on the derivation of rest-frame colours and stellar
population parameters from SED fitting techniques.
The ZFIRE survey extends spectroscopically confirmed $z\sim 2$ samples
across a richer range of
environments, here we make available the first public release of
the data for use by the community.\footnote{\url{http://zfire.swinburne.edu.au}}
\end{abstract}

% delta(z) = zspec-zphot



\section{Introduction}

The rapid development of very deep multi-wavelength imaging surveys from the ground and space in the past decade has greatly enhanced our understanding of  important questions in galaxy evolution particularly through the provision of `photometric redshift' estimates (and hence the evolutionary sequencing of galaxies)  from multi-band spectral energy distribution (SED)
fitting \citep{Whitaker2011,McCracken2012,Skelton2014}. Studies using data from these surveys have led to a more detailed understanding of topics such as the evolution of the galaxy mass function \cite[eg.,][]{Marchesini2010,Muzzin2013,Tomczak2014,Grazian2015}, stellar population properties \cite[eg.,][]{Maseda2014,Spitler2014,Pacifici2015}, evolution of galaxy morphology \cite[eg.,][]{Huertas-Company2015,Papovich2015}, and the growth of the large-scale structure in the universe \citep{Adelberger2005,Wake2011}.

\subsection{Advances with Deep Near-IR Imaging Surveys}

Near-infrared data is vital for this endeavour, both for photometric redshift estimation \citep{Dahlen2013,Rafelski2015} and provision of stellar mass estimates \citep{Brinchmann2000,Muzzin2009}. 
Stellar mass is especially useful for tracking galaxy evolution as it increases monotonically with time, but data at near-infrared wavelengths
are needed to estimate it accurately at high-redshift \citep[][Straatman et al. in press]{Whitaker2011}. New surveys have been made possible by the recent development of relatively wide-field sensitive near infrared (NIR) imagers in 4-8m telescopes such as FourStar \citep{Persson2013} , HAWK-I \citep{Pirard2004}, NEWFIRM \citep{NEWFIRM} and VIRCAM \citep{Dalton2006}. Surveys such as ZFOURGE (Straatman et al., in press), the NEWFIRM medium-band Survey (NMBS) \citep{Whitaker2011}, and ULTRAVISTA \citep{McCracken2012} have obtained deep imaging over relatively large  sky areas (up to 1.5 deg$^2$). The introduction of near-infrared medium-band filters ($\Delta\lambda\sim 1000$\AA) has resulted in photometric redshifts with accuracies of \around2\% \citep{Whitaker2011} and enabled galaxy properties to be accurately derived by SED fitting techniques such as EAZY \citep{Brammer2008} and FAST \citep{Kriek2009}.

These photometric redshift surveys have greatly enhanced our understanding of the universe at $z\sim2$, which is a critical epoch in the evolution of the universe. At this redshift, the universe was only 3 billion years old and was at the peak of cosmic star formation rate activity \citep{Hopkins2006,Lee2015}. We see the presence of massive, often dusty, star-forming galaxies
\citep{Spitler2014,Reddy2015} which were undergoing rapid evolution and the development of a significant
population of massive, quiescent galaxies \citep{vanDokkum2008,Damjanov2009}.
Galaxy clusters have also now been identified at $z\sim2$, and results
indicate that this may be the epoch when environment starts to influence galaxy evolution \citep{Gobat2011,Spitler2012,Yuan2014,Casey2015}. 

\subsection{Need for Spectroscopy}

Even though immense progress on understanding galaxy evolution has been made possible by deep imaging surveys, the spectroscopy of galaxies remains critically important. Spectroscopy provides the basic, precision redshift information that can be used to investigate the accuracy of photometric redshifts derived via SED fitting techniques. The galaxy properties derived via photometry have a strong dependence on the redshifts, and quantifying any systematic biases will help constrain the derived galaxy properties and understand associated errors. 
Spectral emission and absorption lines also provide a wealth of information on physical processes  and kinematics within galaxies \citep{Shapley2009}. Spectroscopy also provides accurate environmental information (for example, the velocity dispersions of proto-clusters; e.g. \cite{Yuan2014})  beyond the resolution of photometric redshifts.

Rest-frame ultraviolet (UV) spectroscopy of galaxies provides information on the properties of massive stars in galaxies and the composition and kinematics of the galaxies' interstellar medium \citep[ISM;][]{Dessauges2010,Quider2010}. 
Rest-frame optical absorption lines are vital to determine the older stellar population properties of the galaxies \citep[eg.,][]{vandeSande2011,Belli2014}. Rest-frame optical emission lines provide information on the state of the ionized gas in galaxies, its density, ionization degree, and metallicity \citep{Pettini2004,Steidel2014,Kacprzak2015,Kewley2016,Shimakawa2015}.

\subsection{Spectroscopy of $z\lesssim1$ Galaxies}

Large-scale spectroscopy is now routine at the low redshift universe. 
Surveys such as the Sloan Digital Sky Survey \citep[][]{York2000}, the 2-Degree Field Galaxy Redshift Survey \citep[][]{Colless2001}, and the Galaxy and Mass Assembly Survey \citep[][]{Driver2009} extensively explored the $z\la 0.2$ universe ($10^5$--$10^6$ galaxies). At  $z\sim 1$
the DEEP2 Galaxy Redshift Survey \citep{Newman2013}, the VIMOS VLT Deep Survey \citep{LeFevre2005}, the VIMOS Public Extragalactic Survey \citep{Garilli2014}, and zCOSMOS \citep{Lilly2007}  have produced large spectroscopic samples ($10^4$--$10^5$ galaxies). 
The large number of galaxies sampled in various environmental and physical conditions by these surveys has placed strong constraints on galaxy models at $z<1$ while revealing rare phases and mechanisms of galaxy evolution \cite[e.g.,][]{Cooper2007,Coil2008,Cheung2012,Newman2013}.

\subsection{Spectroscopy of $z\sim2$ Galaxies}

At a $z\gtrsim1.5$ rest-frame optical features are redshifted to the NIR regime and therefore accessing these diagnostics becomes more challenging. Historically, the spectroscopy of galaxies in these redshifts focussed on the follow up of Lyman break galaxies, which are rest-frame UV selected  using the distribution of the objects in $\cal{U}$, $\cal{G}$, and $\cal{R}$ colour space \citep{Steidel1992}. This technique takes advantage of the discontinuity of the SEDs near the Lyman limit. \citet{Steidel2003} used this technique to target these candidates with multi-object optical spectrographs to obtain rest frame UV spectra for \around1000 galaxies at $z\sim3$. 
Furthermore, $\cal{U}$, $\cal{G}$, and $\cal{R}$ selections can be modified to select similar star-forming galaxies between $1.5<z<2.5$ via their U-band excess flux \citep{Steidel2004}. 
Such sample selections are biased toward UV bright sources and do not yield homogeneous mass complete samples. Surveys such as the Gemini Deep Deep Survey \citep[][]{Abraham2004} and the Galaxy Mass Assembly ultra-deep Spectroscopic Survey \citep[][]{Kurk2013} have attempted to address this by using the IR selection of galaxies (hence much closer to mass-complete samples) before obtaining optical spectroscopy.  
The K20 survey \citep{Cimatti2002} used a selection based on Ks magnitude (Ks$<20$) to obtain optical spectroscopy of extremely dusty galaxies at $z\sim1$. 
These surveys have provided redshift information, but only rest-frame UV spectral diagnostics, and many red galaxies are extremely faint in the rest-UV requiring very long exposure times.

The development of near-IR spectrographs has given us access to rest-frame optical spectroscopy of galaxies at $z\gtrsim1.5$, but the ability to perform spectroscopy of a large number of galaxies has been hindered due to low sensitivity and/or unavailability of multiplexed capabilities. 
For example the MOIRCS Deep Survey \citep{Kajisawa2006} had to compromise between area, sensitivity, number of targets, and resolution due to instrumental limits with MOIRCS in Subaru \citep{Ichikawa2006}. 
The Subaru FMOS galaxy redshift survey \cite{Tonegawa2015}, yielded mostly bright line emitters due to limitations in sensitivity of FMOS \citep{Kimura2010}. 
Furthermore, FMOS does not cover the longer K-band regime which places an upper limit for \Halpha\ detections at $z\sim1.7$. Sensitive long slit spectrographs such as GNIRS  \citep{Elias2006} and XShooter \citep{Vernet2011} have been utilised to observe limited samples of massive galaxies at $z\sim2$.
NIR-grism surveys from the \emph{Hubble Space Telescope (HST)} have yielded large samples such as in the 3DHST survey \citep{Momcheva2015,Treu2015} but have low spectral resolution ($R\sim70-300$) and do not probe wavelengths $>$ 2\micron.

With the introduction of the Multi-object Spectrometer for infrared Exploration (MOSFIRE), a cryogenic configurable multislit system on the 10m Keck telescope \citep{McLean2012}, we are now able to obtain high-quality near-infrared spectra of galaxies in large quantities \citep{Kulas2013,Steidel2014,Kriek2015,Wirth2015}. 
The Team Keck Redshift Survey 2 observed a sample of 97 galaxies at $z\sim2$ to test the performance of the new instrument \citep{Wirth2015} and  investigate the ionization parameters of galaxies at $z\sim2$.
The Keck Baryonic Structure Survey is an ongoing survey of galaxies currently with 179 galaxy spectra, which is primarily aimed to investigate the physical processes between baryons in the galaxies and the intergalactic medium \citep{Steidel2014}. 
The MOSFIRE Deep Evolution Field (MOSDEF) survey is near-infrared selected and aims to observe \around1500 galaxies $1.5<z<3.5$ to study stellar populations, Active Galactic Nuclei, dust, metallicity, and gas physics  using  nebular emission lines and stellar absorption lines \citep{Kriek2015}. 

\subsection{The ZFIRE Survey}

In this paper, we present the ZFIRE survey, which utilizes MOSFIRE to  observe galaxies in rich environments at $z>1.5$ with a complementary sample of field galaxies. A mass/magnitude complete study of rich galaxy environments is essential to overcome selection-bias.
Galaxy clusters are the densest galaxy environments in the universe and are formed via various physical processes \citep{Kravtsov2012}. 
They are a proxy for the original matter density fields of the universe and can be used to constrain fundamental cosmological parameters. Focusing on these rich environments at high-redshift provides access to numerous galaxies with various physical conditions that are rapidly evolving and interacting with their environments. 
These galaxies can be used to study the formation mechanisms of local galaxy clusters in a period where they are undergoing extreme evolutionary processes. Such environments are rare at $z\sim 2$ \citep{Gobat2011,Newman2014,Yuan2014}: for example, we target the \cite{Spitler2012} cluster at $z=2.1$, which was the only such massive structure found in the 0.1 deg$^2$  ZFOURGE survey (and that at only  4\% chance, \citep{Yuan2014}). Hence, a pointed survey on such clusters and their environs is highly complementary to  other field surveys being performed with MOSFIRE.

Here we present the ZFIRE\ survey overview and  first data release. 
We release data for two cluster fields: one at $z=2.095$ \citep{Spitler2012,Yuan2014} and the other at $z=1.62$ \citep{Papovich2010,Tanaka2010}. 
The structure of the paper is as follows: 
in Section \ref{sec:survey}, we describe the ZFIRE survey design, target selection and data reduction.  
In Section \ref{sec:results}, we present our data and calculate the completeness and detection limits of the survey. 
We investigate the accuracy of photometric redshifts  of different surveys that cover the ZFIRE\ fields in Section \ref{sec:photometric_redshifts}. In Section \ref{sec:implications}, we  study the role of photometric redshift accuracy on galaxy physical parameters derived via common SED fitting techniques and how spectroscopic accuracy affects cluster membership identification. 
A brief description of the past/present work and the future direction of the survey is presented in Section \ref{sec:summary}.

We assume a cosmology with H$_0$= 70 km/s/Mpc, $\Omega_\Lambda$=0.7 and $\Omega_m$= 0.3. 
Unless explicitly stated we use AB magnitudes throughout the paper. 
Stellar population model fits assume a \citet{Chabrier2003} initial mass function (IMF), \citet{Calzetti2001} dust law and solar metallicity.  We define \zspec\ as the spectroscopic redshift, \zphoto\ as the photometric redshift, and \zgrism\ as the grism redshift from 3DHST \citep{Momcheva2015}. We express stellar mass (M$_*$) in units of solar mass (M$_\odot$).  
Data analysis was performed using \texttt{iPython} \citep{Perez2007} and \texttt{astropy} \citep{Astropy2013} and \texttt{matplotlib} \citep{Hunter2007} code to reproduce the figures, will be available online\footnote{https://github.com/themiyan/zfire\_survey}.

%%%%%%%%%%%%%%%%%%%%%%%%%%%%%%%%%%%%%%%%%%%%%%%%%%%%%%%%%%%%%%%%%%%%%%%%%%%%%%%%%%%%%%%%%%%%%%%%%%%%%%%%%

\section{ZFIRE\ Observations and Data Reduction}
\label{sec:survey}

The MOSFIRE \citep{McLean2008,McLean2010,McLean2012} operates from 0.97--2.41 microns (i.e. corresponding to atmospheric $YJHK$ bands, one band at a
time) and provides a 6.1$'\times 6.1'$ field of view with a resolving power of $R$\around3500.  It is equipped with a cryogenic configurable slit unit that can include up to 46 slits and be configured in
\around6 minutes.  MOSFIRE has a Teledyne H2RG HgCdTe detector with
2048 $\times$ 2048 pixels ($0''.1798$/pix) and can be used as a multi-object spectrograph and a wide-field imager by removing the masking bars from the field of view.  ZFIRE\ utilizes the multi-object spectrograph capabilities of MOSFIRE.

The galaxies presented in this paper consist of observations of two cluster fields from the Cosmic Evolution Survey (COSMOS) field \citep{Scoville2007} and the Hubble Ultra Deep Survey (UDS) Field \citep{Beckwith2006}.
These clusters are the \cite{Yuan2014} cluster at \zspec=2.095 and IRC 0218 cluster \citep{Papovich2010,Tanaka2010,Tran2015} at \zspec=1.62.
\cite{Yuan2014} spectroscopically confirmed the cluster, which was identified by \citet{Spitler2012} using photometric redshifts and deep Ks band imaging from ZFOURGE.
The IRC 0218 cluster was confirmed independently by \citet{Papovich2010} and \citet{Tanaka2010}.
Field galaxies neighbouring on the sky, or in redshift shells, are also observed and provide a built-in comparison sample.

\subsection{ZFIRE\ Survey Goals and Current Status}

The primary science questions addressed by the ZFIRE\ survey are as follows:

\begin{enumerate}
\item  What are the ISM physical conditions of the galaxies? 
We test the Mappings IV models by using \Halpha, \NII, \Hbeta, \OII, \OIII, and \SII\ nebular emission lines to study the evolution of chemical enrichment and the ISM as a function of redshift \citep{Kewley2016}. 

\item  What is the IMF of galaxies?
We use the \Halpha\ equivalent width as a proxy for the IMF of star-forming galaxies at z\around2 ( T. Nanayakkara et al., in preparation).  
		
\item  What are the stellar and gas kinematics of galaxies?
Using \Halpha\ rotation curves we derive accurate kinematic parameters of the galaxies. Using the Tully-Fisher relation \citep{Tully1977} we track how stellar mass builds up inside dark matter halos to provide a key observational constraint on galaxy formation models \citep[][C. Straatman et al., in preparation]{Alcorn2016}. 

\item  How do fundamental properties of galaxies evolve to $z\sim2$ ?
Cluster galaxies at z\around2 include massive star-forming members that are absent in lower redshift clusters. 
We measure their physical properties and determine how these members must evolve to match the galaxy populations in clusters at $z<$1 \citep{Tran2015, Kacprzak2015}. 

%Furthermore, we will investigate the role of AGN in galaxy clusters and the quenching mechanisms of galaxies in dense high-redshift environments.

\end{enumerate}

Previous results from ZFIRE\ have already been published. 
\citet{Yuan2014} showed that the galaxy cluster identified by ZFOURGE \citep{Spitler2012} at $z=2.095$ is a progenitor for a Virgo like cluster. \citet{Kacprzak2015} found no significant environmental effect on the stellar MZR for galaxies at $z\sim2$.  \citet{Tran2015} investigated \Halpha\ SFRs and gas phase metallicities at a lower redshift of $z\sim1.6$ and found no environmental imprint on gas metallicity but detected quenching of star formation in cluster members. 
\citet{Kewley2016} investigated the ISM and ionization parameters of galaxies at $z\sim2$ to show significant differences of galaxies at $z\sim2$ with their local counterparts.
Here the data used to address the above questions in past and future papers is presented.

\subsection{Photometric Catalogues}

Galaxies in the COSMOS field are selected from the ZFOURGE  survey (Straatman et al. in press) which is 
a 45 night deep Ks band selected photometric legacy survey carried out using the 6.5 meter Magellan Telescopes located at Las Campanas observatory in Chile. 
The survey covers 121 arcmin$\mathrm{^2}$ each in COSMOS, CDFS, and UDS cosmic fields using the near-IR medium-band filters of the FourStar imager \citep{Persson2013}.  
All fields have \emph{HST} coverage from the CANDELS survey \citep{Grogin2011,Koekemoer2011} and a wealth of multi-wavelength legacy data sets \citep{Giacconi2002,Capak2007,Lawrence2007}. 
For the ZFIRE\ survey, galaxy selections were made from the v2.1 of the internal ZFOURGE catalogues. A catalogue comparison between v2.1 and the the updated ZFOURGE public data release 3.1 is provided in the Appendix \ref{sec:ZFOURGE comparison}. 
The v2.1 data release reaches a $5\sigma$ limiting depth of $Ks=25.3$ in FourStar imaging of the COSMOS field \citep{Spitler2012} which is used to select the ZFIRE K-band galaxy sample. 
\emph{HST} WFC3 imaging was used to select the ZFIRE H-band galaxy sample.

EAZY \citep{Brammer2008} was used to derive photometric redshifts by fitting linear combinations of nine SED templates to the observed SEDs\footnote{An updated version of EAZY is used in this analysis compared to what is published by \citet{Brammer2008}. Refer \citet{Skelton2014} Section 5.2 for further information on the changes. The updated version is available at \url{https://github.com/gbrammer/eazy-photoz}.}. 
With the use of medium-band imaging and the availability of multi-wavelength data spanning from UV to Far-IR (0.3-8$\mu$m in the observed frame), ZFOURGE produces photometric redshifts accurate to $1-2\%$ \cite[Straatman et al., in press;][]{Kawinwanichakij2014,Tomczak2014}.

Galaxy properties for the ZFOURGE catalogue objects are derived using FAST \citep{Kriek2009} with synthetic stellar populations from \citet{Bruzual2003}  using a $\chi^2$ fitting algorithm to derive ages, star-formation time-scales, and dust content of the galaxies. 
Full information on the ZFOURGE imaging survey can be found in  Straatman et al. (in press).

The IRC 0218 cluster is not covered by the ZFOURGE survey. Therefore  publicly available UKIDSS imaging \citep{Lawrence2007} of the UDS field is used for sample selection.
The imaging covers 0.77 deg$^2$ of the UDS field and reaches a $5\sigma$ limiting depth of $\rm K_{AB}= 25$  (DR10;  \cite{UDS_DR10}). 
Similar to ZFOURGE, public K-band selected catalogues of UKIDSS were used with EAZY and FAST to derive photometric redshifts and galaxy properties \citep{Quadri2012}.

\subsection{Spectroscopic Target Selection}
\label{sec:sample_def}

In the first ZFIRE\ observing run, the COSMOS field between redshifts $2.0<$\zphoto$<2.2$ was surveyed to spectroscopically confirm the overdensity of galaxies detected by \cite{Spitler2012}. 
The main selection criteria were that the \Halpha\ emission line falls within the NIR atmospheric windows and within the coverage of the MOSFIRE filter set. 
For each galaxy, H and K filters were used to obtain multiple emission lines to constrain the parameters of  interest.

Nebular emission lines such as \Halpha\ are strong in star-forming galaxies and hence it is much quicker to detect them than underlying continuum features of the galaxies. 
Therefore, rest frame UVJ colour selections \citep{Williams2009} were used to select primarily star-forming galaxies in the cluster field for spectroscopic follow up.
While local clusters are dominated by passive populations, it is known that high-$z$ clusters contain a higher fraction of star-forming galaxies \citep{Wen2011,Tran2010,Saintonge2008}.
This justifies our use of K band to probe strong emission lines of star-forming galaxies, but due to the absence of prominent absorption features, which fall in the K band at $z\sim2$, we note that our survey could be incomplete due to missing weak star-forming and/or quiescent cluster galaxies.

\begin{figure*}
\includegraphics[scale=0.60]{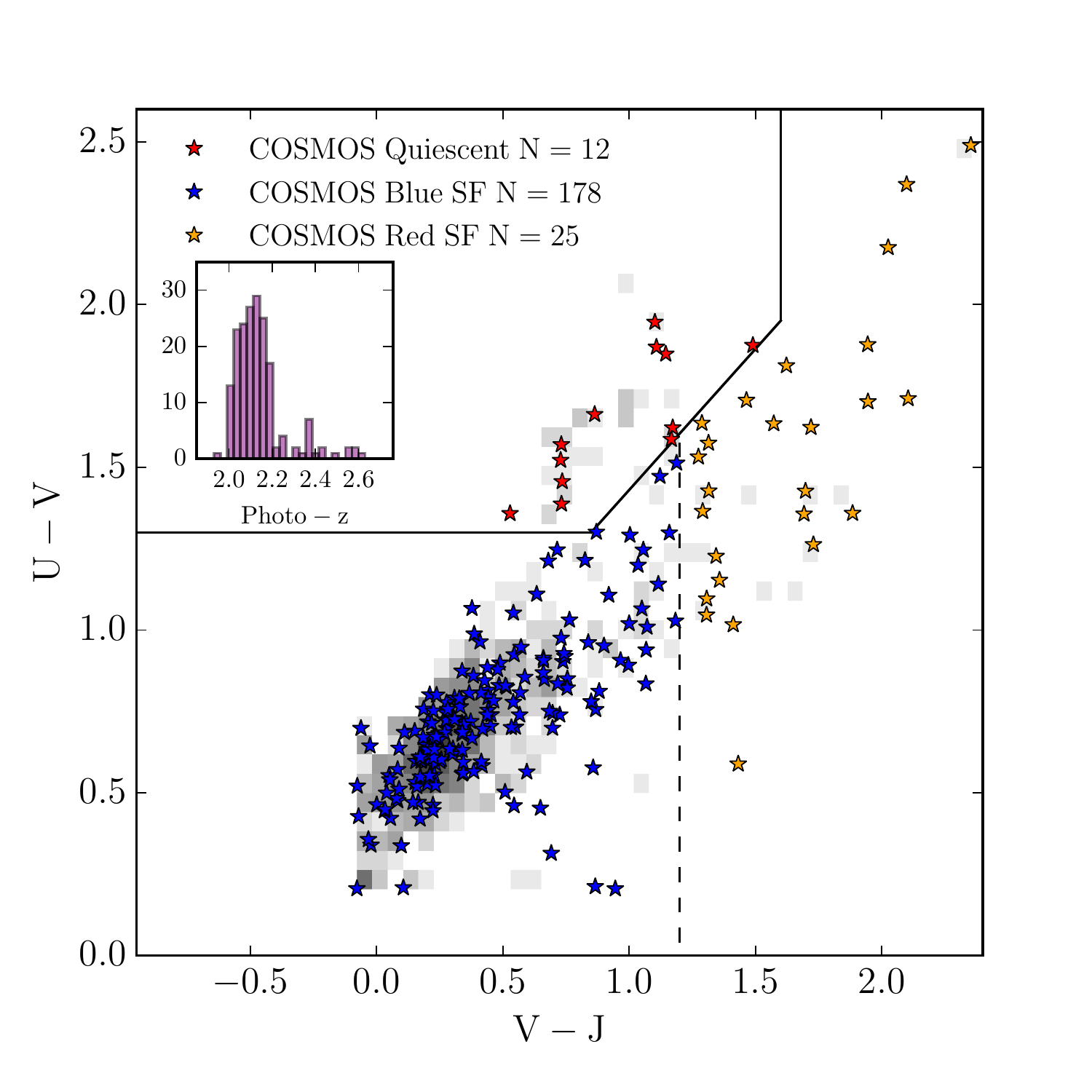}
\includegraphics[scale=0.60]{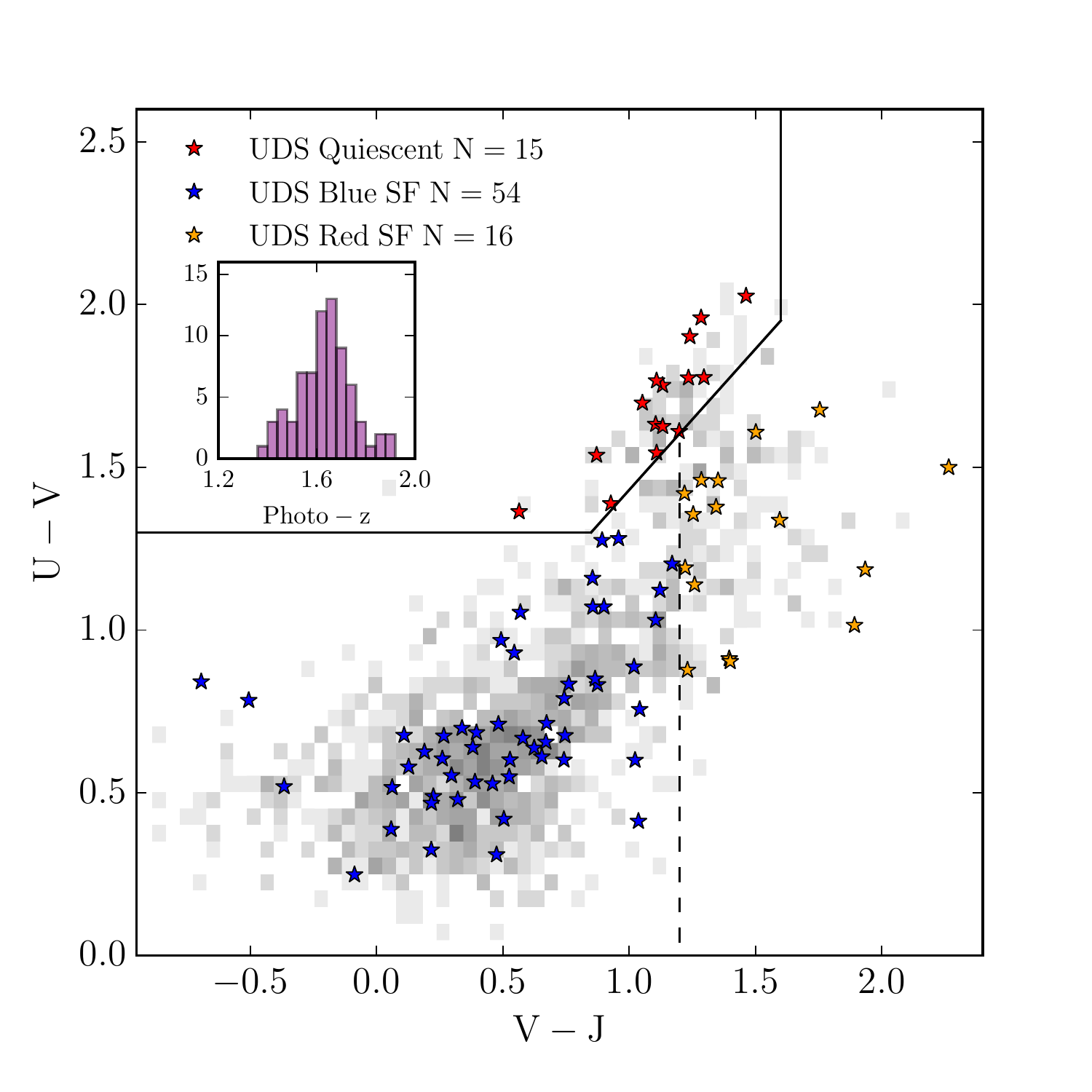}
\caption{ Rest frame UVJ diagram of the galaxy sample selected from ZFOURGE and UKIDSS surveys to be observed.
Quiescent, blue star-forming, and red (dusty) star-forming galaxies are selected using \citet{Spitler2014} criteria which are shown as red, blue, and orange stars, respectively. 
Galaxies above the outlined section are considered to be quiescent. The remaining galaxies are divided to blue and red star-forming galaxies by the dashed vertical line. 
Photometric redshifts are used to derive the rest-frame colours using EAZY. The photometric redshift distribution of the selected sample is shown by the histogram in the inset.
{\bf Left:} the ZFOURGE sample in the COSMOS field selected to be observed by ZFIRE. 
The logarithmic (2D density) greyscale histogram shows the total UVJ distribution of the ZFOURGE galaxies between 1.90$<$\zphoto$<$2.66. 
In the sample selection, priority is given for the star-forming galaxies that lie below the outlined section in the diagram.
{\bf Right:} similar, but now for the UKIDSS sample in the UDS field with galaxies within $10'$ radii from the cluster BCG and at redshifts  $1.57<$\zphoto$<1.67$ shown as the greyscale. 
}
\label{fig:UVJ_selection}
\end{figure*}

The primary goal was to build a large sample of redshifts to identify the underlying structure of the galaxy overdensity, therefore, explicitly choosing  star-forming galaxies increased the efficiency of the observing run. 
Quiescent galaxies were selected either as fillers for the masks or because they were considered to be the brightest cluster galaxies (BCG). 
Rest-frame U$-$V and V$-$J colours of galaxies are useful to distinguish star-forming galaxies from quenched galaxies \citep{Williams2009}.
The rest-frame UVJ diagram and the photometric redshift distribution of the selected sample is shown in the left panel of Figure \ref{fig:UVJ_selection}. 
All rest-frame colours have been derived using photometric redshifts using EAZY with special dustier templates as per \citet{Spitler2014}.  
Out of the galaxies selected to be observed by ZFIRE, \around83\% are (blue) star-forming. The rest of the population comprises  \around11\% dusty (red) star-formers and \around6\% quiescent galaxies. 
For all future analysis in this paper, the \citet{Spitler2014} EAZY templates are replaced with the default EAZY templates in order to allow direct comparison with other surveys. 
More information on UVJ selection criteria is explained in Section \ref{sec:UVJ}.

The COSMOS sample at $z\sim2$ requires K-band observations from MOSFIRE to detect \Halpha\ emission lines. 
A subset of the K-band selected galaxies are then followed up in H-band to retrieve \Hbeta\ and \OIII\ emission lines. 
During the first observing run, object priorities for the galaxies in the COSMOS field were assigned as follows. 
\begin{enumerate}
\item K-band observations for rest frame UVJ selected star-forming K$<$24 galaxies with 2.0$<$\zphoto$<$2.2. 
\item K-band observations for rest frame UVJ selected star-forming K$>$24 galaxies with 2.0$<$\zphoto$<$2.2. 
\item K-band observations for rest frame UVJ selected non-star-forming galaxies with 2.0$<$\zphoto$<$2.2.  
\item Galaxies outside the redshift range to be used as fillers. 
\end{enumerate}
In subsequent observing runs, the following criteria were used to assign priorities. 
\begin{enumerate}
\item H-band observations for galaxies with \Halpha\ and  \NII\ detections from K-band. 
\item H-band observations for galaxies with only \Halpha\ detection for follow up spectroscopic redshift verification with \Hbeta\ and/or  \OIII\ emission lines. 
\item K-band observations for galaxies with only \Halpha\ emission lines for deeper spectroscopic redshift verification and gas phase metallicity study with deeper \NII\ emission lines. 
\end{enumerate}

The UDS sample was selected from  the XMM-LSS J02182-05102  cluster \citep{Papovich2010,Tanaka2010} in order to obtain \OIII, \Halpha\ and \NII\ emission lines. At $z=1.62$, these nebular emission lines are redshifted to J and H-bands. 
Cluster galaxies were specifically targeted to complement with the Keck Low Resolution Imaging Spectrometer (LRIS) observations \citep{Tran2015}. 
Y-band spectra were obtained for a subset of galaxies in the cluster in order to detect \MgII\ absorption features and the D4000 break. 
The UVJ diagram and the photometric redshift distribution of the selected sample is shown by the right panel of Figure \ref{fig:UVJ_selection}. In the selected sample, \around65\% of galaxies are star-forming while dusty star-forming and quiescent galaxies are each \around17\%. 
The highest object priorities for the UDS sample were assigned as follows. 
\begin{enumerate}
\item BCGs of the \citet{Papovich2010} cluster. 
\item LRIS detections with \zspec$\sim$1.6 by \cite{Tran2015}. 
\item Grism spectra detections with $z_{\mathrm{grism}}\sim1.6$ \citep[3DHST][]{Momcheva2015}
\item Cluster galaxy candidates within R$<1$ Mpc and \zphoto$\sim1.6$ \citep{Papovich2010}. 
\end{enumerate} 
For further information on target selection, refer to \citet{Tran2015}.

\subsection{Slit Configurations with MAGMA}
\label{sec:mask_design}

MOSFIRE slit configurations are made through the publicly available MOSFIRE Automatic GUI-based Mask Application (MAGMA\footnote{http://www2.keck.hawaii.edu/inst/mosfire/magma.html}) slit configuration design tool. 
The primary purpose of MAGMA is to design slit configurations to be observed with MOSFIRE and to execute the designed slit configurations in real time at the telescope. 
Once the user specifies a target list and priorities for each of the objects, the software will dither the pointing over the input parameters (which can be defined by the user) to determine the most optimized slit configuration. 
%An optimized configuration is defined as the configuration with the highest total priority score, which is calculated by adding the assigned priorities of the objects selected to the slit configuration. 
%The dithering process is done iteratively and the user is provided with the overall highest priority score and the associated slit configuration with the object list.  
%The number of iterations can be defined by the user. Higher the number of iterations the more optimized the slit configuration is expected to be. 

The slit configurations can then be executed during  MOSFIRE observing. With MAGMA, the physical execution of the slit configurations can be done within $<$15 minutes. 
For the objects in the COSMOS field \around10,000 iterations were used to select objects from a target list compromising of \around2000 objects. 
\citet{vanderWel2012} used \emph{HST} imaging to derive position angles of galaxies in the CANDELS sample using GALFIT \citep{Peng2010b}. 
The number of slits within $\pm30^{\circ}$ of the galaxy major axis were maximized using position angles of \citet{vanderWel2012} catalogue by cross-matching it with ZFOURGE.

Due to the object prioritization, a subset of galaxies was observed in multiple observing runs. These galaxies were included in different masks and hence have different position angles. When possible, position angles of these slits were deliberately varied to allow coverage of a different orientation of the galaxy.

\subsection{MOSFIRE Observations}

\begin{figure*}
\includegraphics[scale=0.61]{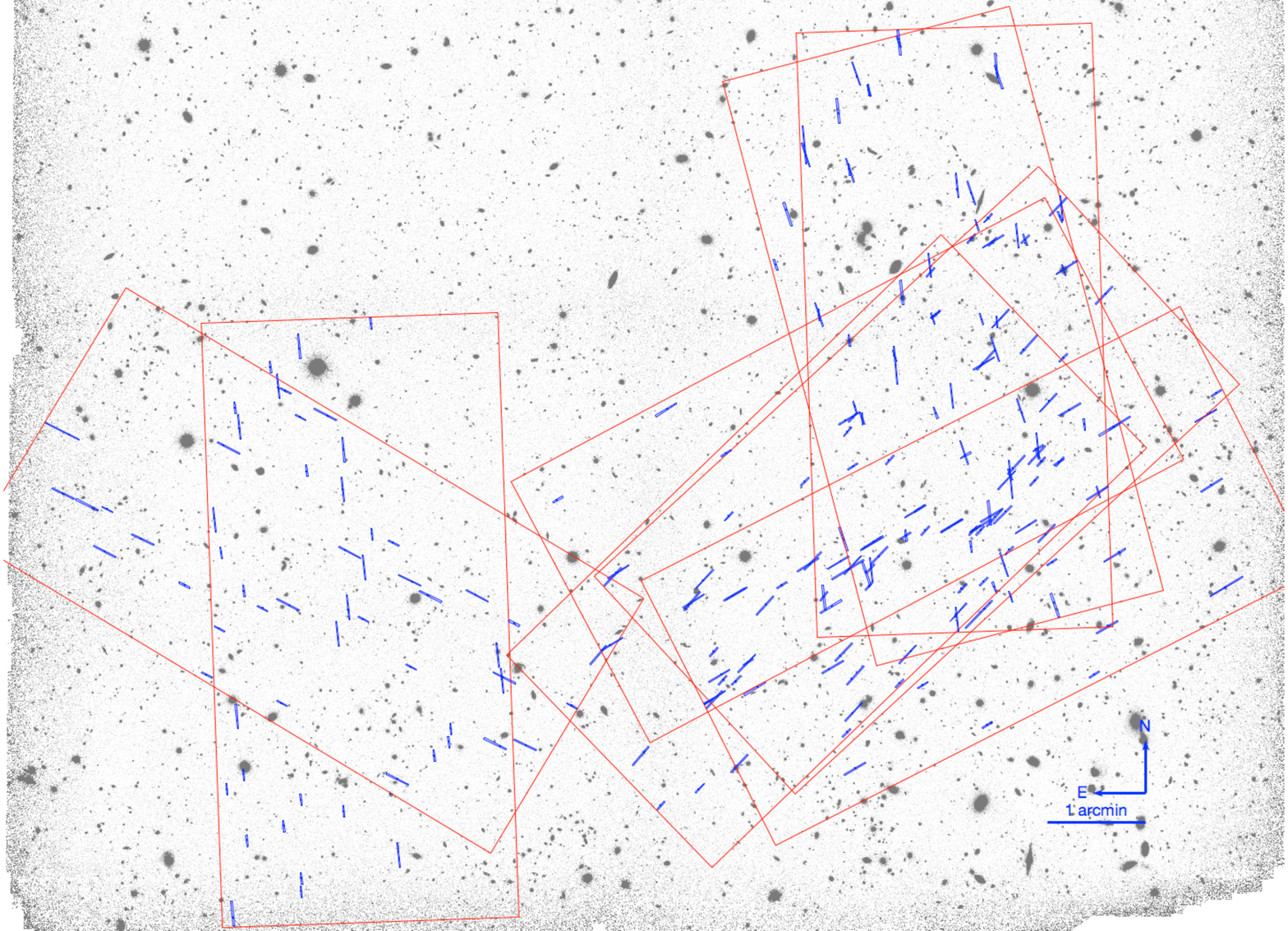}
\caption{MOSFIRE slit configurations for the 6 K-band masks in the COSMOS field. 
The blue lines show each individual slit.  
Each slit in a mask is expected to target a single galaxy. However, some galaxies are targeted in multiple masks. 
The red boxes are the individual masks. 
The inverse greyscale image is from the Ks imaging from FourStar obtained as a part of the ZFOURGE survey.}
\label{fig:masks}
\end{figure*}

Between 2013 and 2016 15 MOSFIRE nights were awarded to the ZFIRE program  by a 
combination of Swinburne University (Program IDs- 2013A\_W163M, 2013B\_W160M, 2014A\_W168M, 2015A\_W193M, 2015B\_W180M), Australian National University (Program IDs- 2013B\_Z295M, 2014A\_Z225M, 2015A\_Z236M, 2015B\_Z236M), and NASA (Program IDs- 2013A\_N105M, 2014A\_N121M) telescope time allocation committees. 
Data for 13 nights observed between 2013 and 2015 are released with this paper, where six nights resulted in useful data collection. 
Observations during 2013 December resulted in two nights of data in excellent conditions, while four nights in 2014 February were observed in varying conditions. Exposure times and observing conditions are presented in Table \ref{tab:observing_details}. 
With this paper, data for 10 masks observed in the COSMOS field and four masks observed in the UDS field are released. An example of on-sky orientations of slit mask designs used for K-band observations in the COSMOS field is shown in Figure \ref{fig:masks}. 
Standard stars were observed at the beginning, middle, and end of each observing night. 

The line spread functions were calculated using Ne arc lamps in the K-band, and were found to be \around2.5 pixels. The partial first derivative for the wavelength (CD1\_1) in Y, J, H, and K-bands are respectively 1.09 \AA/pixel, 1.30 \AA/pixel, 1.63 \AA/pixel, and 2.17 \AA/pixel. 

$0.7''$ width slits were used for objects in science masks and the telluric standard, while, for the flux standard star a slit of width $3''$ was used to minimize slit loss. On average, $\sim$ 30 galaxies were included per mask. A flux monitor star was included in all of the science frames to monitor the variation of the seeing and atmospheric transparency. In most cases only frames that had a FWHM of $\lesssim0''.8$ was used for the flux monitor stars. A standard 2 position dither pattern of ABBA was used.\footnote{For more information, see: \url{http://www2.keck.hawaii.edu/inst/mosfire/dither\_patterns.html\#patterns}} 

\begin{deluxetable*}{llllrrr}
\tabletypesize{\scriptsize}
\tablecaption{ ZFIRE\ Data Release 1: Observing details}  
\tablecomments{ This table presents information on all the masks observed by ZFIRE\ between 2013 and 2015 with the integration times and observing conditions listed.  
\label{tab:observing_details}}
\tablecolumns{6}
\tablewidth{0pt} 
\startdata
\hline \hline \\ [+1ex]
Field  & Observing & Mask & Filter & Exposure & Total integra-     & Average \\
	   & Run       & Name &        & Time (s) & -tion Time (h) 		& Seeing ($''$) \\ [+1ex]  \hline \\ [+1ex]

COSMOS & Dec2013 & Shallowmask1 (SK1)    & K & 180 & 2.0  & 0$''$.70\\ 
COSMOS & Dec2013 & Shallowmask2 (SK2)   & K & 180 & 2.0  & 0$''$.68\\
COSMOS & Dec2013 & Shallowmask3 (SK3)   & K & 180 & 2.0  & 0$''$.70\\ 
COSMOS & Dec2013 & Shallowmask4 (SK4)   & K & 180 & 2.0  & 0$''$.67\\ 

COSMOS & Feb2014 & KbandLargeArea3 (KL3) & K & 180 & 2.0  & 1$''$.10\\ 
COSMOS & Feb2014 & KbandLargeArea4 (KL4) & K & 180 & 2.0  & 0$''$.66\\ 

COSMOS & Feb2014 & DeepKband1      (DK1) & K & 180 & 2.0   & 1$''$.27\\ 
COSMOS & Feb2014 & DeepKband2      (DK2) & K & 180 & 2.0   & 0$''$.70\\ 

COSMOS & Feb2014 & Hbandmask1      (H1) & H & 120 & 5.3 & 0$''$.90\\ 
COSMOS & Feb2014 & Hbandmask2      (H2) & H & 120 & 3.2 & 0$''$.79\\ 

UDS    & Dec2013 & UDS1            (U1H) & H & 120 & 1.6 & 0$''$.73\\ 
UDS    & Dec2013 & UDS2            (U2H) & H & 120 & 1.6 & 0$''$.87\\ 
UDS    & Dec2013 & UDS3            (U3H) & H & 120 & 0.8 & 0$''$.55\\ 

UDS    & Dec2013 & UDS1            (U1J) & J & 120 & 0.8 & 0$''$.72\\ 
UDS    & Dec2013 & UDS2            (U2J) & J & 120 & 0.8 & 0$''$.90\\ 
UDS    & Dec2013 & UDS3            (U3J) & J & 120 & 0.8 & 0$''$.63\\ 

UDS    & Feb2014 & uds-y1          (UY) & Y & 180 & 4.4 & 0$''$.80\\ 

\end{deluxetable*}

%The COSMOS field was observed in 6 masks in the K-band with $\sim$ 2 hours of on source integration time with 180s exposures and 2 masks in the H-band with $\sim$ 5.3 \& 3.2 hours of on source integration time with 120s exposures. 

%UDS field was observed in 3 masks in J and H-bands with 120s exposures and 1 mask in the Y-band with 180s exposures. The J-band masks and 1 H-band mask was observed for 0.8 hours per mask while the remaining 2 H masks were observed for 1.6 hours each. The Y-band mask was observed for $\sim$ 4.4 hours. 

\subsection{MOSFIRE Spectroscopic Reduction}
\label{data_reduction}

The data were reduced in two steps. 
Firstly, a slightly modified version of the publicly available 2015A MOSFIRE DRP release \footnote{A few bug fixes were applied along with an extra function to implement barycentric corrections to the spectra. This version is available at \url{https://github.com/themiyan/MosfireDRP\_Themiyan}.} was used to reduce the raw data from the telescope.
Secondly, a custom made IDL package was used to apply telluric corrections and flux calibrations to the data and extract 1D spectra. Both are described below. 

Extensive tests were performed to the MOSFIRE DRP while it was in a beta stage, and multiple versions of the DRP were used to test the quality of the outputs. 
The accuracy of the error spectrum generated by the DRP was investigated by comparing the noise we expect from the scatter of the sky values with the DRP noise. 
The following steps are currently performed by the modified MOSFIRE DRP. 
\begin{enumerate}
\item Produce a pixel flat image and identify the slit edges. 
\item For K-band: remove the thermal background produced by the telescope dome. 
\item Wavelength calibrate the spectra. This is performed using the sky lines. For K-band: due to the lack of strong sky lines at the red end of the spectra, a combination of night sky lines along with Neon and/or Argon\footnote{As of version 2015A, using both Ar and Ne lamps together with sky line wavelength calibration is not recommended. See the MOSFIRE DRP github issues page for more details.} arc lamp spectra are used to produce per pixel wavelength calibration.
\item Apply barycentric corrections to the wavelength solution. 
\item Remove the sky background from the spectra. This is done in two steps. Firstly, the different nod positions of the telescope are used to subtract most of the background. 
Secondly, any residual sky features are removed following the prescription by \citet{Kelson2003}.
\item Rectify the spectra.
\end{enumerate}
All the spectra from the DRP were calibrated to vacuum wavelengths with a typical residual error of $<$ 0.1 \AA. 

The customized IDL package was used to continue the data reduction process using outputs of the public DRP. The same observed standard star was used to derive telluric sensitivity and flux calibration curves  to be applied to the science frames as follows.
\begin{enumerate}
\item The 1D standard star spectrum was extracted from the wavelength calibrated 2D spectra.
\item Intrinsic hydrogen absorption lines in the stellar atmosphere  were removed from the telluric A0 standard by fitting Gaussian profiles and then  interpolating over the filled region. 
\item The observed spectrum was ratioed to a theoretical black body function corresponding to the temperature of the star.
%A black body transmission function is fit to the spectra to remove the intrinsic black body shapes of the stellar spectra. \kg{Don't understand this part. Don't you mean the observed spectrum is divided by a theoretical BB coresponding to the temperature of the star?}
\item The resulting spectrum was then normalised and smoothed to be used as the sensitivity curve, i.e., the wavelength-dependent sensitivity that is caused by the atmosphere and telescope-instrument response. 
\item The sensitivity curve was used on the flux standard star to derive the flux conversion factor by comparing it to its 2MASS magnitude \citep{Skrutskie2006}.
\end{enumerate}
These corrections are applied to the 2D science frames to produce telluric corrected, flux calibrated spectra. 
Further information is provided in Appendix \ref{sec:MOSFIRE cals}.  
The derived response curves that were applied to all data include corrections for the MOSFIRE response function, the telescope sensitivity, and atmospheric absorption. 
If the mask were observed in multiple nights, the calibrated 2D spectra were co-added by weighting by the variance spectrum.  Extensive visual inspections were performed to the 2D spectra to identify possible emission line-only detections and to flag false detections due to, e.g. sky line residuals.

To extract 1D spectra, Gaussian extractions were used to determine the FWHM of the spatial profile. If the objects were too faint compared to the sky background, the profile from the flux monitor star of the respective mask was used to perform the extraction. 
The same extraction procedure was performed for any secondary or tertiary objects that fall within any given slit.
Depending on how object priorities were handled, some objects were observed during multiple observing runs in different masks. There were 37 such galaxies.  
Due to variations in the position angles between different masks, these objects were co-added in 1D after applying the spectrophotometric calibration explained in Section \ref{sec:sp calibration}.

\subsection{Spectrophotometric Flux Calibration}
\label{sec:sp calibration}

\subsubsection{COSMOS Legacy Field}

Next zero-point adjustments were derived for each mask to account for any atmospheric transmission change between mask and standard observations. Synthetic slit aperture magnitudes were computed from the ZFOURGE survey to calibrate the total magnitudes of the spectra, which also allowed us to account for any slit-losses due to the $0''.7$ slit-width used during the observing.  
The filter response functions for FourStar \citep{Persson2013} were used to integrate the total flux in each of the 1D calibrated spectra.  

For each of the masks in a respective filter, first,  all objects with a photometric error $>0.1$ mag were removed. 
Then, a background subtracted Ks and F160W (H-band) images from ZFOURGE were used with the seeing convolved from $0''.4$ to $0''.7$ to match the average Keck seeing. 
Rectangular apertures, which resemble the slits with various heights were overlaid in the images to integrate the total counts within each aperture. 
Any apertures that contain multiple objects or had bright sources close to the slit edges were removed. 
Integrated counts were used to calculate the photometric magnitude to compare with the spectroscopy. 
A slit-box aligned with similar PA to the respective mask with a size of $0''.7 \times 2''.8$ was found to give the best balance between the spectrophotometric comparison and the number of available slits with good photometry per mask.

Next, the median offset between the magnitudes from photometry and spectroscopy were calculated by selecting objects with a photometric magnitude less than 24 in the respective filters.  
This offset was used as the scaling factor and was applied to all spectra in the mask. Typical offsets for K and H bands were $\sim \pm0.1$ mag. 
We then performed 1000 iterations of bootstrap re-sampling of the objects in each mask to calculate the scatter of the median values. We parametrized the scatter using normalized absolute median deviation (\NMAD) which is defined as 1.48$\times$ median absolute deviation. 
The median \NMAD\ scatter in K and H-bands for these offsets are \around0.1 and \around0.04 mag, respectively.

\begin{figure}
\includegraphics[scale=0.57]{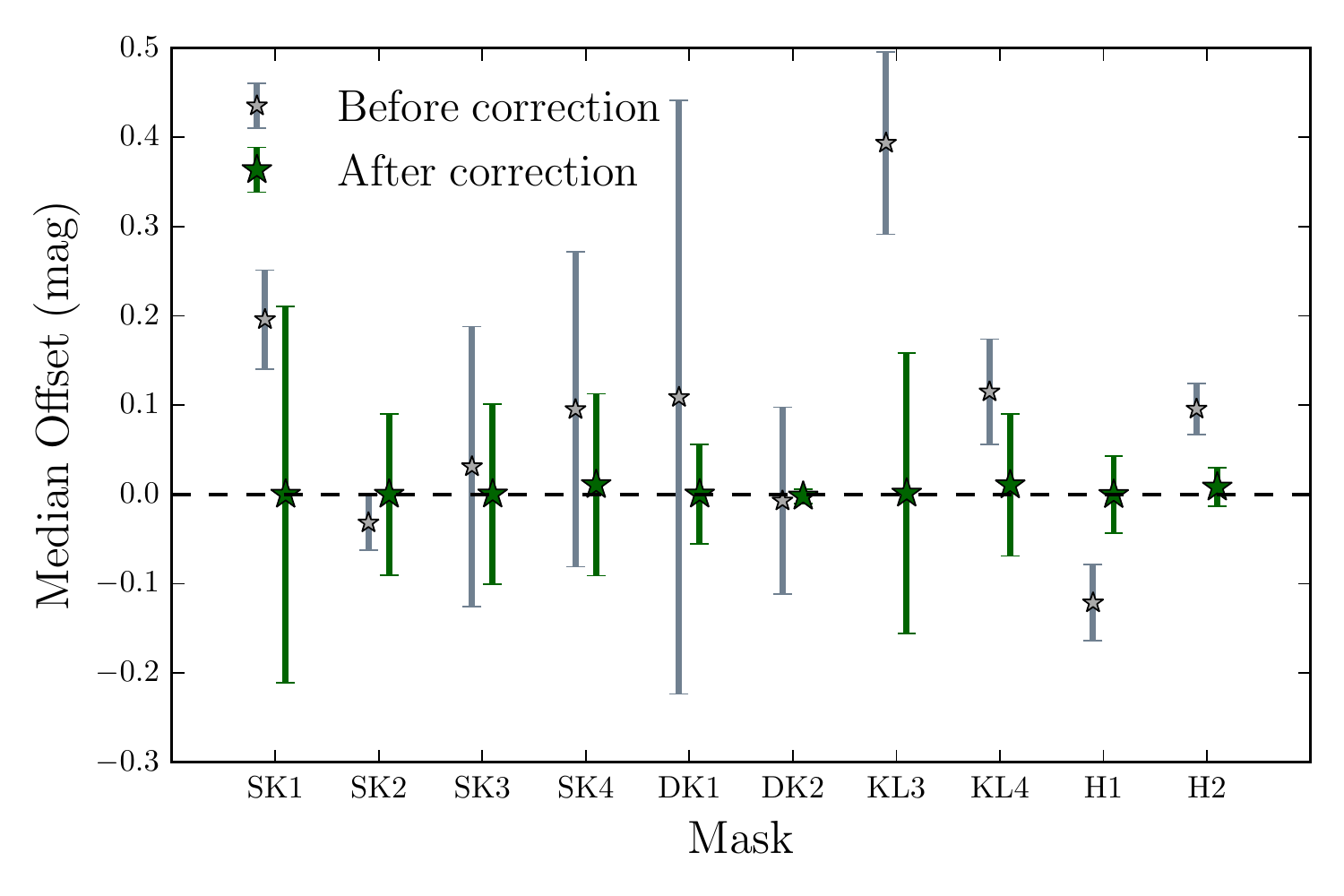}
\includegraphics[scale=0.57]{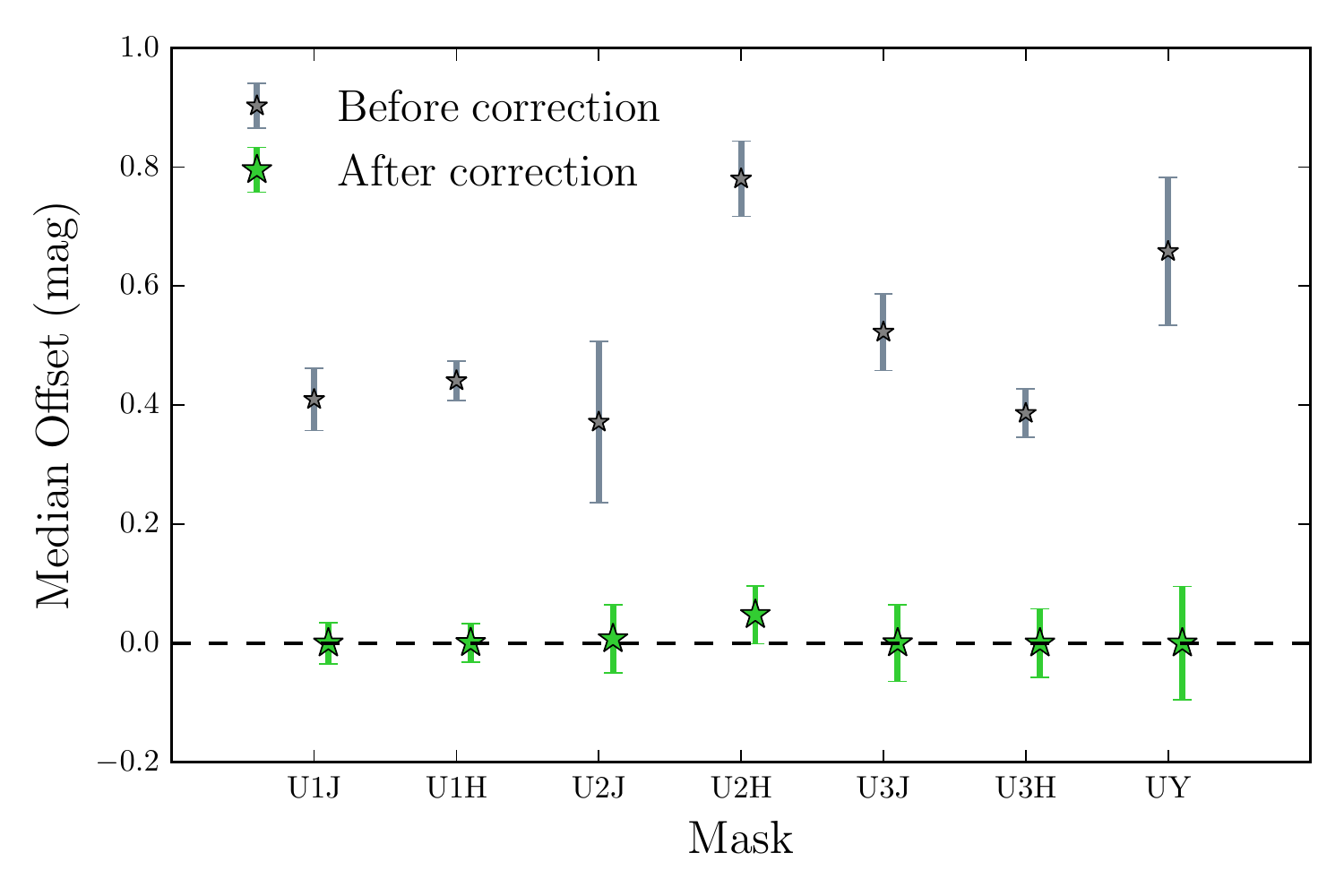}
\caption{ Spectrophotometric calibration of the ZFIRE masks. The median offsets between spectroscopic flux and the photometric flux before and after the scaling process is shown in the figure. Filter names correspond to the names in Table \ref{tab:observing_details}.
The grey stars denote the median offsets for the standard star flux calibrated data before any additional scaling is applied. 
The median mask sensitivity factors are applied to all objects in the respective masks to account for slit loss. The green stars show the median offsets after the flux corrections are applied. The errors are the \NMAD\ scatter of the median offsets calculated via bootstrap re-sampling of individual galaxies. 
{\bf Top:} all COSMOS masks. Photometric data are from a slit-box aligned with similar PA to the respective mask with a size of $0''.7 \times 2''.8$.
{\bf Bottom:} all UDS masks. Photometric data are total fluxes from UKIDSS.}
\label{fig:scaling_values}
\end{figure}

The median offset values per mask before and after the scaling process with its associated error is shown in the top panel of Figure \ref{fig:scaling_values}. Typical offsets are of the order of $\lesssim0.1$ mag which are consistent with expected values of slit loss
and the small amount of cloud variation seen during the observations. 
The offset value after the scaling process is shown as green stars with its bootstrap error. 

The scaling factor was applied as a multiple for the flux values for the 2D spectra following Equation \ref{eq:scale_single_masks},  
\begin{subequations}
\label{eq:scale_single_masks}
\begin{equation}
F_i = f_i \times \hbox{scale}_{\mathrm{mask}} 
\end{equation}
\begin{equation}
\Sigma_i = \sigma_i \times \hbox{scale}_{\mathrm{mask}}
\end{equation}
\end{subequations}
where $\mathrm{f_i}$ and $\mathrm{\sigma_i}$ are, respectively, the flux and error per pixel before scaling and scale$\mathrm{_{mask}}$ is the scaling factor calculated.

1D spectra are extracted using the same extraction aperture as before. 
The bootstrap errors after the scaling process is \around0.08 mag (median) for the COSMOS field,  which is considered to be the final uncertainty of the spectrophotometric calibration process.  Once a uniform scaling was applied to all the objects in a given mask, the agreement between the photometric slit-box magnitude and the spectroscopic magnitude increased.

As aforementioned, if an object was observed in multiple masks in the same filter, first the corresponding mask scaling factor was applied and then co-added optimally in 1D such that a higher weight was given to the objects, which came from a mask with a lower scaling value (i.e. better transmission). The procedure is shown in equation \ref{eq:1D_scale_and_coadd}, 
\begin{subequations}
\label{eq:1D_scale_and_coadd}
\begin{equation}
 F_i = \frac{\sum\limits_{j=1}^n (P_j/\sigma_{ji})^2 (F_{ji}/P_j)}{\sum\limits_{j=1}^n(P_j/\sigma_{ji})^2 }
\end{equation}
\begin{equation}
\sigma_i^2 = \frac{\sum \limits_{j=1}^n \big\{(P_j/\sigma_{ji})^2 (F_{ji}/P_j)\big\}^2} {\big\{\sum \limits_{j=1}^n(P_j/\sigma_{ji})^2\big\}^2 }
\end{equation}
\end{subequations}
where $P$ is the 1/scale value, $i$ is the pixel number, and $j$ is the observing run. Further examples for the spectrophotometric calibration process are shown in Appendix \ref{sec:MOSFIRE cals}.

\subsubsection{UDS Legacy Field}

The filter response functions for WFCAM \citep{Casali2007} was used to integrate the total flux in each of the 1D calibrated spectra in the UDS field. 
The {\it total} photometric fluxes from the UKIDSS catalogue were used to compare with the integrated flux from the spectra since images were not 
available to simulate slit apertures. 
To calculate the median offset a magnitude limit of 23 was used.  
This magnitude limit was brighter than the limit used for COSMOS data since the median photometric magnitude of the UDS data are \around0.5 mag brighter than COSMOS.

Typical median offsets between photometric and spectroscopic magnitudes were \around0.4 magnitude.
the lower panel of Figure \ref{fig:scaling_values} shows the median offset values per mask before and after the scaling process with its associated error. 
The median of the bootstrap errors for the UDS masks after scaling is \around0.06 mag. 

Comparing with the COSMOS offsets, the UDS values are heavily biased toward a positive offset. 
This behaviour is expected for UDS data because the broadband total fluxes from the UKIDSS data are used, and therefore the flux expected from the finite MOSFIRE slit should be less than the total flux detected from UKIDSS. 
Since UDS objects are not observed in multiple masks in the same filter, only Equation \ref{eq:scale_single_masks} is applied to scale the spectra. \\

\subsection{Measuring Emission Line Fluxes}
\label{sec:line_fits}

A custom made IDL routine was used to fit nebular emission lines on the scaled 1D spectra. This was done by fitting Gaussian profiles to user defined emission lines. 
The code identifies the location of the emission line in wavelength space and calculates the redshift.  

In emission line fitting, if there were multiple emission lines detected for the same galaxy in a given band, the line centre and velocity width were kept to be the same. Emission lines with velocity structure were visually identified and were fit with multiple component Gaussian fits.  
If the line was narrower than the instrumental resolution, the line width was set to match the instrument resolution.  
The code calculated the emission line fluxes (f) by integrating the Gaussian fits to the emission lines. The corresponding error for the line fluxes ($\sigma$(f)) were calculated by integrating the error spectrum using the same Gaussian profile. The code further fits a 1$\sigma$ upper level for the flux values (f$_{limit}$). The signal-to-noise ratio (SNR) of the line fluxes was defined as the line flux divided by the corresponding error for the line flux.

%%%%%%%%%%%%%%%%%%%%%%%%%%%%%%%%%%%%%%%%%%%%%%%%%%%%%%%%%%%%%%%%%%%%%%%%%%%%%%%%%%%%%%%%%%%%%%%%%%%%%%%%

\section{Properties of ZFIRE Galaxies}
\label{sec:results}

\subsection{Spectroscopic Redshift Distribution}
\label{sec:Q flags}

Using nebular emission lines, 170 galaxy redshifts were identified for the COSMOS sample and 62 redshifts were identified for the UDS field. 
A combination of visual identifications in the 2D spectra and emission line fitting procedures explained in Section \ref{sec:line_fits} were used to identify these redshifts. 
The redshift quality is defined using three specific flags: 
\begin{itemize}
\item Q$_z$ Flag $=$ 1: These are objects with no line detection with SNR $<5$. These objects are not included in our final spectroscopic sample. 
\item Q$_z$ Flag $=$ 2: These are objects with one emission line with SNR $>$ 5 and a $|$\zspec $-$ \zphoto $|$ $> 0.2$.
\item Q$_z$ Flag $=$ 3: These are objects with more than one emission line identified with SNR $>$ 5 or one  emission line identified with SNR $>$ 5 with a $|$\zspec $-$ \zphoto $|$ $< 0.2$.
\end{itemize}

\begin{figure}
\includegraphics[scale=0.85]{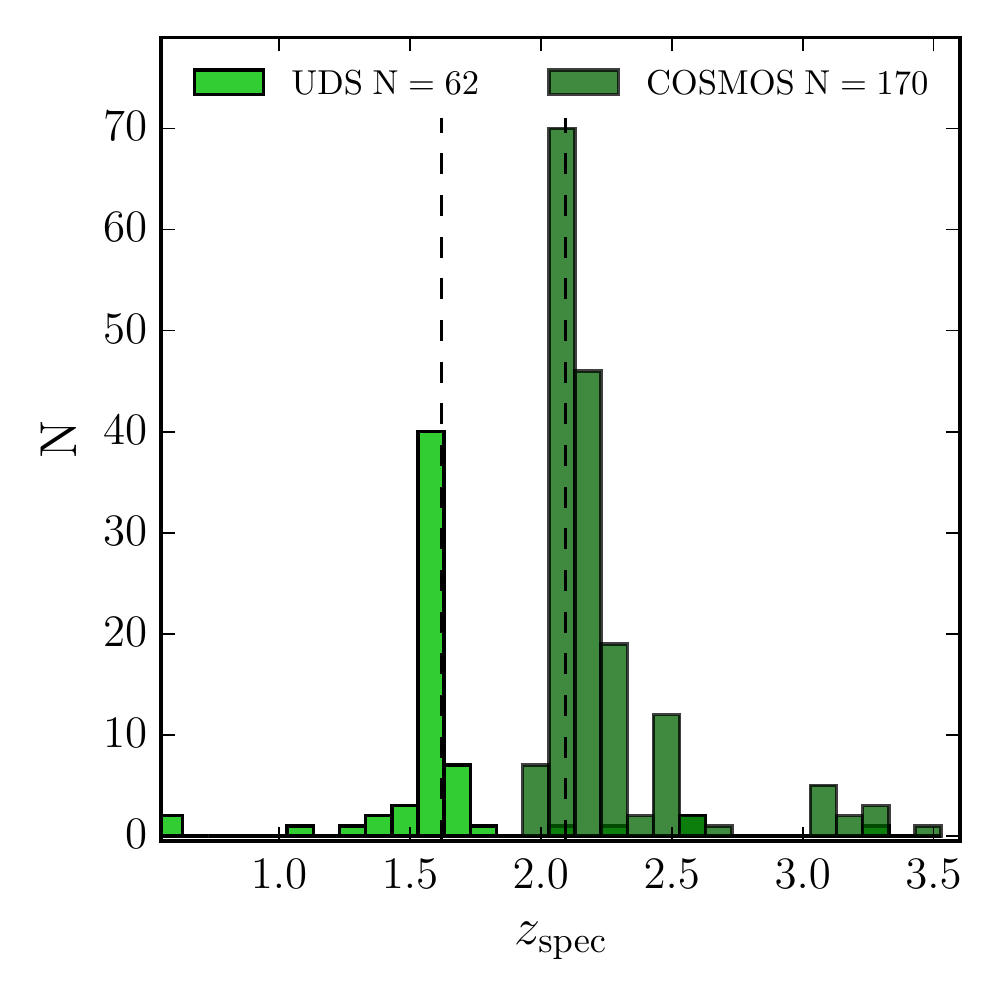}
\caption{Redshift distribution of the ZFIRE data release. All detected galaxies with Q$_z$=2 and Q$_z$=3 from UDS (light green) and COSMOS (dark green) are shown in the figure. The two dashed vertical lines at x=1.620 and x=2.095 shows the location of the IRC 0218 cluster \citep{Tran2015} and the COSMOS cluster \citep{Yuan2014}, respectively.}
\label{fig:zspec}
\end{figure}

The redshift distribution of all ZFIRE\ Q$_z$=2 and Q$_z$=3 detections are shown in Figure \ref{fig:zspec}. 62 galaxy redshifts were detected in the UDS field, out of which 60 have a Q$_z$ of 3 and 2 have a Q$_z$ of 2. Similarly, for the COSMOS field, there are 161  Q$_z$=3 objects and 9 Q$_z$=2 objects. 

The systematic error of the redshift measurement was estimated by comparing Q$_z$=3 objects with a SNR $>$ 10 in both H and K-bands in the COSMOS field. 
\citet{Yuan2014} showed that the agreement between the redshifts in the two bands is $\Delta z$(median) = 0.00005 with a rms of $\Delta z$(rms) = 0.00078. Therefore, the error in redshift measurement is quoted as $\Delta z$(rms) = 0.00078/$\sqrt 2$= 0.00055, which corresponds to  $\sim\mathrm{53km~s^{-1}}$ at $z=2.1$.
This is $\sim$2 times the spectral resolution of MOSFIRE, which is $\sim\mathrm{26km~s^{-1}}$ \citep{Yuan2014}. However, for the \citet{Yuan2014} analysis barycentric corrections were not applied to the redshifts and H and K masks were observed on different runs. Once individual mask redshifts were corrected for barycentric velocity, the rest-frame velocity uncertainty  decreased to $\sim\mathrm{15km~s^{-1}}$.

\begin{figure*}
\includegraphics[scale=0.85]{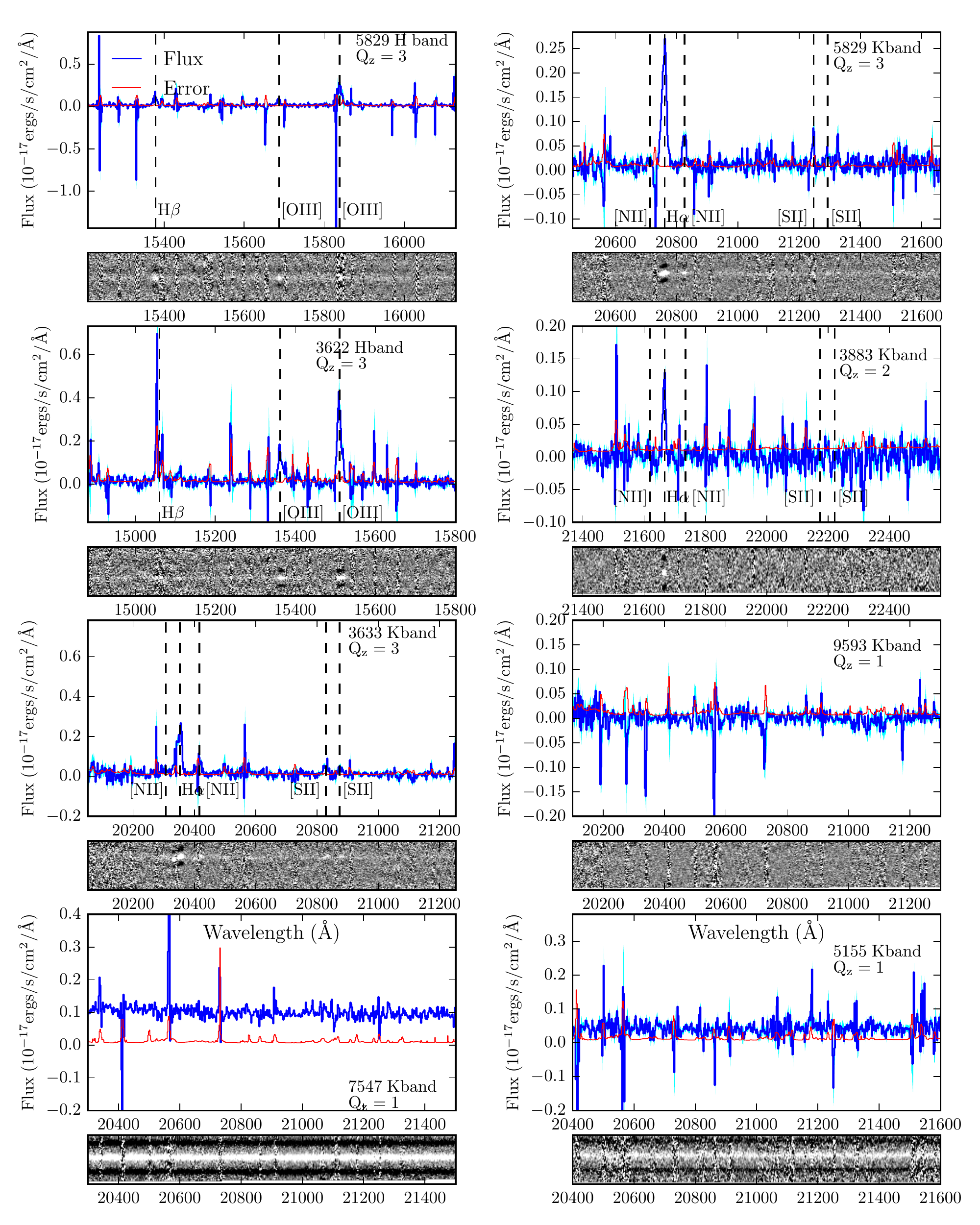}
\caption{Example MOSFIRE H and K-band spectra from the COSMOS field. 
In the 1D spectra, the flux is shown in blue and the corresponding error in red. The 1$\sigma$ scatter of the flux value parametrized by the error level is highlighted around the flux value in cyan.
Each 1D spectra are accompanied by the corresponding 2D spectra covering the same wavelength range. 
Each panel shows the name of the object, the wavelength it was observed in, and the redshift quality of the object. Vertical dashed lines show where strong optical emission lines ought to lie given the spectroscopic redshift.}
\label{fig:spectra}
\end{figure*}

A few example spectra are shown in Figure \ref{fig:spectra}. Object 5829 is observed in both H and K-bands with strong emission lines detected in both instances. Object 3622 has strong H-band detections, while 3883 has only one emission line detection. Therefore, 3883 is assigned a Q$_z$ of 2. The 2D spectrum of object 3633 shows two emission line detections around \Halpha\ at different y pixel positions, which occur due to multiple objects falling within the slit. Object 9593 shows no emission line or continuum detection.
Objects 7547 and 5155 have strong continuum detections with no nebular emission lines. These galaxies were selected to be the BCGs of the D and A substructures by \citet{Yuan2014} and \citet{Spitler2012}, respectively, and have absorption line redshifts from \citet{Belli2014}.

The ZFIRE data release catalogue format is given in Table \ref{tab:catalogue}.
An overview of the data presented is provided in the table, along with the 1D spectra, which is available online at \url{zfire.swinburne.edu.au}.
Galaxy stellar mass and dust extinction values are from ZFOURGE, but for Q$_z>1$ galaxies these values are rederived using the spectroscopic redshifts with FAST.
The ZFIRE-COSMOS galaxy sample comprises  both field and cluster galaxies selected in the Ks band with an 80\% mass completeness down to $\log_{10}($\mass$)>9.30$ (Figure \ref{fig:detection_limits}). 

The survey selection for this data release was done using the ZFOURGE internal catalogues, and therefore the results presented here onwards could vary slightly from the ZFOURGE public data release.  
For the 2016 ZFOURGE public data release, the catalogue was upgraded by including pre-existing public K-band imaging for the source detection image.
This increased the amount of galaxies in the COSMOS field by \around 50\%, which was driven by the increase of fainter smaller mass galaxies. In Appendix \ref{sec:ZFOURGE comparison}, a comparison between the internal ZFOURGE catalogue and the public data release version is shown.

\begin{deluxetable*}{ || l | l || }
\tabletypesize{\scriptsize}
\tablecaption{ The ZFIRE\ v1.0 data release }
%\tablenotemark{a}
%\tablenotetext{a}{}
\tablecomments{ This table presents an overview of the data available online.  
All galaxy properties and nebular emission line values of the galaxies targeted by ZFIRE between 2013 to 2015 are released with this paper.
\label{tab:catalogue}}
\tablecolumns{2}
\tablewidth{0pt} 
\tablewidth{0pt}
\startdata
\hline 
&   \\ 
ID												&	Unique ZFIRE\ identifier.								\\ [+1ex]
RA 												&	Right ascension (J2000)		\\ [+1ex]
DEC												&	Declination (J2000)  		\\ [+1ex]
Field											&	COSMOS or UDS			\\ [+1ex]
\Ks \tablenotemark{a}							&  \Ks\ magnitude from ZFOURGE \\ [+1ex]
$\mathrm{\sigma}$\Ks			    			&	Error in \Ks\ magnitude.    								\\ [+1ex]
\zspec					    					&	ZFIRE\ spectroscopic redshift. 							\\ [+1ex]
$\sigma$(\zspec)								&	Error in spectroscopic redshift.					\\ [+1ex]
Q$_z$											& 	ZFIRE\ redshift quality flag (see Section \ref{sec:Q flags})      \\ [+1ex]
Cluster\tablenotemark{b}						&   Cluster membership flag   		\\ [+1ex]
Mass\tablenotemark{c}							&	Stellar mass  from FAST.    		\\ [+1ex]
Av              								&	Dust extinction from  FAST.     	\\ [+1ex]
AGN\tablenotemark{d}							&	AGN flag. \\[+1ex]
\Halpha	\tablenotemark{e}						&   Emission line \Halpha\ flux from ZFIRE\ spectrum   \\ [+1ex]
$\sigma$(\Halpha)\tablenotemark{f}				&	Error in \Halpha\ flux.      						\\ [+1ex]
 \Halpha$_{\mathrm{limit}}$\tablenotemark{g}	&	1$\sigma$ upper limit for the \Halpha\ flux detection   \\ [+1ex]
 \NII \tablenotemark{e}							&	Emission line \NII\ flux (6585\AA) from ZFIRE\ spectrum   \\ [+1ex]
$\sigma$(\NII) \tablenotemark{f}				&	Error in \NII\ flux     							\\ [+1ex]
\NII$_{\mathrm{limit}}$\tablenotemark{g}		&	1$\sigma$ upper limit for the \NII\ flux detection   	\\ [+1ex]
\Hbeta \tablenotemark{e}						&	Emission line \Hbeta\ flux from ZFIRE\ spectrum 	\\ [+1ex]
$\sigma$(\Hbeta) \tablenotemark{f}				&	Error in \Hbeta\ flux 								\\ [+1ex]
\Hbeta$_{\rm limit}$\tablenotemark{g}			&	1$\sigma$ upper limit for the \Hbeta\ flux detection	\\ [+1ex]
\OIII \tablenotemark{e}							&	Emission line \OIII\ flux (5008\AA) from ZFIRE\ spectrum 	\\ [+1ex]
$\sigma$(\OIII)	\tablenotemark{f}				&	Error in \OIII\ flux 								\\ [+1ex]
\OIII$_{\rm limit}$\tablenotemark{g}			&	1$\sigma$ upper limit for the \OIII\ flux detection.  \\
& 

\tablenotetext{a}{Magnitudes are given in the AB system.}	
\tablenotetext{b}{Cluster=True objects are spectroscopically confirmed cluster members in either the COSMOS \citep{Yuan2014} or UDS \citep{Tran2015} fields.}
\tablenotetext{c}{Stellar mass (M$_*$) is in units of $\mathrm{log_{10}}$\msol\ as measured by FAST.}
\tablenotetext{d}{AGNs are flagged following \citet{Cowley2016} and/or \citet{Coil2015} selection criteria.}
\tablenotetext{e}{The nebular emission line fluxes (along with errors and limits) are given in units of $10^{-17}ergs/s/cm^2$.}
\tablenotetext{f}{The error of the line fluxes  are from the integration of the error spectrum within the same limits used for the emission line extraction.}
\tablenotetext{g}{Limits are $1\sigma$ upper limits from the Gaussian fits to the emission lines.}
\end{deluxetable*}

\subsection{Spectroscopic Completeness}
\label{sec:completeness}

The main sample of galaxies in the COSMOS field were selected in order to include \Halpha\ emission in the MOSFIRE K-band, which corresponds to a redshift range of $1.90<$\zphoto$<2.66$. Due to multiple objects in the slits and object priorities explained in Section \ref{sec:mask_design}, there were nine galaxies  outside this redshift range. 

\begin{figure}[h!]
\includegraphics[trim = 10 0 5 5, clip, scale=0.9]{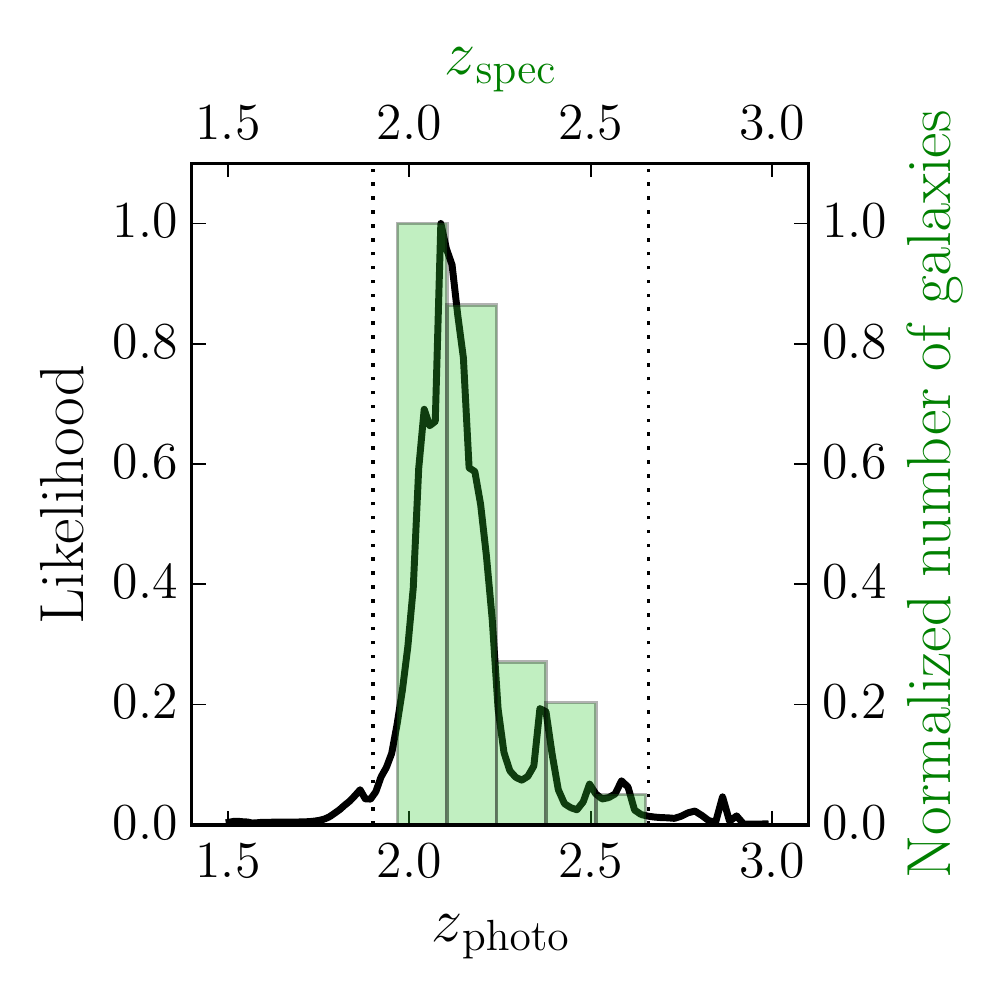}
\caption{Stacked probability distribution functions of the photometric redshifts for galaxies targeted in the ZFIRE-COSMOS field (shown by the black solid line). 
The black dotted lines show the redshift limits for \Halpha\ detection in the K-band. 
The wavelength coverage is corrected by the slit positions for each of the galaxies and the total probability that falls within the detectable range is calculated to be \around73\%. 
The actual \Halpha\ detection in the COSMOS field is \around73\%.
The bias toward z=2.1 is due to the object priorities weighting heavily towards the cluster galaxies. 
The green histogram shows the distribution of \zspec\ values for galaxies with \Halpha\ detections in K-band in the COSMOS field. 
}
\label{fig:completeness}
\end{figure}

We assess completeness against an expectation computed using the photometric redshift likelihood functions ($P(z)$) from EAZY, i.e. the  expected number of galaxies with \Halpha\ within the bandpass in the ZFIRE-COSMOS sample, taking account of the slightly different wavelength coverage of each slit. 
There were 203 galaxies targeted in the K-band. Of the galaxies, 10 had spectroscopic redshifts that were outside the redshift range of interest ($1.90<$\zspec$<2.66$).
The remaining 193 $P(z)$s of the detected and non-detected galaxies were stacked. 
Figure \ref{fig:completeness} shows the average $P(z)$ of the stacked 193 galaxies. 
If the \Halpha\ emission line falls on a sky line, the emission line may not be detected. Therefore, in the $P(z)$ of each of the galaxies' sky line regions parametrized by the MOSFIRE K-band spectral resolution was masked out ($\pm$5.5\AA).  
We then calculate the area of the $P(z)$ that falls within detectable limits in K-band of the object depending on the exact wavelength range of each slit.  
Since each $P(z)$ is normalized to 1, this area gives the probability of an \Halpha\ detection in K-band for a given galaxy. 
The probability to detect all 193 galaxies is calculated to be \around73\%. 
141 galaxies are detected with \Halpha\ SNR $>5$ which is a \around73\% detection rate. 
As seen by the overlaid histogram in Figure \ref{fig:completeness}, the detected redshift distribution of the ZFIRE-COSMOS sample is similar to the expected redshift distribution from $P(z)$.
%Invoking a lower detection threshold of  \Halpha\ SNR $>3$, we get 135 \Halpha\ detected galaxies increasing the detection rate to ~73\%. conf=5 galaxy is removed: shows a positive \Halpha\ detection due to high continuum level but there is no evident emission line.

\begin{figure*}
\includegraphics[trim = 10 0 5 5, clip, scale=0.6]{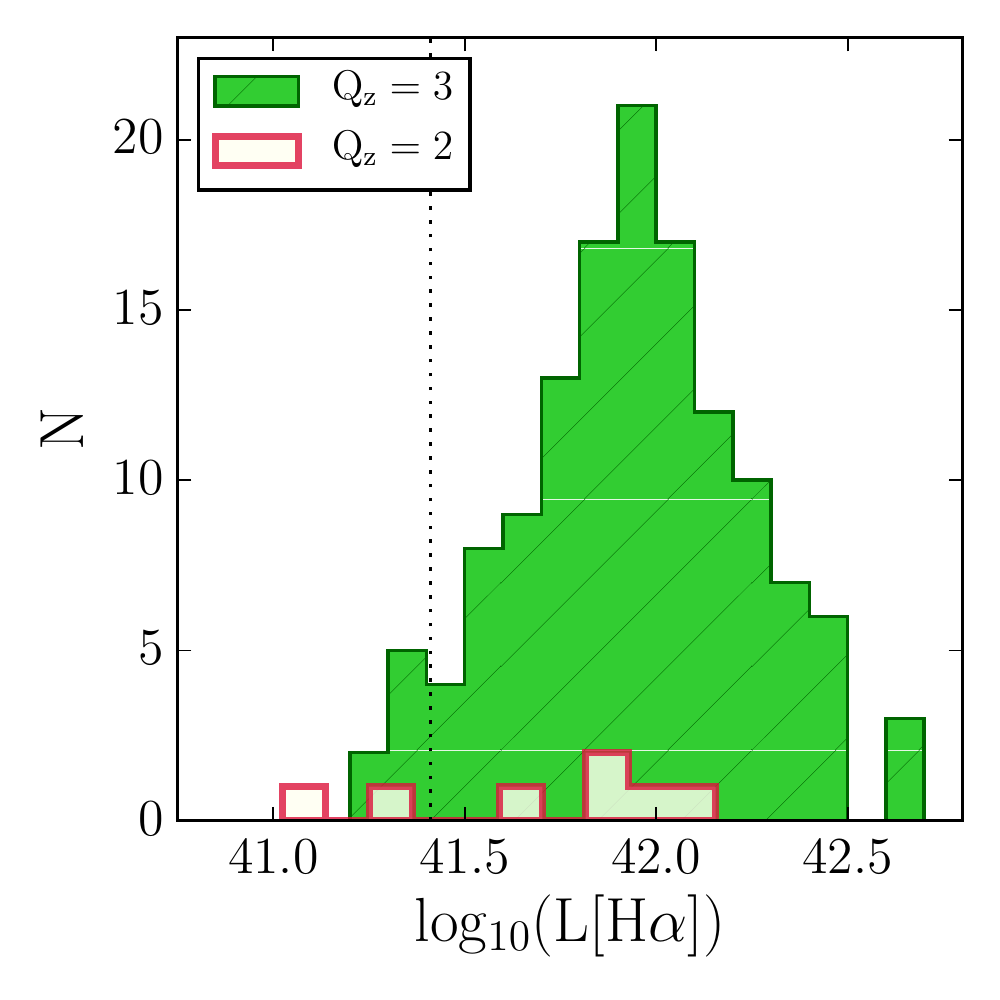}
\includegraphics[trim = 10 0 5 5, clip, scale=0.6]{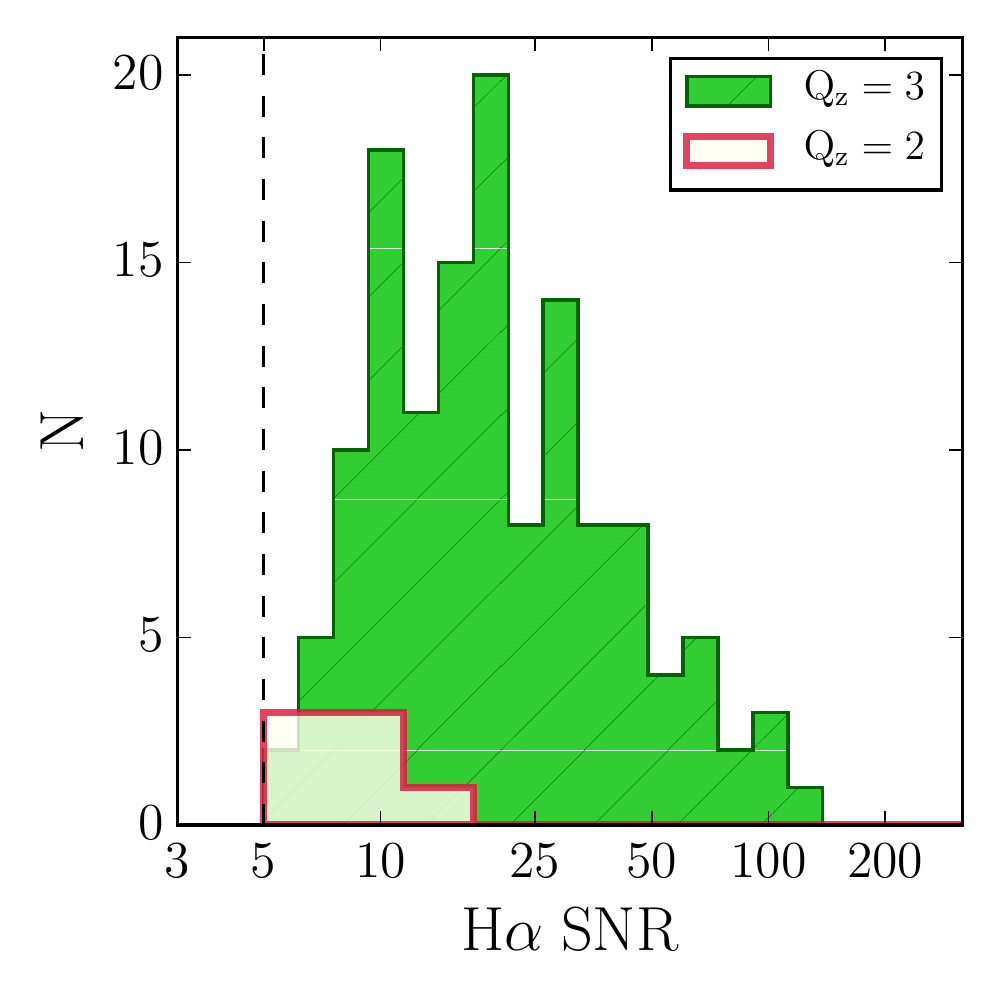}
\includegraphics[trim = 10 0 5 5, clip, scale=0.6]{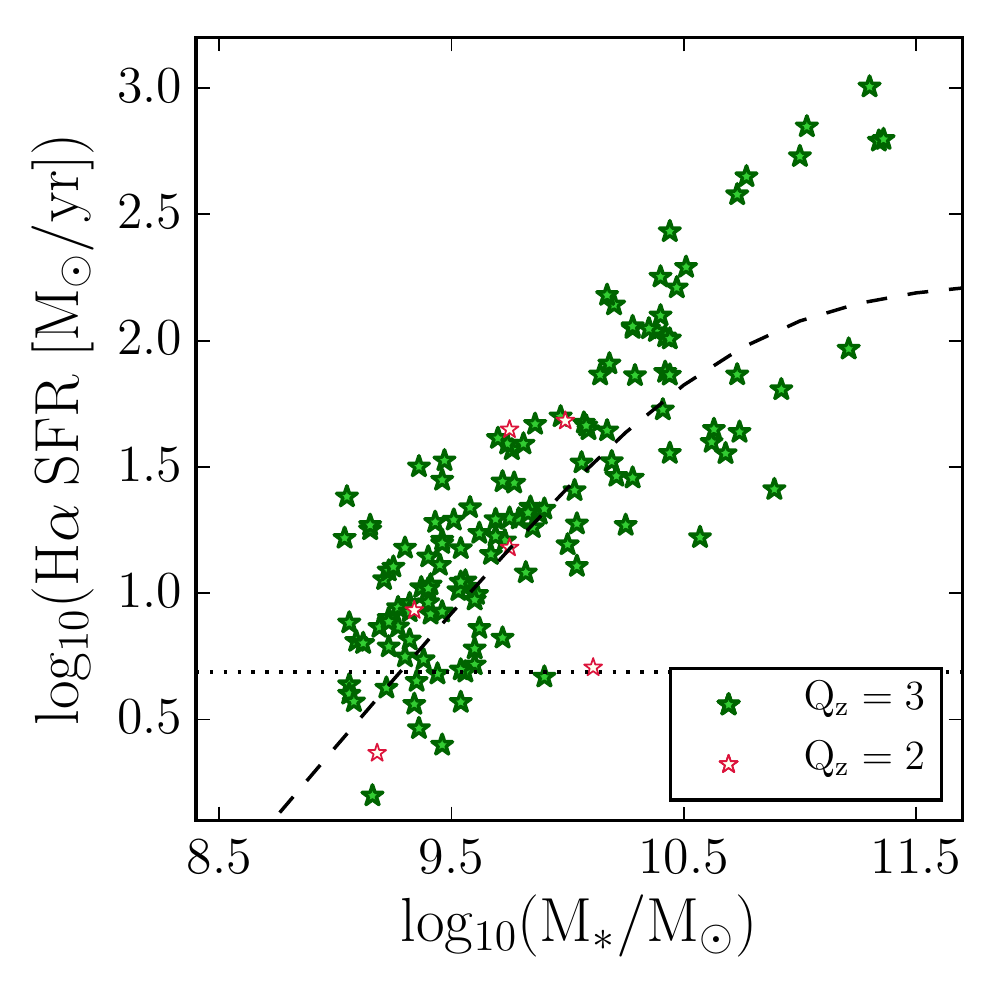}
\caption{ {\bf Left:} the distribution of \Halpha\ luminosity of all ZFIRE-COSMOS galaxies in log space. The green histogram (with horizontal lines) is for galaxies with a quality flag of 3, while the ivory histogram is for galaxies with a quality flag of 2.
The vertical dotted line is the \Halpha\ SFR for a typical \Halpha\ SNR of \around 5 at z=2.1. 
{\bf Middle:} similar to the left figure, but the distribution of \Halpha\ SNR of all ZFIRE-COSMOS detections are shown. The dashed vertical line is SNR = 5, which is the \Halpha\ detection threshold for ZFIRE. 
{\bf Right:} the \Halpha\ SFR vs. stellar mass distributions for the objects shown in the left histograms. The stellar masses and dust extinction values are derived from FAST. The  dashed line is the star-forming main sequence from \citet{Tomczak2014}. The horizontal dotted line is the \Halpha\ SFR for a typical \Halpha\ SNR of \around 5 at z=2.1.
}
\label{fig:Halpha}
\end{figure*}

Figure \ref{fig:Halpha} shows the \Halpha\ luminosity (left) and SNR distribution (middle) of the ZFIRE-COSMOS galaxies with \Halpha\ detections. 
The detection threshold is set to SNR $\geq$ 5 which is shown by the vertical dashed line in the centre panel. There are 134 galaxies in the Q$_{z}$=3 sample, 7 in the Q$_{z}$=2 sample. 

The \Halpha\ luminosity in Figure \ref{fig:Halpha} (left panel) is peaked $\sim10^{42}$ergs/s. From the SNR distribution it is evident that the majority of galaxies detected have a \Halpha\ SNR $>10$, with the histogram peaking \around SNR of 20.
Normally astronomical samples are dominated by low SNR detections near the limit. 
It is unlikely that objects with SNR$<$20 are missed. Our  interpretation of  this distribution is that because  the sample is mass-selected the drop off of low flux \Halpha\ objects is because the region below the stellar mass-SFR main sequence \citep{Tomczak2014} at $z\sim 2$ is probed. 
This is shown in Figure \ref{fig:Halpha} where we make a simple conversion of \Halpha\ to SFR assuming the \citet{Kennicutt1998}  conversion and stellar extinction values from FAST
which we convert to nebula extinction using the  \citet{Calzetti2000} prescription  with  $R_V=4.05$. 
It is indeed evident that the ZFIRE-COSMOS sample limits do probe the limits of the galaxies in the star-forming main sequence at $z\sim2$ with a $3\sigma$ \Halpha\ SFR detection threshold at $\sim4$ \msol/yr. A more detailed analysis of the \Halpha\ main sequence will be presented in a future paper (K. Tran et al. in preparation). 

%#################

\subsection{Magnitude and Stellar Mass Detection Limits}

\begin{figure}[h!]
\includegraphics[trim = 10 10 10 5, clip, scale=0.58]{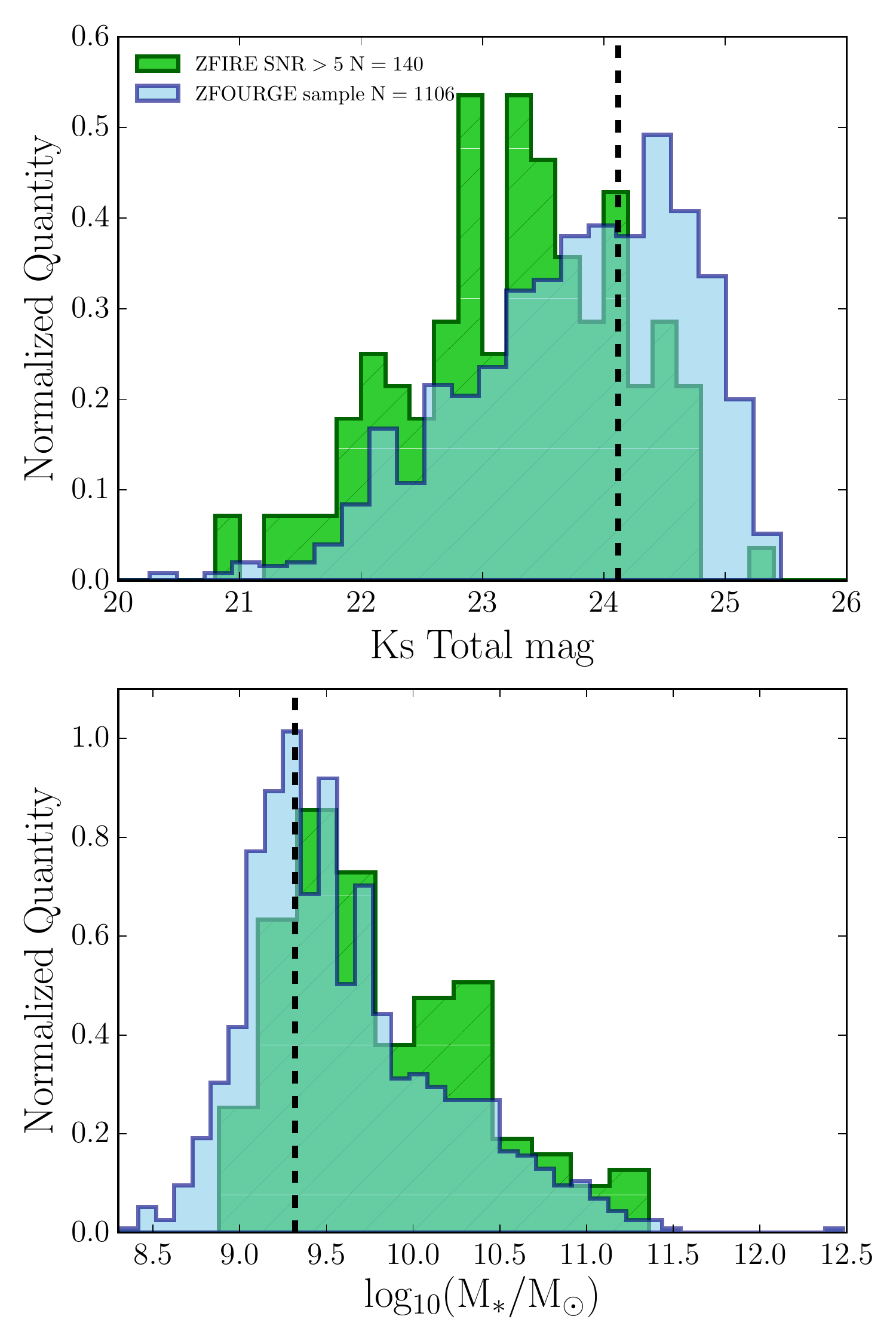}
\caption{The Ks magnitude and mass distribution of the $1.90<z<2.66$ galaxies from ZFOURGE (cyan) overlaid with the ZFIRE (green) detected sample for the COSMOS field. The ZFOURGE distribution is derived using the photometric redshifts and spectroscopic redshifts (when available). The ZFIRE histogram uses the spectroscopic redshifts. 
The histograms are normalized for area. UVJ quiescent galaxies (only 1 in ZFIRE) are removed from both the samples. 
{\bf Top:} Ks magnitude distribution. The black dashed line (Ks=24.11) is the limit in which 80\% of the detected sample lies below. 
{\bf Bottom:} stellar mass distribution of the galaxies in log space as a fraction of solar mass.  Masses are calculated using FAST and spectroscopic redshifts are used where available. The black dashed line ($\mathrm{Log}_{10}($\mass$)=9.3$) is the limit down to where the detected sample is 80\% mass complete.  
}
\label{fig:detection_limits}
\end{figure}

The ZFIRE-COSMOS detection limits in Ks magnitude and stellar mass are estimated using ZFOURGE photometry. 
Out of 141 objects with \Halpha\ detections (Q$_{z}$=2 or Q$_{z}$=3) and $1.90<$\zspec$<2.66$, galaxies identified as UVJ quiescent are removed since the spectroscopic sample does not significantly sample these (see Section \ref{sec:UVJ}). The remaining sample comprises 140 UVJ blue (low dust attenuation) and red (high dust attenuation) star-forming galaxies. 
Similarly, galaxies from the ZFOURGE survey are selected with redshifts between $1.90<$\zspec$<2.66$ and all UVJ quiescent galaxies are removed. The Ks magnitude and the stellar mass distributions of the remaining 1106 ZFOURGE galaxies with the selected ZFIRE sample are compared in Figure \ref{fig:detection_limits}.  

The top panel of Figure \ref{fig:detection_limits} demonstrates that the \Halpha\ detected galaxies reach Ks$>$24. 
80\% of the detected ZFIRE-COSMOS galaxies have Ks$\leq$24.11. The ZFOURGE input sample reaches deeper to 
Ks$\leq$24.62 (80\%-ile). The photometric detection completeness limit of ZFOURGE is discussed in detail in Straatman
et al. (2014), but we note that at $K=24.62$, 97\% of objects are detected. It is important to understand if the distribution in Ks of the spectroscopic sample is biassed relative to the photometric sample. A two-sample K-S test for Ks$\leq$24.1 is performed to find a $p$ value of 0.03 suggesting that there is no significant bias between the samples. 

Similarly, the mass distribution of the \Halpha\ detected sample is investigated in the bottom panel of Figure  \ref{fig:detection_limits}. Galaxies are detected down to $\log_{10}($\mass$)\sim9$.
80\% of the \Halpha\ detected galaxies have a stellar masses $\log_{10}($\mass$)>9.3$. A K-S test on the two distributions for galaxies $\log_{10}($\mass$)>9.3$ gives a $p$ value of 0.30 and therefore, similar to the Ks magnitude distributions, the spectroscopic sample shows no bias in stellar mass compared to the ZFOURGE photometric sample. 

This shows that the ZFIRE-COSMOS detected sample of UVJ star-forming galaxies has a similar distribution in magnitude and stellar mass as the ZFOURGE distributions except at the very extreme ends. 
Removing UVJ dusty galaxies from the star-forming sample does not significantly change this conclusion. 

A final test is to evaluate the photometric magnitude at which continuum emission in the spectra can be typically detected. To estimate this, a constant continuum level is fit to blank sky regions across the whole $K$-band spectral range. This shows that  the $2\sigma$ spectroscopic continuum detection limit for the ZFIRE-COSMOS sample is Ks$\simeq 24.1$ ($0.05 \times 10^{-17} \mathrm{erg/s/cm^2/\AA}$). More detailed work on this will be presented in the IMF analysis (T. Nanayakkara et al., in preparation).

\subsection{Rest frame UVJ colours}
\label{sec:UVJ}

The rest-frame UVJ colours are used to assess the stellar populations of the detected galaxies. 
In rest frame U$-$V and V$-$J colour space, star-forming galaxies and quenched galaxies show strong bimodal dependence \citep{Williams2009}. Old quiescent stellar populations with strong 4000\AA\ and/or Balmer breaks show redder U$-$V colours and bluer V$-$J colours, while effects from dust contribute to redder V$-$J colours. 

\begin{figure}[h!]
\includegraphics[trim=12.5 10 0 0, clip, scale=0.625]{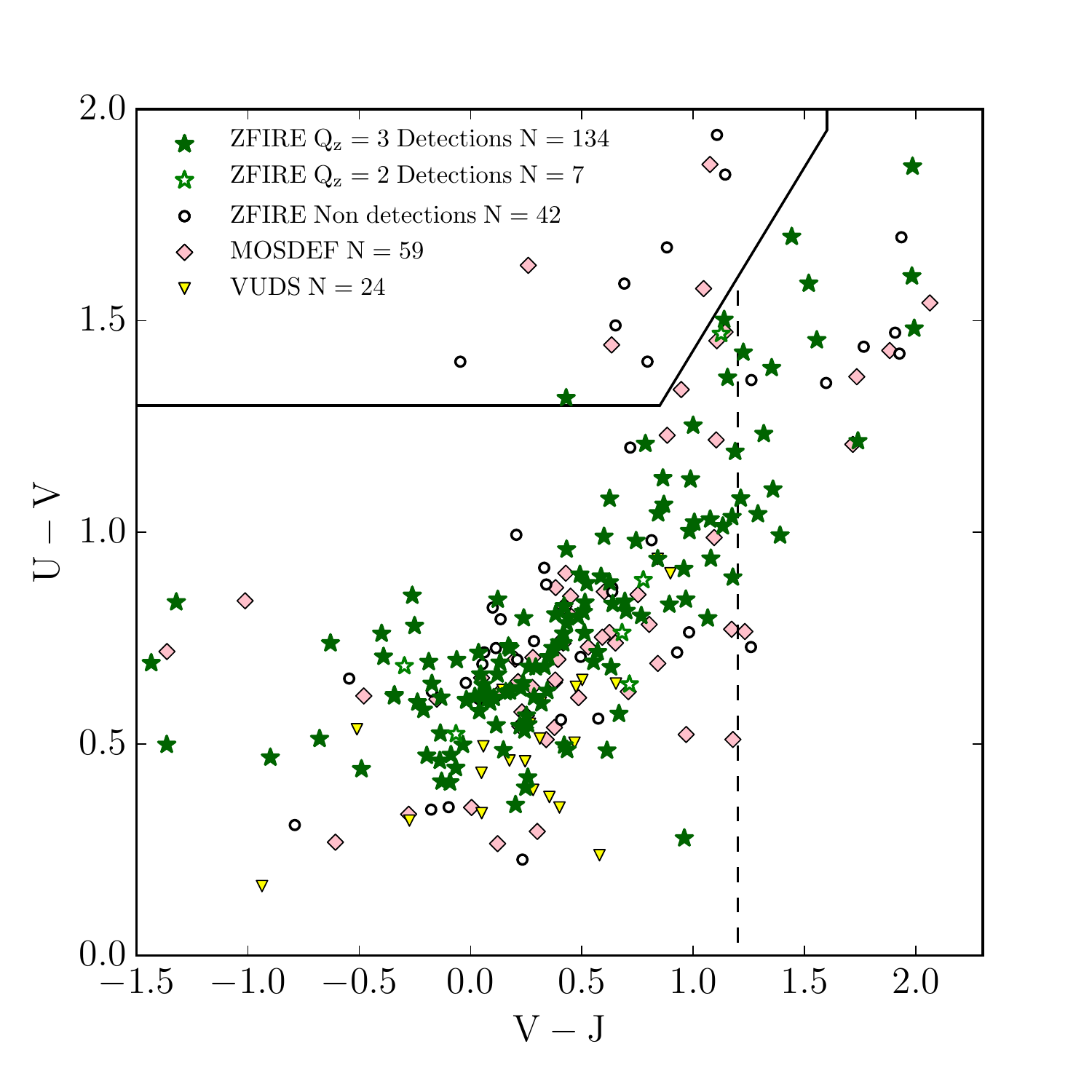}
\caption{ The rest frame UVJ diagram of the ZFIRE-COSMOS sample with redshifts $1.90<z<2.66$. 
Quiescent, star-forming, and dusty star-forming galaxies are selected using \citet{Spitler2014} criteria. 
The green stars are ZFIRE detections (filled$\rightarrow$Q$_z=3$, empty$\rightarrow$Q$_z=2$) and the black circles  are the non-detections. 
Pink diamonds and yellow triangles are MOSDEF and VUDS detected galaxies respectively, in the same redshift bin with matched ZFOURGE counterparts.  
Rest frame colours are derived using spectroscopic redshifts where available. 
}
\label{fig:UVJ}
\end{figure}

Figure \ref{fig:UVJ} shows the UVJ selection of the COSMOS sample, which lies in the redshift range between $1.99<$\zspec$<2.66$. 
The selection criteria are adopted from \citet{Spitler2014} and are as follows.
Quiescent galaxies are selected by (U$-$V)$>$1.3 , (V$-$J)$<$1.6, (U$-$V) $>$ 0.867$\times$(V$-$J)$+$0.563. 
Galaxies which lie below this limits are considered to be star-forming. 
These star-forming galaxies are further subdivided into two groups depending on their dust content. Red galaxies with (V$-$J)$>$1.2 are selected to be dusty star-forming galaxies, which correspond to A$_{v}\gtrsim$1.6. Blue galaxies with (V$-$J)$<$1.2 are considered to be relatively unobscured. MOSFIRE detected galaxies are shown as green stars while the non-detections (selected using \zphoto\ values) are shown as black filled circles. 

The total sampled non-detections are \around23\% for this redshift bin. 
\around82\% of the blue star-forming galaxies and \around70\% of the dusty star-forming galaxies were detected, but only 1 quiescent galaxy was detected out of the potential 12 candidates in this redshift bin. 
Galaxies in the red sequence are expected to be quenched with little or no star formation and hence without any strong \Halpha\ features; therefore the low detection rate of the quiescent population is expected. \citet{Belli2014} has shown that \around8 hours of exposure time is needed to get detections of continua of quiescent galaxies with J\around22 using MOSFIRE. 
The prominent absorption features occur in the H-band at $z\sim2$. ZFIRE currently does not reach such integration times per object in any of the observed bands and none of the quiescent galaxies show strong continuum detections.  We note that this is a bias of the ZFIRE survey, which may have implications on the identification of weak star-forming and quiescent cluster members by \citet{Yuan2014}.

For comparison MOSDEF and VUDS detections in the COSMOS field with matched ZFOURGE candidates are overlaid in Figure \ref{fig:UVJ}. All rest-frame UVJ colours for the spectroscopic samples are derived from photometry using the spectroscopic redshifts.  The MOSDEF sample, which is mainly H-band selected, 
primarily includes star-forming galaxies independently of the dust obscuration level. 
VUDS survey galaxies are biased toward blue star-forming galaxies, which is expected because it is an optical spectroscopic survey. This explains why their spectroscopic sample does not include any rest-frame UVJ selected dusty star-forming or quiescent galaxies.

\subsection{Spatial distribution}
\label{sec:spatial}

\begin{figure*}
\includegraphics[trim = 10 20 10 5, clip, scale=1.00]{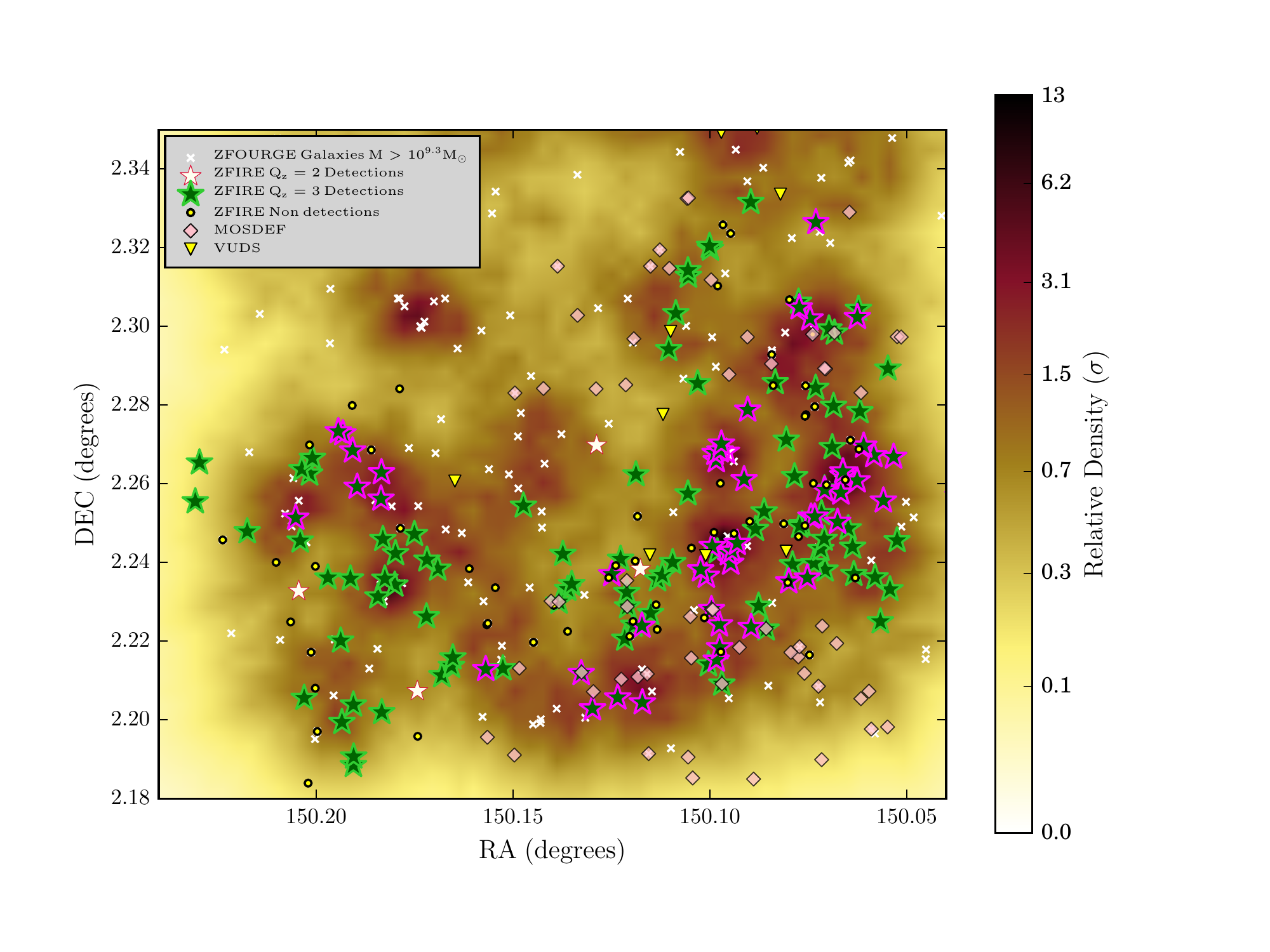}
\caption{ Spatial distribution of the ZFIRE-COSMOS sample. 
Galaxies that fall within $2.0<z<2.2$ are used to produce the underlying seventh nearest neighbour density map. The units are in standard deviations above the mean of redshift bins (see Section~\ref{sec:spatial}).The white crosses are the ZFOURGE galaxies with M$>$10$^{9.34}$\msol, which is the 80\% mass completeness of the ZFIRE\ detections. 
Spectroscopically detected galaxies with redshifts between $1.90<z_\mathrm{spec}<2.66$ have been overlaid on this plot.
The stars are ZFIRE-COSMOS detections (green filled$\rightarrow$Q$_z=3$, white filled $\rightarrow$Q$_z=2$) and the black circles are the non-detections. Galaxies outlined in bright pink are the confirmed cluster members by \citet{Yuan2014}. 
The light pink filled diamonds are detections from the MOSDEF survey. Yellow triangles are from the VUDS survey.  
}
\label{fig:detection_map}
\end{figure*}

The COSMOS sample is primarily selected from a cluster field. The spatial distribution of the  field is shown in Figure \ref{fig:detection_map}. 
(The ZFOURGE photometric redshifts are replaced with our spectroscopic values where available.)  A redshift cut between $2.0<z<2.2$ is used to select galaxies in the cluster redshift range. 
Using necessary ZFOURGE catalogue quality cuts there are  378 galaxies within this redshift window.
Following \citet{Spitler2012}, these galaxies are used to produce a seventh nearest neighbour density map.
Similar density distributions are calculated to the redshift window immediately above and below $2.0<z<2.2$. These neighbouring distributions are used to  calculate the mean and the standard deviation of the densities. The density map is plotted in units of standard deviations above the mean of the densities of the neighbouring bins similar to \citet{Spitler2012}. Similar density maps were also made by \citet{Allen2015}.

The figure shows that ZFIRE has achieved a thorough sampling of the underlying density
structure at $z\sim2$ in the COSMOS field.  Between
$1.90<z_\mathrm{spec}<2.66$, in the COSMOS field the sky density of
ZFIRE is 1.47 galaxies/arcmin$^2$. For MOSDEF and VUDS it is 1.06
galaxies/arcmin$^2$ and 0.26 galaxies/arcmin$^2$, respectively.  A
detailed spectroscopic analysis of the cluster from
ZFIRE redshifts has been published in  \citet{Yuan2014}.

\begin{figure*}
\includegraphics[scale=0.87]{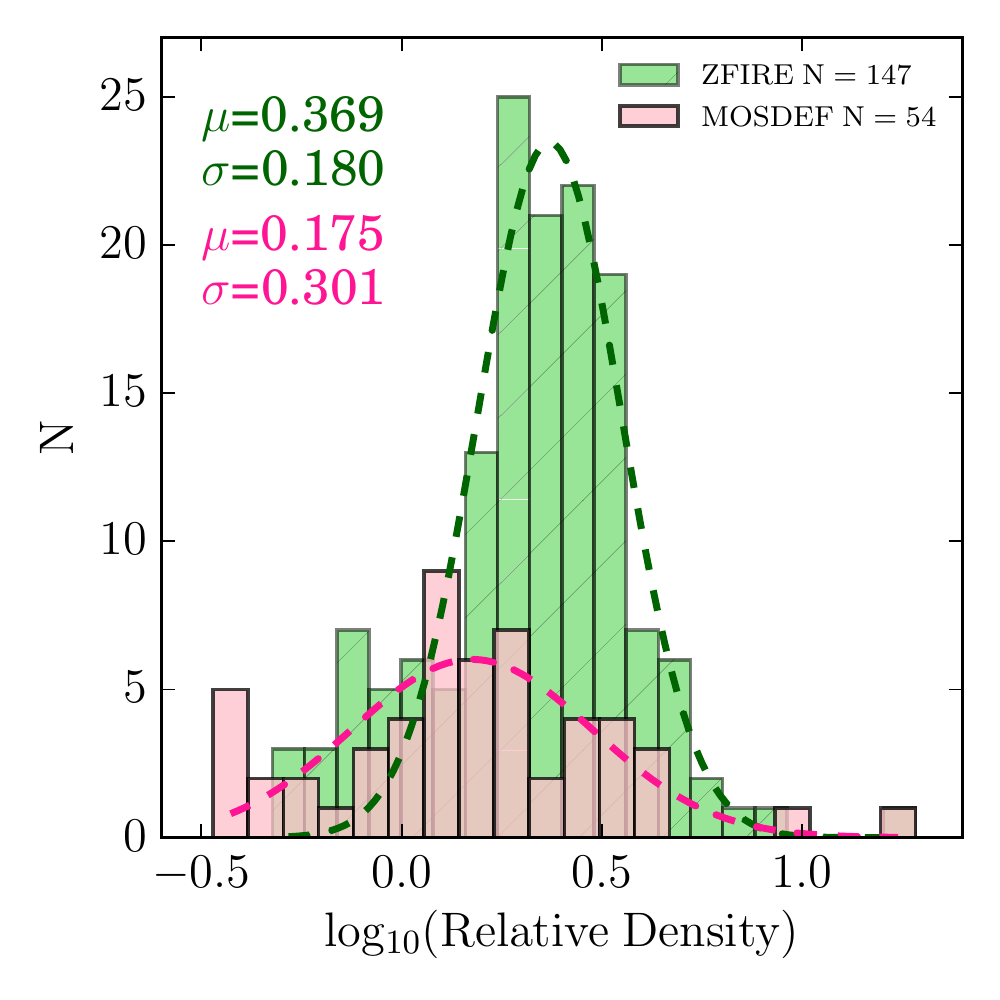}
\includegraphics[scale=0.87]{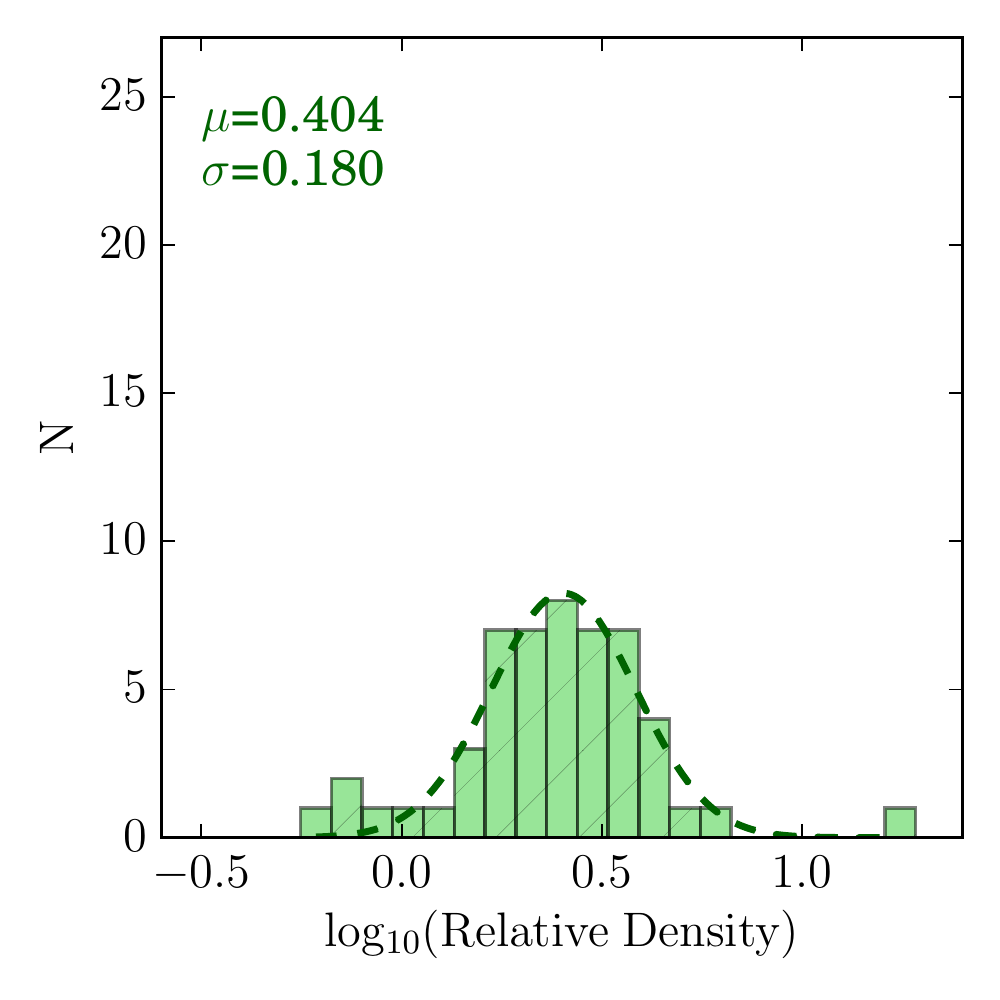}
\caption{ {\bf Left:} the relative galaxy density distribution of the galaxies with confident redshift  detections in the COSMOS field. Galaxies with $1.90<z_\mathrm{spec}<2.66$  in ZFIRE (green) and MOSDEF (pink) surveys are shown in the histogram. The density calculated is similar to what is shown in Figure \ref{fig:detection_map}.
Gaussian fits have been performed to both the samples. The density of the ZFIRE sample is distributed in logarithmic space around $\mu=0.369$ and $\sigma=0.180$, which is shown by the green dashed line. Similarly, the fit for the MOSDEF sample shown by the pink dashed line has $\mu=0.175$ and $\sigma=0.301$. Compared to MOSDEF, ZFIRE probes galaxies in richer environments. 
{\bf Right:} similar to the left plot but only the confirmed cluster members by \citet{Yuan2014} are shown in the histogram. The normalisation is lower because  the cluster identification of \citet{Yuan2014} came from a smaller earlier sample. (MOSDEF has only detected two cluster members and hence only the ZFIRE sample is shown in the figure.) The Gaussian best-fit parameters shown by the green dashed line has $\mu=0.404$ and $\sigma=0.180$. 
}
\label{fig:density_hist}
\end{figure*}

Figure \ref{fig:density_hist} shows the relative density distribution of the $1.90<$\zspec$<2.66$ galaxies. The MOSDEF sample is overlaid on the left panel and a Gaussian best-fit functions are fit for both ZFIRE (cluster and field) and MOSDEF samples. It is evident from the distributions, that in general ZFIRE galaxies are primarily observed in significantly higher density environments (as defined by the Spitler et al. metric) compared to MOSDEF. 
Because of the explicit targeting of `cluster candidate' fields, this is expected. 
In the right panel, the density distribution of the confirmed
cluster members of \citet{Yuan2014} is shown.

%------------------------------------------------------------
\section{Comparing ZFIRE Spectroscopic Redshifts to the Literature}
\label{sec:photometric_redshifts}

The new spectroscopic sample, which is in well-studied deep fields is ideal to test the redshift accuracy of some of the most important photometric redshift surveys, including the ZFOURGE survey from which it is selected.

\subsection{Photometric Redshifts from ZFOURGE and UKIDSS}

\begin{figure*}
\includegraphics[trim = 15 0 5 5, clip, scale=0.62]{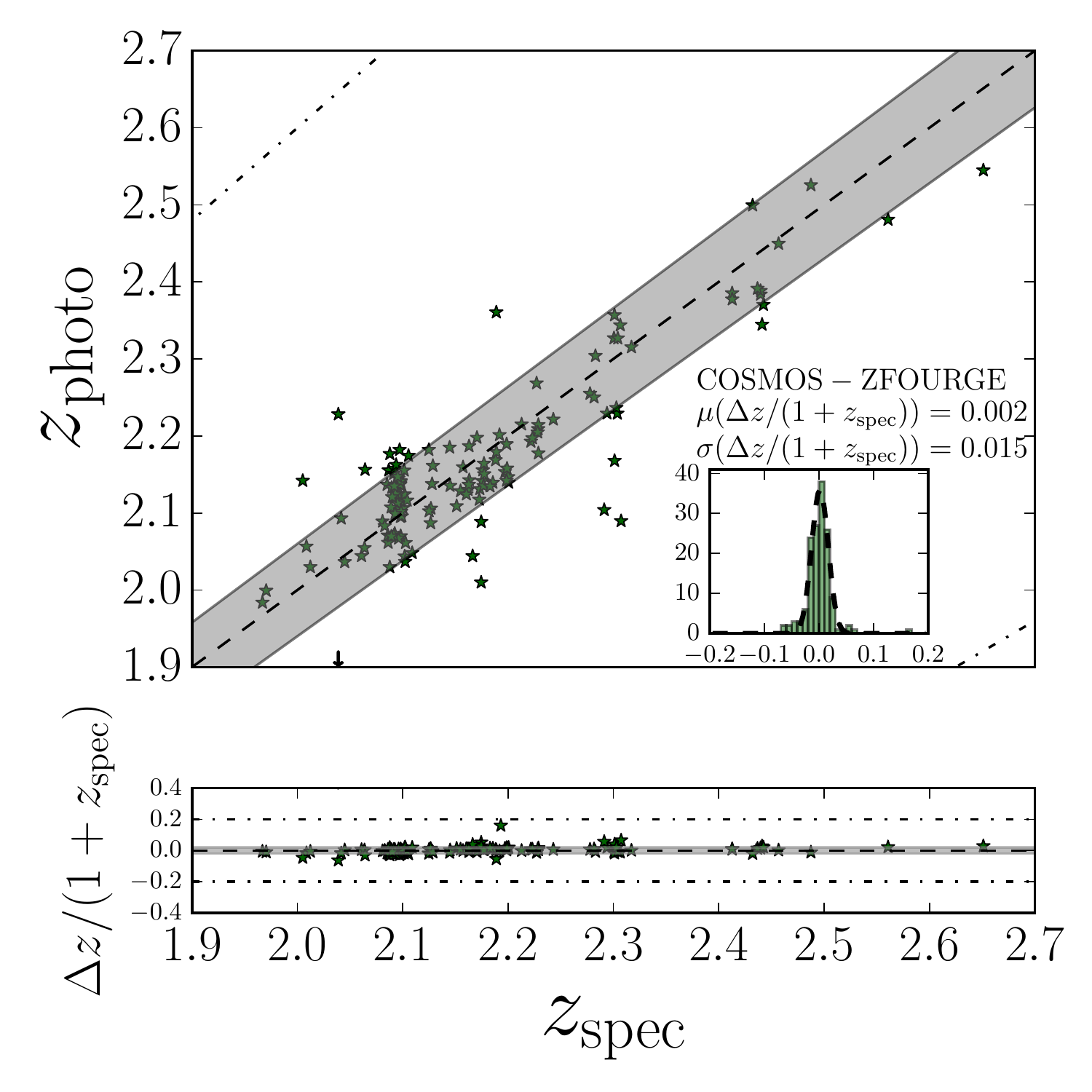}
\includegraphics[trim = 15 0 5 5, clip, scale=0.62]{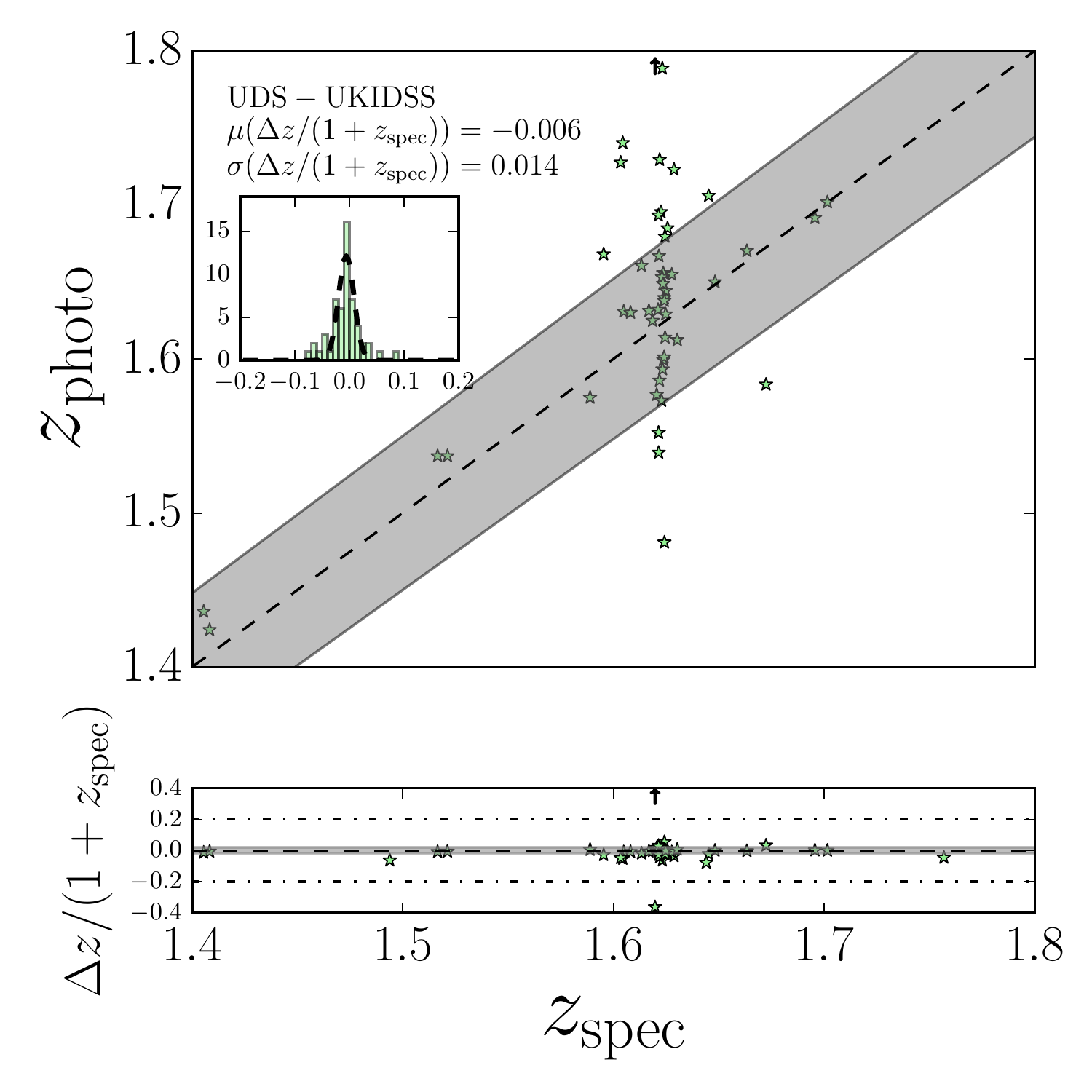}
\caption{ Comparison between the photometrically derived redshifts from ZFOURGE and UKIDSS with the ZFIRE Q$_{z}$=3 spectroscopic redshifts. 
{\bf Upper left:} $z_{\mathrm{photo}}$ vs. $z_{\mathrm{spec}}$ for the COSMOS field. $z_{\mathrm{photo}}$ values are from ZFOURGE v3.1 catalogue. 
The black dashed line is the one-to-one line. The grey shaded region represents the 2\% error level expected by the photometric redshifts (Straatman et al., in press). 
The dashed dotted line  shows the $\mid$$\Delta z$/(1+$z_{\mathrm{spec}}$)$\mid$ $>$ 0.2  drastic outlier cutoff.
The inset histogram shows the histogram of the  $\Delta z$/(1+$z_{\mathrm{spec}}$) values and Gaussian fits as described in the text.
Only galaxies with $1.90<z_{\mathrm{spec}}<2.70$ are shown in the figure.  
{\bf Lower left:} similarly for the residual  $\Delta z / (1+z_{\mathrm{spec}})$ between photometric and spectroscopic redshifts plotted against the spectroscopic redshift. 
{\bf Right:} similar to left panels but for the UDS field. $z_{\mathrm{photo}}$ values are from UKIDSS. 
}
\label{fig:specz_photoz}
\end{figure*}

The comparison of photometric redshifts and the spectroscopic redshifts for the ZFIRE-COSMOS sample is shown by the left panel of Figure \ref{fig:specz_photoz}. The photometric redshifts of the v3.1 ZFOURGE catalogue are used for this purpose because they represent the best calibration and photometric-redshift performance of the imaging. 
For the 42 detected secondary objects in the slits, 25 galaxies are identified with Ks selected ZFOURGE candidates.
Deep HST F160W band selected catalogues from ZFOURGE show probable candidates for eight these galaxies. 
Five galaxies cannot be confidently identified. HST imaging shows unresolved blends for four of these galaxies, which are listed as single objects in ZFOURGE. 
Only galaxies uniquely identified in ZFOURGE are shown in the figure.

Straatman at al., (in press) has determined that photometric redshifts are accurate to $<$2\% based on previous spectroscopic redshifts.  
Results from ZFIRE\ agree within this estimate. 
This error level is shown as a grey shaded region in Figure \ref{fig:specz_photoz} (left panel). Defining $\Delta z=\mathrm{z_{spec}-z_{photo}}$ (which will be used throughout this paper)
galaxies with $|\Delta z$/(1+\zspec)$|>$ 0.2 are considered to be ``drastic outliers''. There is one  drastic outlier in the Q$_{z}$=3 sample.
The advantage of medium-band NIR imaging relies on probing the D4000 spectral feature at $z>1.6$ by the J1, J2, and J3 filters, which span  \around 1--1.3\micron. 
Drastic outliers may arise due to blue star-forming galaxies having power-law-like SEDs, which do not have a D4000 breaks \citep{Bergh1963}, leading to uncertain photometric redshifts at $z\sim2$ and also from confusion between Balmer and Lyman breaks. Furthermore, blending of multiple sources in ground based imaging can also lead to drastic outliers. 
%With new confidence levels: there are no drastic outliers
%The single Q$_{z}$=3 drastic outlier  is due to the latter, where \emph{HST} imaging shows possible multiple targets being which identified as a single object by FourStar imaging.   

The inset in Figure \ref{fig:specz_photoz} (left panel) is a histogram that shows the residual for the Q$_{z}$=3 sample. 
A Gaussian best fit is performed for these galaxies (excluding  drastic outliers). The $\sigma$ of the Gaussian fit is considered to be the the accuracy of the photometric redshift estimates for a typical galaxy. The Q$_{z}$=3 sample is bootstrapped 100 times with replacement and the \NMAD\ scatter is calculated, which is defined as the error on  $\sigma$. 
The photometric redshift accuracy of the ZFOURGE-COSMOS sample is $1.5\pm0.2\%$ which is very high. 
The bright Ks $<23$ Q$_{z}$=3 galaxies show better redshift accuracy, but are within error limits of the redshift accuracy of the total sample. 
Furthermore, the Q$_{z}$=3 blue and red star-forming galaxies (as shown by Figure \ref{fig:UVJ}) also show similar redshift accuracy within error limits. 
The Q$_{z}$=2 ZFOURGE-COSMOS sample comprises 8 galaxies with a redshift accuracy of 14$\pm$12\%.

In Figure \ref{fig:specz_photoz} (right panel), a similar redshift analysis is performed to investigate the accuracy of the UKIDSS photometric redshift values with the ZFIRE-UDS spectroscopic sample. For the Q$_{z}$=3 objects, there are four drastic outliers (which give a rate of $\sim7\%$) and the accuracy is calculated to be 1.4$\pm$0.8\%. There are 12 Q$_{z}$=2 objects with one drastic outlier (which gives a rate of $\sim14\%$) and an accuracy of $3\pm12\%$. 
UKIDSS, which does not contain medium-band imaging has a comparable accuracy to the ZFOURGE medium-band survey. This is likely to arise from the lower redshifts probed by UKIDSS compared to ZFOURGE.

\subsection{Photometric Redshifts from NMBS and 3DHST}

\begin{figure*}
\includegraphics[trim = 15 0 5 5, clip, scale=0.62]{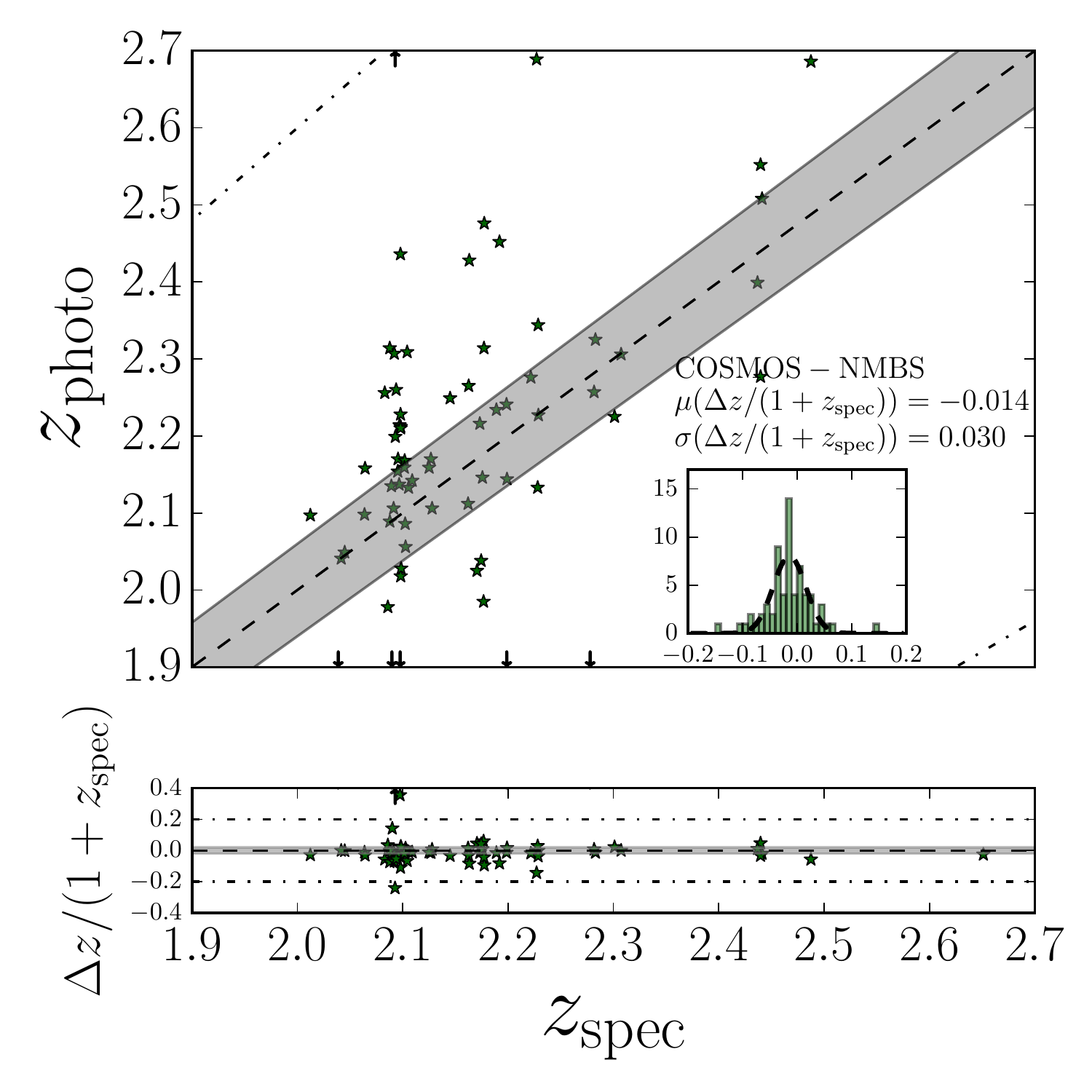}
\includegraphics[trim = 15 0 5 5, clip, scale=0.62]{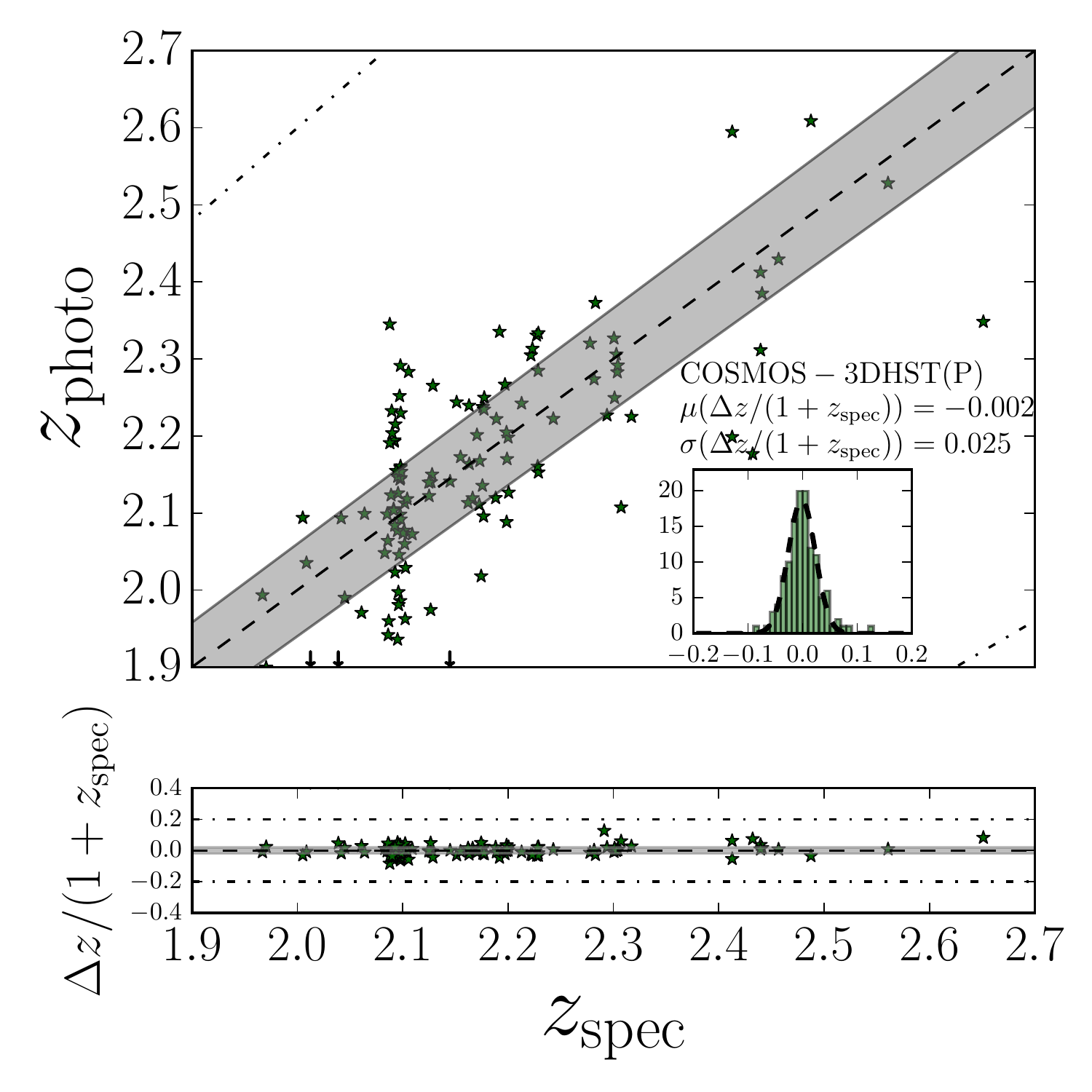}
\caption{ Comparison between photometric redshifts derived by NMBS and 3DHST photometric \citep{Skelton2014} with the ZFIRE\ spectroscopic sample. 
Lines and inset figures are similar to Figure \ref{fig:specz_photoz}. }
\label{fig:photo_z_comp}
\end{figure*}

Figure \ref{fig:photo_z_comp} shows a redshift comparison for the 3DHST photometric redshift input sample \citep{Skelton2014} and NMBS \citep{Whitaker2011} surveys with the ZFIRE Q$_{z}$=3 spectroscopic redshifts. 3DHST comes from the photometric data release  of \cite{Skelton2014}. 
The catalogues are compared to ZFOURGE by matching objects within a 0$''$.7 radius.
The ZFOURGE survey is much deeper than NMBS, so comparison to NMBS is only possible for a smaller number of brighter objects. 3DHST and ZFOURGE are similarly deep, with much better overlap.
The residuals between the photometric redshifts and spectroscopic redshifts are calculated using the same methods as for ZFOURGE.

Table \ref{tab:photo_z_comparision} shows the Gaussian best-fit values, redshift accuracies, and the drastic outlier fractions of all comparisons.  
All surveys resulted in high accuracy for the photometric redshifts. In particular, at $z\sim2$ some comparisons can be made between the ZFOURGE, 3DHST, and NMBS surveys. NMBS has the worst performance, both in scatter, bias, and outlier fraction, presumably because of the shallower data set, which also includes fewer filters (no HST-CANDELS data).  
NMBS  samples brighter objects, and in ZFOURGE  such bright objects have better photometric redshift performance compared to the main sample (for galaxies with $K<23$ photometric redshift accuracies for ZFOURGE and NMBS are respectively, $1.3\pm0.2\%$ and $2\pm1$). 
3DHST fares  better in all categories.  ZFOURGE performs the best of the three in this comparison.
This is attributed to the much better seeing and depth of ZFOURGE NIR medium-band imaging, which is consistent with the findings of Straatman et al., (in press). 
%For NMBS  and 3DHST surveys, the ZFIRE derived redshift accuracies are slightly less accurate than what is expected from the relevant survey findings \citep{Whitaker2011,Skelton2014}. 

\begin{deluxetable*}{lrrcccccc}
\tabletypesize{\scriptsize}
\tablecaption{ 
Photometric (P)/Grism (G) redshift comparison results for ZFIRE Q$_{z}$=3 galaxies.
\label{tab:photo_z_comparision}}
\tablecolumns{8}
\tablewidth{0pt} 
\tablehead{
\colhead{Survey}&
\colhead{ N (Q$_z=3$)\tablenotemark{a}}&
\colhead{ $\mu$ ($\Delta z$/(1+$z_{\mathrm{spec}}$))} &
\colhead{ $\sigma$ ($\Delta z$/(1+$z_{\mathrm{spec}}$))} &
\colhead{ $z_{\mathrm{err}}$\tablenotemark{b}} &
\colhead{$\Delta z_{\mathrm{err}}$\tablenotemark{c}} &
\colhead{ Drastic Outliers \tablenotemark{d}} &
\colhead{ N$\mathrm{_{Q_z=3}\ Ks<23}$ \tablenotemark{e}}&
}
\startdata
ZFOURGE (P)-Total  			& 147		&   0.002 & 0.016  & 1.5\%   & $\pm$0.2\%  & $0.7\%$ & 53 \\ 
ZFOURGE (P)-Ks $<23$         & 53		&   0.004 & 0.013  & 1.3\%   & $\pm$0.2\%  & $2.0\%$ & -- \\ 
& & & & & & &\\ 
\hline
& & & & & & &\\ 
NMBS    (P)  & 67   	&  -0.014 & 0.030  & 3.0\%   & $\pm$0.8\%  & 10.0\% & 48 \\
3DHST   (P)  & 127  	&  -0.002 & 0.025  & 2.5\%   & $\pm$0.3\%  & 3.2\% & 49 \\
3DHST   (P+G)  & 64  	    &  -0.001 & 0.009  & 0.9\%   & $\pm$0.2\%  & 4.7\% & 43 \\
& & & & & & &\\ 
UKIDSS  (P)  & 58   	&  -0.006 & 0.014  & 1.4\%   & $\pm$0.8\%  & 7.0\% & 38 \\
\tablenotetext{a}{The number of spectroscopic objects matched with each photometric/grism catalogue.}
\tablenotetext{b}{The accuracy of the photometric redshifts. }
\tablenotetext{c}{The corresponding bootstrap error for the redshift accuracy.}
\tablenotetext{d}{Drastic outliers defined as  $\Delta z$/(1+$z_{\mathrm{spec}}$) $>0.2$. They are given as a percentage of the total matched sample $(N)$ for each photometric/grism catalogue. Limits correspond to having $<1$ outlier.}
\tablenotetext{e}{The number of bright galaxies with Ks$<$23.}
\end{deluxetable*}

\subsection{Grism Redshifts from 3DHST}

\begin{figure}
\includegraphics[trim = 15 0 5 5, clip, scale=0.59]{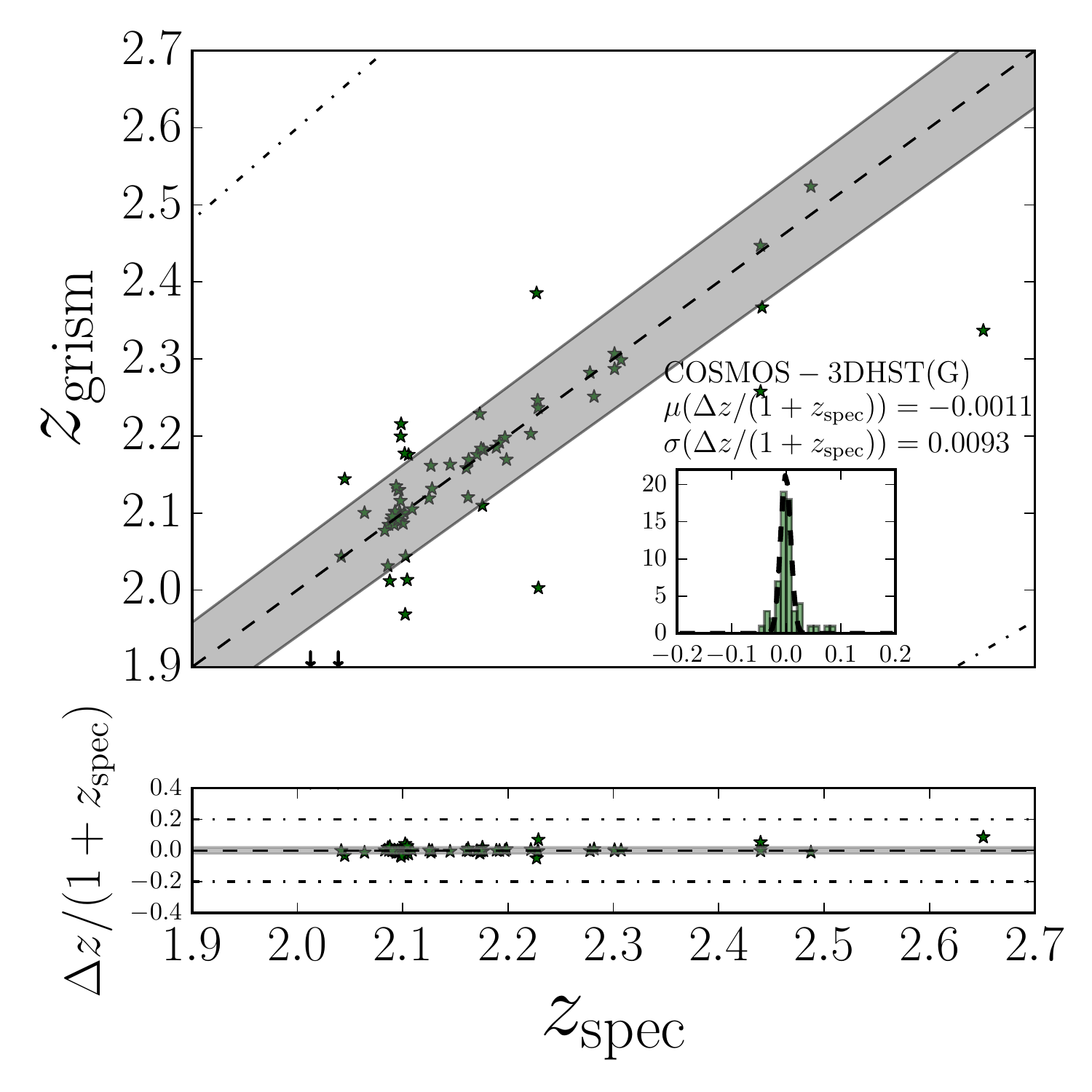}
\caption{ Spectroscopic redshift comparison between ZFIRE and 3DHST grism + photometric redshifts. 
This figure is similar to Figure \ref{fig:specz_photoz} with the exception of all photometric redshifts being replaced with the 3DHST \citet{Momcheva2015} data. The Gaussian fit to $\Delta$z/(1+z) has a $\mu$=-0.0011 and $\sigma$=0.009$\pm$0.001. Only galaxies with 1.90$<z_{\mathrm{spec}}<$2.70 are shown in the figure. 
}
\label{fig:specz_3DHST_grism}
\end{figure}

3DHST grism data is used to investigate the improvement of redshift accuracy with the introduction of grism spectra to the SED fitting technique. \citet{Momcheva2015} uses a combination of grism spectra and multi-wavelength photometric data to constrain the redshifts of the galaxies. 
\citet{Momcheva2015} states that 3DHST grism data quality has been measured by two independent users. All objects, which are flagged to be of good quality by both of the users are selected to compare with the ZFIRE sample. 
This gives 175 common galaxies out of which 123 have Q$_{z}$=3 and 64 of them pass the 3DHST grism quality test.
The \zgrism\ vs. \zspec\ distributions of these 64 galaxies are shown in Figure \ref{fig:specz_3DHST_grism}.  
There are three drastic outliers, which have been identified as low-redshift galaxies by 3DHST grism data with \zgrism$<0.5$. ZFIRE \zspec\ of these outliers are $>$2.

Comparing with the 3DHST redshifts derived only via pure photometric data, it is evident that the introduction of grism data increases the accuracy of the redshifts by \around$\times$3 to an accuracy of $0.9\pm0.1$\%.  The \zgrism\ accuracy is lower than the \around0.4\% accuracy computed by \citet{Bezanson2016} for grism redshifts. We note that \citet{Bezanson2016} is performed for galaxies with $H_{F160W}<24$ and that the ZFIRE-COSMOS sample probes much fainter magnitudes.

\subsection{Spectroscopic Redshifts from MOSDEF and VUDS}
\label{sec:specz_comparisions}

The final comparison is with other public spectroscopic redshifts in these fields. Galaxies from MOSDEF \citep{Kriek:2014fk} and VUDS \citep{Cassata2015} surveys are matched with the ZFIRE sample within a 0$''$.7 aperture. 

The MOSDEF overlap comprises  84 galaxies in the COSMOS field with high confidence redshift detections, out of which 74 galaxies are identified with matching partners from the ZFOURGE survey. In the ZFOURGE matched sample, 59 galaxies are at redshifts between 1.90$<$\zspec$<2.66$. 
7 galaxies are identified to be in common between ZFIRE and MOSDEF detections. 
The RMS of the scatter between the spectroscopically derived redshifts is \around0.0007.
This corresponds to a rest frame velocity uncertainty of  \around67 km s$^{-1}$, which is attributed to barycentric redshift corrections not being applied for the MOSDEF sample. 
We note that barycentric velocities should be corrected as a part of the wavelength solution by the DRP for each observing night, and therefore we are unable to apply such corrections to the MOSDEF data. Considering ZFIRE data, once the barycentric correction is applied we find, by analysing repeat observations in K band, that our redshifts are accurate to $\pm13$ km s$^{-1}$.

Similarly, the VUDS COSMOS sample comprises 144 galaxies with redshift detections $>3\sigma$ confidence, out of which 76 galaxies have ZFOURGE detections. In the ZFOURGE matched sample, 43 galaxies lie within $1.90<$\zspec$<2.66$. 
There are two galaxies in common between ZFIRE and VUDS detections and redshifts agree within 96 km s$^{-1}$ and 145 km s$^{-1}$. The redshift confidence for the matched two galaxies are $<\mathrm{2}\sigma$ in the VUDS survey, while the ZFIRE has multiple emission line detections for those galaxies.  Furthermore, the VUDS survey employs VIMOS in the low-resolution mode ($R\sim200$) in its spectroscopy leading to absolute redshift accuracies of $\sim200$ km s$^{-1}$. Therefore, we expect the ZFIRE redshifts of the matched galaxies to be more accurate than the VUDS redshifts.

\begin{figure*}
\includegraphics[trim = 0 0 10 5, clip, scale=0.90]{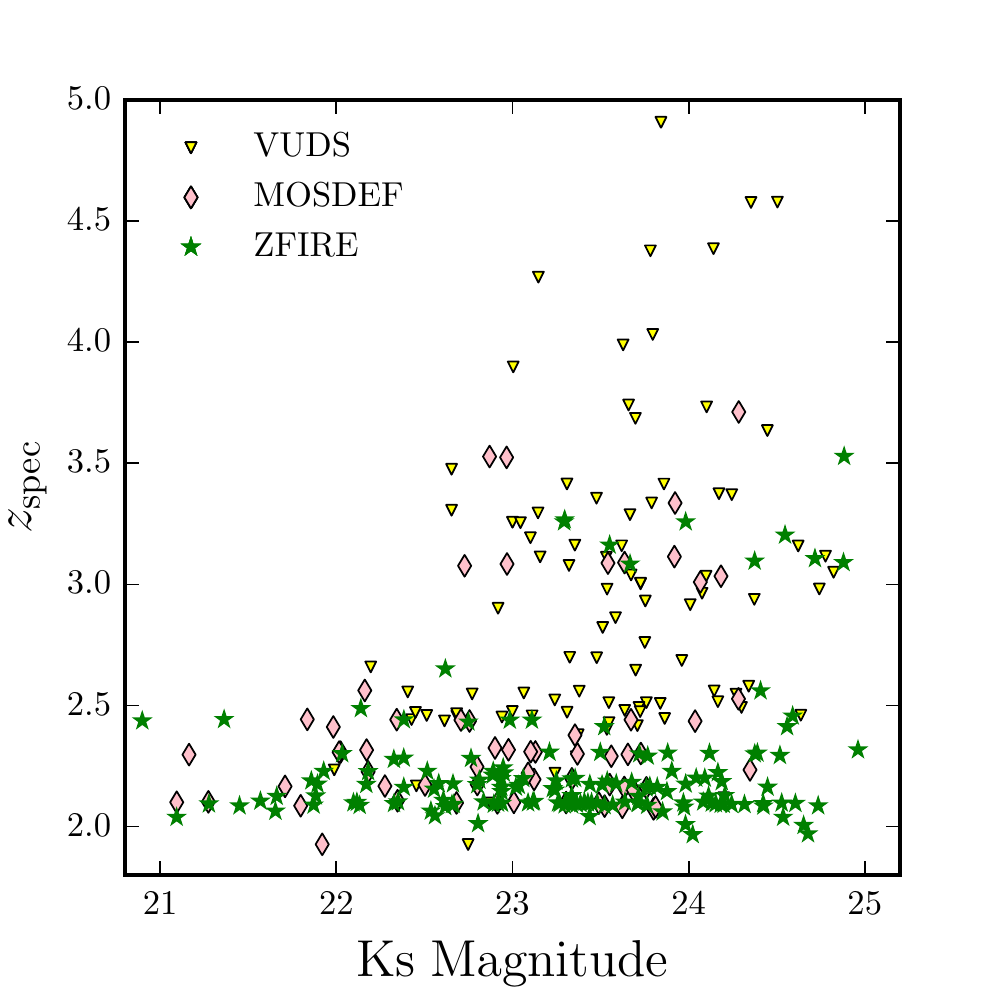}
\includegraphics[trim = 0 0 10 5, clip, scale=0.90]{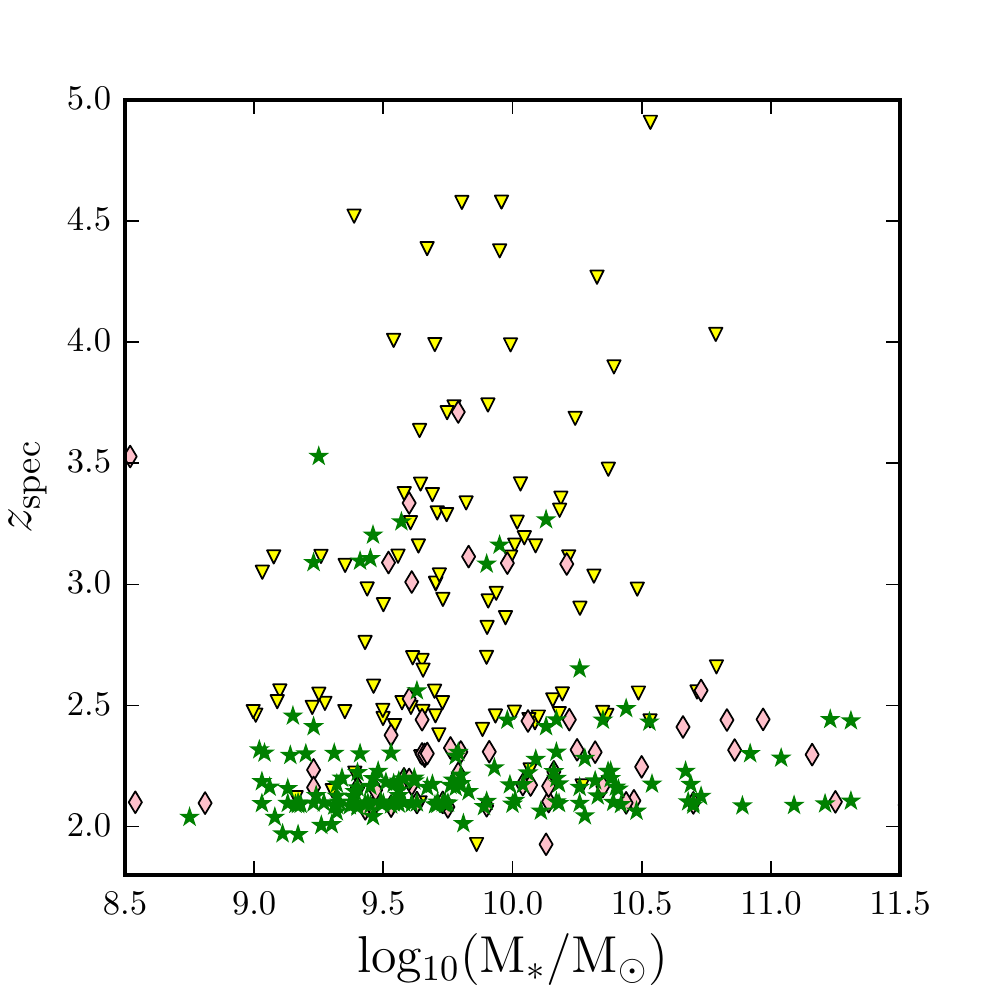}
\caption{ Redshift comparison as a function of Ks magnitude and stellar mass. Along with the ZFIRE\ Q$_{z}$=3 detections, the MOSDEF and VUDS samples are shown for comparison. 
For the MOSDEF sample, only galaxies with identified ZFOURGE detections are shown.  All VUDS galaxies with $\mathrm{z_{spec}}>1.8$ with $>3\sigma$ detections are shown.  
Note that VUDS observes galaxies in the optical regime, while ZFIRE\ and MOSDEF observes in the NIR. 
{\bf Left:} $\mathrm{z_{spec}}$ vs. Ks magnitude for the spectroscopically detected galaxies. The VUDS sample is plotted as a function of K magnitude. 
{\bf Right:} $\mathrm{z_{spec}}$ vs. stellar mass for the same samples of galaxies.  
}
\label{fig:survey_depth_comp}
\end{figure*}

Figure \ref{fig:survey_depth_comp} shows the distribution of the redshifts of the ZFIRE sample as a function of Ks magnitude and stellar mass. ZFIRE detections span a wide range of Ks magnitudes and stellar masses at $z\sim2$. The subset of galaxies observed at $z\sim3$ are fainter and are of lower mass. MOSDEF and VUDS samples are also shown for comparison. VUDS provides all auxiliary stellar population parameters, which are extracted from the CANDELS survey and hence all data are included. However, MOSDEF only provides the spectroscopic data and thus, only galaxies with identified ZFOURGE counterparts are shown in the figure, which is $\sim$90\% of the MOSDEF COSMOS field galaxies  with confident redshift detections. 

In Figure \ref{fig:survey_depth_comp}, MOSDEF detections follow a similar distribution to ZFIRE. Since both the surveys utilize strong emission lines in narrow NIR atmospheric passbands, similar distributions are expected.  VUDS, however, samples a different range of redshifts as it uses  optical spectroscopy. We note the strong \zspec=2.095 overdensity due to the cluster in the ZFIRE sample, but not in the others.

%------------------------------------------------------------

\section{Broader Implications}
\label{sec:implications}

The large spectroscopic sample presented can be used to assess the fundamental accuracy of galaxy physical parameters (such as stellar mass, SFR, and galaxy
SED classification) commonly derived from photometric redshift surveys. It can also be used to understand the performance of the previous
cluster selection that was done.

\subsection{Galaxy Cluster Membership}

The completeness and purity of galaxy cluster membership of the $z=2.1$ cluster based on photometric redshifts is next investigated and compared with spectroscopic results.
First,  photometric redshifts are used to compute a seventh nearest neighbour density map as shown in Figure \ref{fig:detection_map}. 
Any galaxy that lies in a region with density $>3\sigma$ is assumed to be a photometric cluster candidate.
From the ZFOURGE photometric redshifts in the COSMOS field (coverage of $\sim 11'\times11'$) for $2.0<$\zphoto$<2.2$, there are 66 such candidates. All of these galaxies have been targeted to obtain spectroscopic redshifts. 
\citet{Yuan2014} cluster galaxies are chosen to be within $3\sigma$ of the Gaussian fit to the galaxy peak at $z = 2.095$. 
Only 25 of the photometric candidates are identified to be a part of the \citet{Yuan2014} cluster, which converts to \around38\% success rate. 
The other 32 spectroscopically confirmed cluster galaxies at $z=2.095$ from Yuan et al. are not selected as cluster members using photometric redshifts, $i.e.$ membership identification based on photometric redshifts and seventh nearest
neighbour is \around56\% incomplete. 

\citet{Yuan2014} finds the velocity dispersion of the cluster structure to be $\sigma_{\mathrm{v1D}}= 552\pm52$ km s$^{-1}$, while the photometric redshift accuracy of ZFOURGE at $z=2.1$ is $\sim4500$ km s$^{-1}$. Therefore, even high-quality photometric redshifts such as from ZFOURGE, we are unable to precisely identify cluster galaxy members, which demonstrates that  spectroscopic redshifts are crucial for identifying and studying cluster galaxy populations at $z\sim2$.

\subsection{Luminosity, Stellar Mass, and Star Formation Rate}
\label{sec:M-SFR-dz}

\begin{figure}
\includegraphics[trim = 10 10 10 10, clip, scale=0.58]{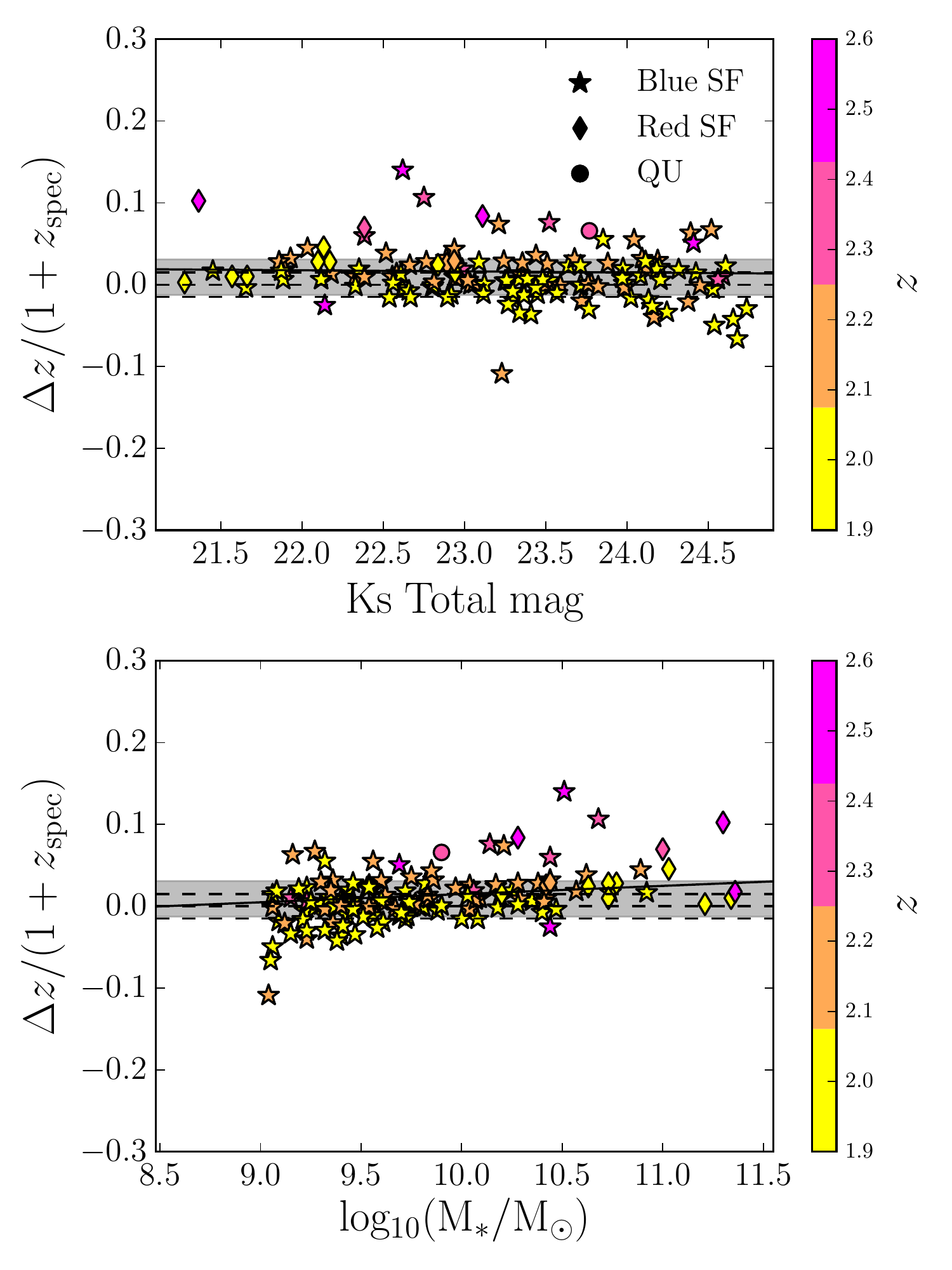}
\caption{ Photometric redshift accuracies as a function of Ks magnitude and stellar mass. 
All Q$_z=3$ ZFIRE-COSMOS galaxies with redshifts between $1.90<z<2.66$ have been selected. 
All galaxies are divided into blue star-forming, red (dusty) star-forming, and quiescent galaxies, which are shown with different symbols. Galaxies are further sub-divided into redshifts and are colour coded as shown. 
{\bf Top:} $\Delta z/(1+z_\mathrm{spec}$) vs. Ks total magnitude from ZFOURGE.  
{\bf Bottom:} similar to above but with stellar mass on the x-axis. 
The median  $\Delta z/(1+z_\mathrm{spec}$) is 0.009.
The grey shaded region in both the plots shows the \NMAD\ of the $\Delta z/(1+z_\mathrm{spec}$) scatter (0.022) around the median of the selected galaxies. The solid lines are the least squares best-fit lines for the data. 
}
\label{fig:delta_z_vs_param}
\end{figure}

An important question in utilising photometric redshifts is whether their accuracy depends on key galaxy properties such as luminosity, stellar mass, and/or SFR. This could lead
to biases in galaxy evolution studies.
The Ks total magnitudes and stellar masses from ZFOURGE (v2.1 catalogue) are used for this comparison, which is shown in Figure \ref{fig:delta_z_vs_param}.
The redshift error is plotted as a function of Ks magnitude and stellar mass for all Q$_z$=3 ZFIRE galaxies. The sample is binned into redshift bins and further subdivided into star-forming, dusty star-forming, and quiescent galaxies depending on their rest-frame UVJ colour.

The least squares best-fit lines for the Ks magnitude and stellar mass are $y=-0.001 (\pm0.003)x+0.05 (\pm0.06)$ and $y=0.010 (\pm0.005)x-0.08 (\pm0.05)$, respectively.
Therefore, it is evident that there is a slight trend in stellar mass in determining the accuracy of photometric redshifts with more massive galaxies showing positive offsets for $\Delta z/(1+$\zspec$)$.
However, the relationship of $\Delta z/(1+$\zspec$)$ with Ks magnitude is not statistically significant.
The typical \NMAD\ of $\Delta z/(1+$\zspec$)$ is 0.022 with a median of 0.009. 
Note that the $\Delta z/(1+$\zspec$)$ scatter parametrized here is different from the \zphoto\ vs. \zspec\ comparison in Figure \ref{fig:specz_photoz} for the ZFOURGE sample. We use the ZFOURGE catalogue version 2.1 for the  $\Delta z/(1+$\zspec$)$ vs. mass, magnitude comparison while for the \zphoto\ vs. \zspec\ comparison,  we use v3.1. Furthermore, the scatter here is calculated using \NMAD\ , while in Figure \ref{fig:specz_photoz} a Gaussian function is fit to the $\Delta z/(1+$\zspec$)$ after removing the drastic outliers. The changes in \zphoto\ between v2.1 and v3.1 is driven by the introduction of improved SED templates. This comparison is expanded on in Appendix \ref{sec:ZFOURGE comparison}.

\begin{figure}
\includegraphics[trim = 10 10 10 10, clip, scale=0.58]{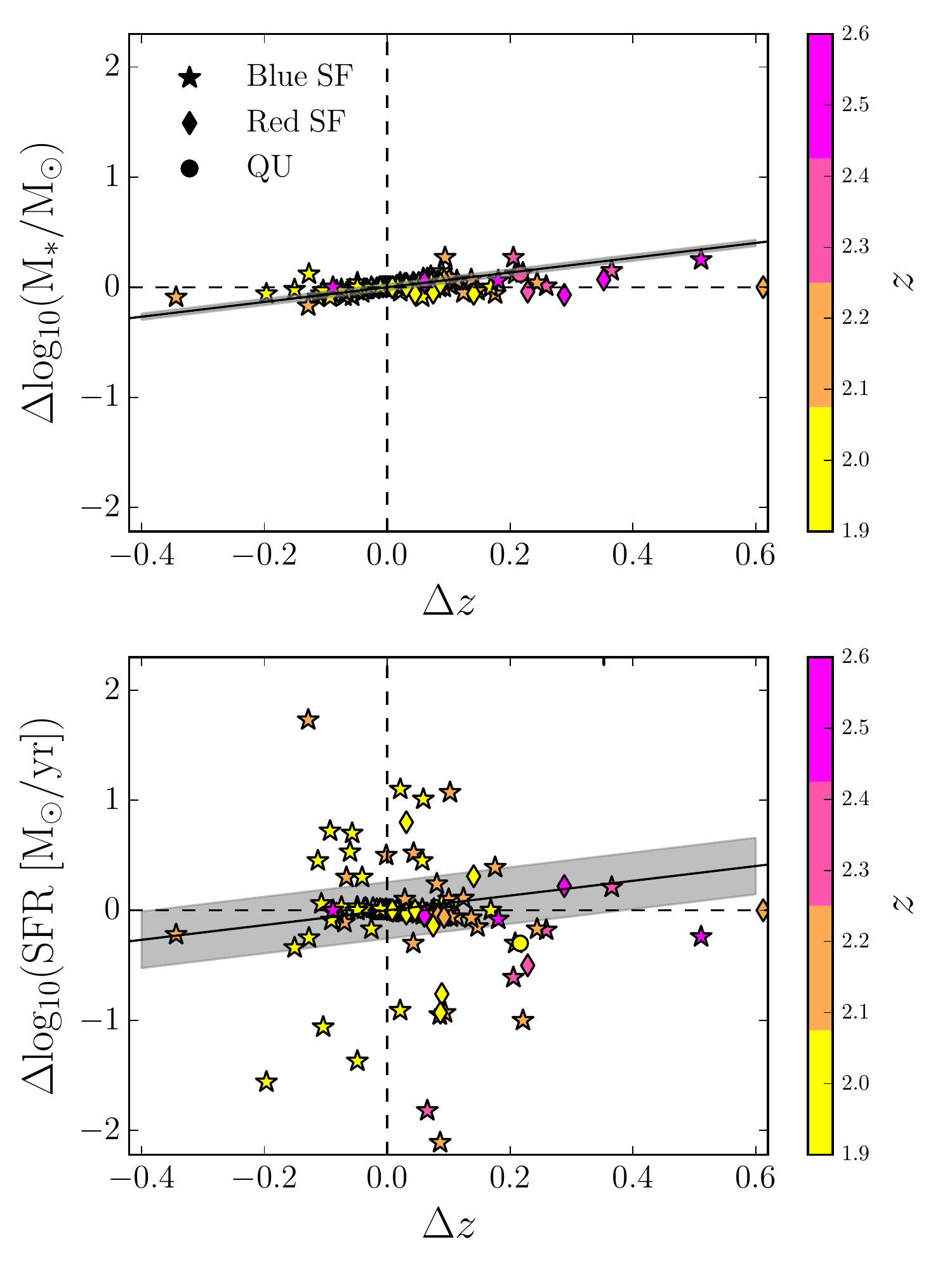}
\caption{ Effect of $\Delta z$ on galaxy stellar mass and dust extinction derived by FAST.
All ZFIRE-COSMOS galaxies with redshifts between $1.90<z<2.66$ have been selected. All galaxies are divided into blue star-forming, red (dusty) star-forming and quiescent galaxies which are shown as different symbols. Galaxies are further sub-divided into redshifts and are colour coded as shown. 
The diagonal solid lines are Equation (\ref{eq:delta_m_z}), which is the simplified theoretical expectation for mass/SFR  correlation with redshift error.
The grey shaded regions corresponds to the $\sigma$ value of the best-fit Gaussian functions that describes the deviation of the observed values from the theoretical expectation.
{\bf Top:} $\Delta\log_{10}$Mass vs. $z_\mathrm{spec}-z_\mathrm{photo}$. 
{\bf Bottom:} similar to top but with $\Delta\log_{10}$(SFR) on the y axis. 
}
\label{fig:delta_param_vs_delta_z}
\end{figure}

There should be a  dependency of galaxy properties derived via SED fitting techniques on $\Delta z$. Figure \ref{fig:delta_param_vs_delta_z} shows the change of stellar mass and SFR (both calculated using FAST using either photometric or spectroscopic redshifts) as a function of $\Delta z$. To first order, an analytic calculation of the expected residual can be made.
SED fitting techniques estimate galaxy stellar masses from luminosities and
mass-to-light ratios. The luminosity calculated from the flux will depend on the redshift used, and hence the mass and redshift change should correlate.
Ignoring changes in mass to light ratios and K-correction effects, from the luminosity distance change we expect
\begin{subequations}
\begin{equation}
\frac{d[\log_{10}(M)]}{dz} = \frac{2}{D_L} \left(\frac{dD_L}{dz}\right)_{z=2}
\end{equation}
where $M$ is the stellar mass of the galaxy  and $D_L$ is the luminosity distance. Evaluating for $z=2$, with $D_L=15.5$ Gpc:
\begin{equation}
\label{eq:delta_m_z}
\Delta \log_{10}(M) = 0.67 \Delta z
\end{equation} 
\end{subequations} 

Equation (\ref{eq:delta_m_z}) is plotted in Figure \ref{fig:delta_param_vs_delta_z}.  The top panel of the figure shows that the mass and redshift changes correlate approximately as expected with a \NMAD\ of 0.017 dex. SED SFRs are also calculated from luminosities, albeit with a much greater  
weight to the UV section of the SED, and thus should scale similarly to mass. 
The \NMAD\ scatter around this expectation is 0.086 dex, which is higher than the mass scatter  with a much greater number of outliers. To fully comprehend the role of outliers in the scatter, we fit a Gaussian function to the deviation of $\Delta\log_{10}$(Mass) and $\Delta\log_{10}$(SFR) for each galaxy from its  theoretical expectation. The $\Delta\log_{10}$(SFR) shows a larger scatter of $\sigma=0.2$ in the Gaussian fit compared to the $\sigma=0.03$ of $\Delta\log_{10}$(Mass). 
It is likely that the higher scatter in $\Delta\log_{10}$(SFR)  is because the  rest-frame UV luminosity is much more sensitive to the star formation history and dust extinction encoded in the best-fit SED than the stellar mass.

It is evident that photometric-redshift derived stellar masses are robust against the typical redshift errors, however, caution is warranted when using SED based SFRs with
photometric redshifts because they are much more sensitive to small redshift changes 
 (in our sample \around26\% of galaxies have $|\Delta\log_{10}$SFR$|>0.3$ even though the photometric redshifts have good precision). 
Studies that investigate galaxy properties solely relying on photometric redshifts may result in inaccurate conclusions about inherent galaxy properties and therefore, it is imperative that  they are supported by spectroscopic studies. It should be noted that previous ZFOURGE papers have extensively used photometric redshift derived stellar masses (for example, the mass function evolution of \citet{Tomczak2014}), which we find to be reliable, but not SED-based SFRs. Most commonly, the best-fit SEDs are used to derive the UV+IR fluxes in order to derive SFRs, since SFRs derived directly via FAST templates \cite[eg.,][]{Maraston2010} are degenerate with age, metallicity, and dust law. See \citet{Conroy2013} for a review on this topic.

\subsection{Rest-Frame UVJ Colours}

ZFOURGE rest frame UVJ colours are derived using photometric redshifts. 
UVJ colours from \zphoto\ are commonly used to identify the evolutionary stage of a galaxy \citep{Williams2009}. Here we investigate the effect of photometric redshift
accuracy on the UVJ colour derivation of galaxies. 

\begin{figure}
\includegraphics[trim = 0 0 0 0, clip, scale=0.925]{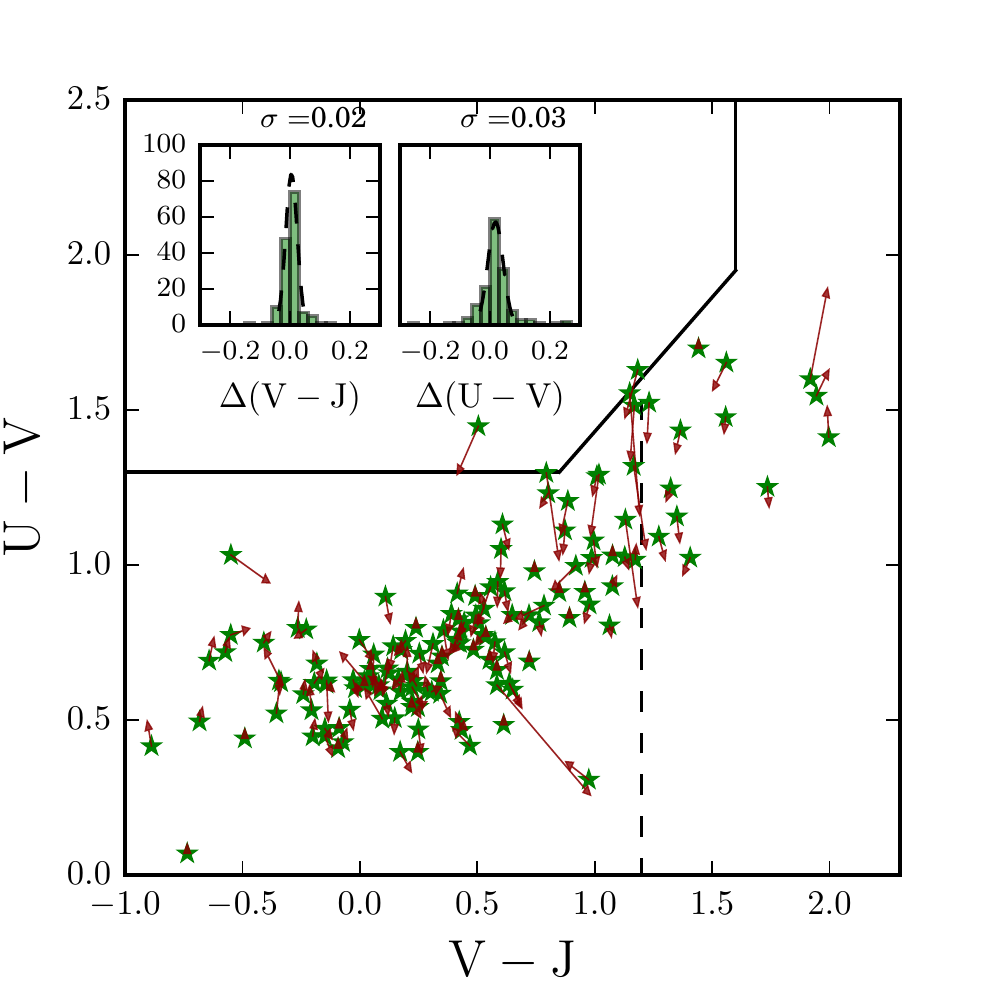}
\caption{ Effect of $\Delta z$ on rest frame UVJ colours. 
All ZFIRE-COSMOS galaxies are shown in the redshift bin $1.90<z<2.66$. 
The green stars are rest frame UVJ colours derived using photometric redshifts from EAZY. The rest frame colours are re-derived using spectroscopic redshifts from ZFIRE. The brown arrows denote the change of the position of the galaxies in the rest frame UVJ colour space when $z_{\mathrm{spec}}$ is used. The large arrows (one of which moves outside the plot range) are driven by $\Delta z$ outliers. 
The two inset histograms show the change in (V$-$J) and (U$-$V) colours for these sample of galaxies. Gaussian fits with $\sigma$ of 0.02 and 0.03 are performed, respectively, for the (V$-$J) and (U$-$V) colour differences.  
}
\label{fig:delta_UVJ}
\end{figure}

Figure \ref{fig:UVJ} shows the rest frame UVJ colours of Q$\mathrm{_z}$=3 objects re-derived using spectroscopic redshifts from the same SED template library. 
Figure \ref{fig:delta_UVJ} shows the change of location of the galaxies in rest frame UVJ colour when ZFIRE redshifts are used to re-derive them (the lack of quiescent galaxies overall is a bias in the ZFIRE\ sample selection as noted earlier).
Only one to two galaxies change their classifications from the total sample of 149. 
The inset histograms show the change of (U$-$V) and (V$-$J) colours. Gaussian functions are fit to the histograms to find that the scatter in (U$-$V) colours ($\sigma$=0.03) to be higher than that of (V$-$J) colours ($\sigma$=0.02) and (U$-$V)  has a greater number of outliers. 
The conclusion is that the U$-$V rest-frame colours are more sensitive to redshift compared to V$-$J colours by \around50\%, which may contribute to a selection bias in high-redshift samples. This sensitivity of the UV part of the SED is in accordance with the results of Section~\ref{sec:M-SFR-dz}.

To further quantify the higher sensitivity of U magnitude on redshift, Gaussian fits are performed on the $\Delta$U, $\Delta$V, and $\Delta$J magnitudes of the ZFIRE galaxies, by calculating the difference of the magnitudes computed when using \zphoto\ and \zspec. $\Delta$U shows a larger scatter of $\sigma=0.04$, while $\Delta$V and $\Delta$J show a scatter of $\sigma=0.01$. This further validates our conclusion that the UV part of the SED has larger sensitivity to redshift.

%%%%%%%%%%%%%%%%%%%%%%%%%%%%%%%%%%%%%%%%%%%%%%%%%%%%%%%%%%%%%%%%%%%%%%%%%%%%%%%%%%%%%%%%%%%%%%%%%%%%%%%%

\section{Summary}
\label{sec:summary}

Here we present the ZFIRE survey of galaxies in rich environments and our first public data release.  A detailed description of the data reduction used by ZFIRE is provided.  The use of a flux standard star along with photometric data from ZFOURGE and UKIDSS has made it possible to flux calibrate the spectra to $\lesssim10$\% accuracy.  The ZFIRE-COSMOS sample spans a wide range in Ks magnitude and stellar mass and secures redshifts for UVJ star-forming galaxies to Ks=24.1 and stellar masses of $\log_{10}($\mass$)>9.3$.  We show that selecting using rest-frame UVJ colours is an effective method for identifying \Halpha-emitting galaxies at $z\sim2$ in rich environments. Redshifts have been measured for 232 galaxies of which 87 are identified as members of the rich clusters we have targeted in COSMOS and UDS fields.

Photometric redshift probability density functions from EAZY are used to show that the expected \Halpha\ detections are similar to the ZFIRE detection rate in the COSMOS field. In the COSMOS field, the ZFIRE survey has detected \around80\% of the targeted star-forming galaxies. We also show that the density structure discovered by \citet{Spitler2012} has been thoroughly sampled by ZFIRE.

Using spectroscopic redshifts from ZFIRE with ZFOURGE and other public photometric survey data, we investigated the accuracies of photometric redshifts. The use of medium-band imaging in SED fitting techniques can result in photometric redshift accuracies of $\sim1.5\%$. ZFIRE calculations of photometric redshift accuracies are consistent with the expectations of the ZFOURGE survey (Straatman at al., in press) but are slightly less accurate  than the NMBS \citep{Whitaker2011} and 3DHST \citep{Skelton2014} survey results. The higher redshift errors can be attributed to sampling differences, which arises from the deeper NIR medium-band imaging in ZFOURGE compared to the other surveys (i.e. overlapping galaxies tend to be fainter than typical in the respective galaxies in NMBS). 
If we select a brighter subset of NMBS (Ks $<23$) we find that the redshift accuracy increases by 30\%.

Using UKIDSS, \citet{Quadri2012} shows that the photometric redshift accuracy is dependent on redshift and that at higher redshifts the photometric redshift error is higher. Between UKIDSS at $z\sim1.6$  and ZFOURGE at $z\sim2$ the photometric redshift accuracies are similar. Therefore, the use of medium-band imaging in ZFOURGE has resulted in more accurate redshifts at $z\sim2$, due to finer sampling of the D4000 spectral feature by the J1, J2, and J3 NIR medium-band filters. The introduction of medium-bands in the K band in future surveys may allow photometric redshifts to be determined to higher accuracies at $z\gtrsim4$. 

The importance of spectroscopic surveys to probe the large-scale structure of the universe is very clear. For the COSMOS \citet{Yuan2014} cluster, we compute a 38\% success rate (i.e., 38\% of galaxies in $3\sigma$ overdensity regions are identified spectroscopically as cluster galaxies) and a 56\% incompleteness (56\% of spectroscopic cluster galaxies are not identified from data based on purely photometry) using the best photometric redshifts (with seventh nearest neighbour algorithms) to identify clustered galaxies. 

We find a systematic trend in photometric redshift accuracy, where massive galaxies give higher positive offsets up to $\sim$0.05 for $\Delta z/(1+z_\mathrm{spec}$) values as a function of galaxy stellar mass. However, it is not evident that there is any statistically significant trend for a similar relationship with galaxy luminosity. 
Results also suggest that the stellar mass and SFR correlates with redshift error. This is driven by the change in the calculated galaxy luminosity as a function of the assigned redshift and we show that the values correlate approximately with the theoretical expectation. SFR shows larger scatter compared to stellar mass in this parameter space, which can be attributed to the stronger weight given to UV flux, which is very sensitive to the underlying model, in the derivation of the SFR.

This stronger correlation of the UV flux with redshift error is further evident when comparing the change in (U$-$V) and (V$-$J) colour with change in redshift. When rest-frame U,V, and J colours are re-derived using spectroscopic redshifts, our results show a stronger change in (U$-$V) colour compared to the (V$-$J) colour. Therefore, a redshift error may introduce an extra selection bias on rest-frame UVJ selected galaxies.  Further studies using larger samples of quiescent and dusty star-forming galaxies at $z\sim2$ are needed to quantify this bias.

Clearly the use of photometric redshifts can lead to biases even when using the same SED template set. However, it is important to acknowledge the underlying uncertainties that lie in deriving galaxy properties even with spectroscopic redshifts. 
Future work could consider the role of SED templates used in SED fitting techniques. Generally the templates used are empirically derived, which limits the capability to understand the inherent properties of the observed galaxies. With the use of physically motivated models such as MAGPHYS \citep{daCunha2008}, more statistically meaningful relationships between different physical parameters of the observed galaxies could be obtained. Improving such models to include photo-ionization of galaxies the in future will allow us to directly make comparisons of star-forming galaxies at $z\sim2$, which will be vital to study the inherent galaxy properties. 

Furthermore, the accuracy of underlying assumptions used in SED fitting techniques such as the IMF, dust properties, and star formation histories at $z\sim2$ should be investigated. These assumptions are largely driven by observed relationships at $z\sim0$, and if the galaxies at higher redshifts are proven to be inherently different from the local populations, results obtained via current SED fitting techniques may be inaccurate.  Future work should focus on the physical understanding of the galaxy properties at $z\gtrsim2$ with large spectroscopic surveys to better constrain the galaxy evolution models. The recent development of sensitive NIR integral field spectrographs with multiplexed capabilities will undoubtedly continue to add a wealth of more information on this topic over the next few years.

The ZFIRE survey will continue focusing on exploring the large spectroscopic sample of galaxies in rich environments at $1<z<3$ to investigate galaxy properties in rich environments.  Upcoming papers
include analyses of the IMF (T. Nanayakkara et al. 2016, in preparation), kinematic scaling relations (\citet{Alcorn2016}; C. Straatman et al. 2016, in preparation), the mass--metallicity fundamental plane \citep{Kacprzak2016}, and galaxy growth in cluster and field samples (K. Tran et al., in preparation).

\acknowledgements

The data presented herein were obtained
at the W.M. Keck Observatory, which is operated as a scientific
partnership among the California Institute of Technology, the
University of California and the National Aeronautics and Space
Administration. The Observatory was made possible by the generous
financial support of the W.M. Keck Foundation. 
The authors wish to recognize and acknowledge the very significant cultural role and
reverence that the summit of Mauna Kea has always had within the
indigenous Hawaiian community.  We are most fortunate to have the
opportunity to conduct observations from this mountain and we hope we
will be able to continue to do so.
We thank Nick Konidaris and the Keck observatory support staff for the
extensive and generous help given during the observing and data
reduction.  We thank Gabriel Brammer for providing us the updated EAZY
and help with several issues.  T.N., K.G., and G.G.K. acknowledge
Swinburne-Caltech collaborative Keck time, without which this survey
would not have been possible.  K.G acknowledges the support of the
Australian Research Council through Discovery Proposal awards
DP1094370, DP130101460, and DP130101667.  G.G.K. acknowledges the support
of the Australian Research Council through the award of a Future
Fellowship (FT140100933).  K.T. acknowledges the support of the
National Science Foundation under Grant \#1410728.  This work was
supported by a NASA Keck PI Data Award administered by the NASA
Exoplanet Science Institute.

Facilities: \facility{Keck:I (MOSFIRE)}

\bibliographystyle{apj}
%\bibliography{../../../papers/bibliography.bib}
\bibliography{bibliography}

\clearpage
\newpage

\appendix

\section{A: MOSFIRE calibrations}
\label{sec:MOSFIRE cals}

\subsection{Telluric Corrections}

\begin{figure*}[h!]
\includegraphics[trim = 10 10 10 10, clip, scale=0.45]{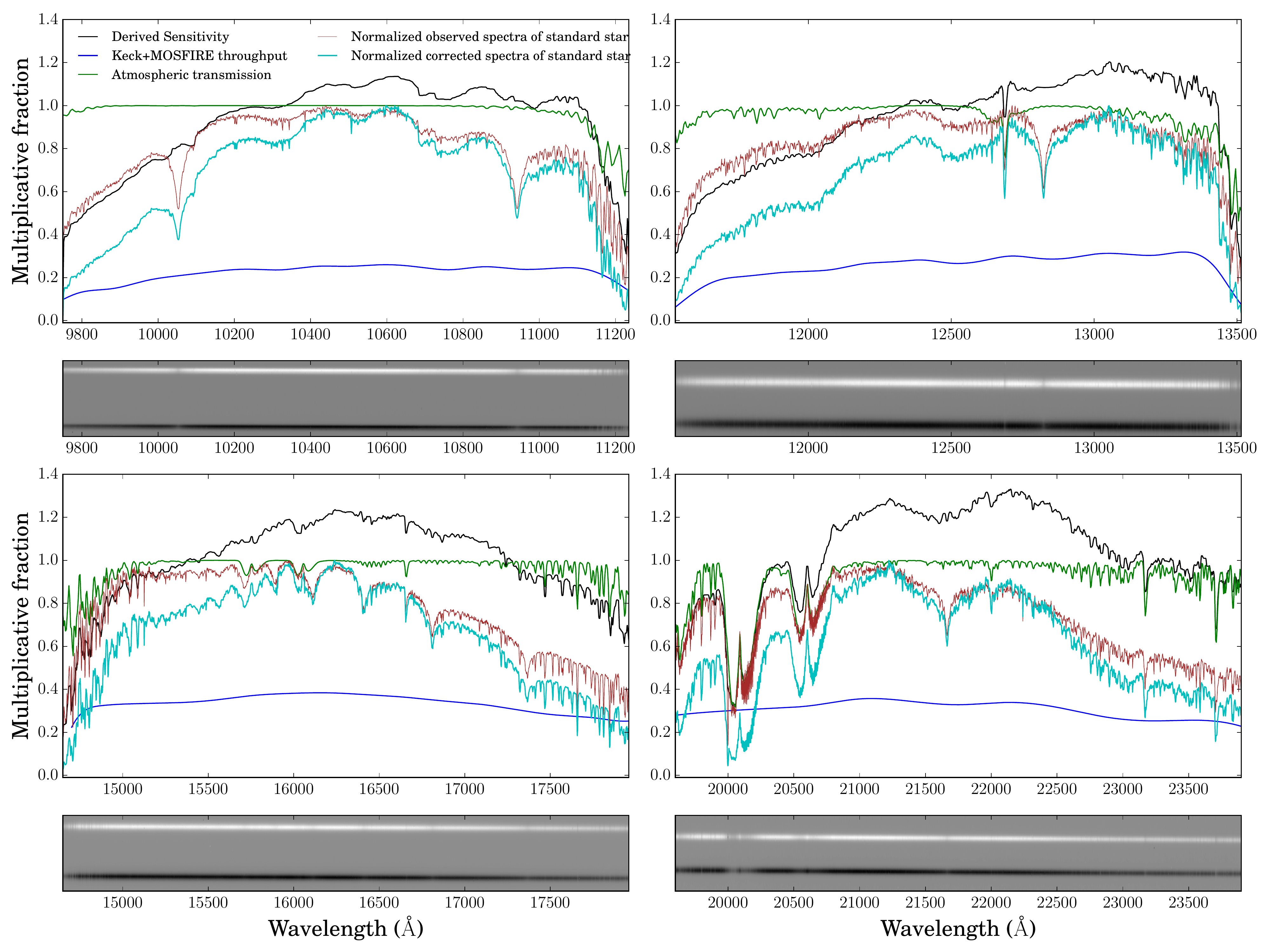}
\caption{Example set of derived sensitivity curves for MOSFIRE filters. From left to right, in the top panels we show the Y and J-bands and in the bottom panels we show the H and K bands. Pre-ship spectroscopic throughput for MOSFIRE is shown in blue. This takes into account the instrument response and the telescope throughput and \cite{McLean2012} shows that the these predictions agree extremely well with the measured values. The green line is the measured atmospheric transmission provided by the University of Hawai'i (private communication). 
The normalized spectra of the observed 1D standard stars before any corrections are applied are shown in brown. 
We remove the stellar atmospheric hydrogen lines and fit the spectra by a blackbody emission curve.
We use this derived spectra as a sensitivity curve (shown in black) and multiply our galaxy spectra by this to apply telluric corrections. 
We multiply the observed standard star spectra with the derived sensitivity curve to obtain a telluric corrected normalized standard star spectrum, which is shown in cyan. 
Each panel is accompanied with a 2D spectra of the standard star as given by the DRP. The black and white lines are the negative and positive images. Strong telluric features can be seen in regions where the intensity of the 2D spectra drops rapidly.
All 1D curves are normalized to a maximum value of 1.}
\label{fig:sensitivity}
\end{figure*}

Additional figures related to the MOSFIRE data reduction process are shown in this section. 
Figure \ref{fig:sensitivity} shows an example set of derived sensitivity curves and the normalized 1D spectra applied to all observed bands. 

\clearpage

\subsection{Spectrophotometric Calibrations}

\begin{figure}[h!]
\includegraphics[scale=1.20]{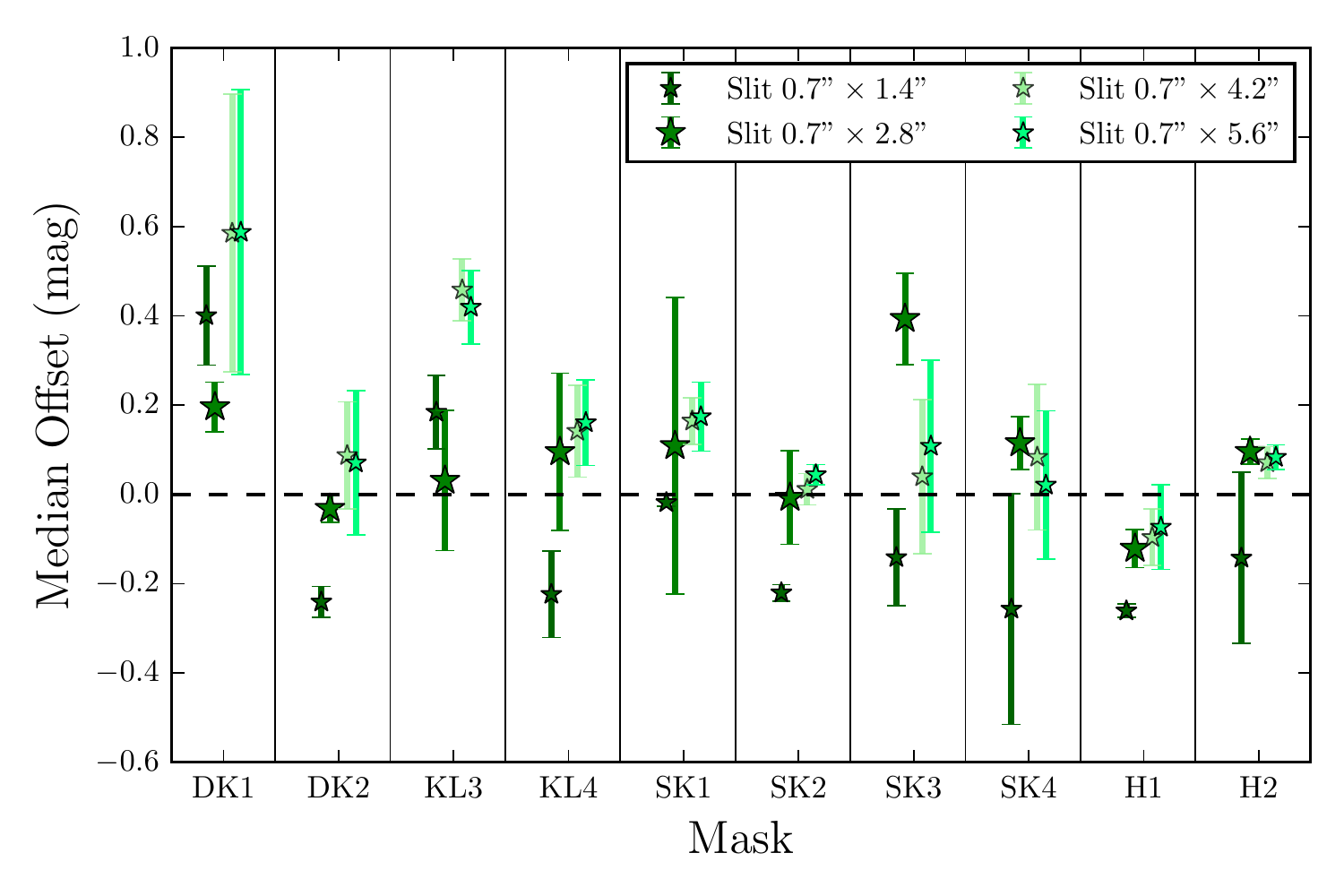}
\caption{ The median offset values for different aperture sizes for the COSMOS field masks. This figure is similar to Figure \ref{fig:scaling_values} top panel, but shows the median offset values computed for all slit-box like aperture sizes considered in our spectrophotometric calibration process. 
Filter names correspond to the names in Table \ref{tab:observing_details}.
The green stars in different shades for a given mask relates to the median offset between spectroscopic magnitude of the objects in the mask to its photometric magnitude computed using ZFOURGE and HST imaging with varying aperture sizes. 
The errors are the \NMAD\ scatter of the median offsets calculated via bootstrap re-sampling of individual galaxies. 
The vertical lines are for visual purposes to show data points belonging to each mask. 
}
\label{fig:scaling_values_varying_slit_boxes}
\end{figure}

As mentioned in Section \ref{sec:sp calibration}, for the COSMOS field, we overlaid synthetic slit apertures with varying slit heights on the ZFOURGE imaging to count the integrated flux within each aperture. The main purpose of the process was to account for the light lost due to the finite slit size. 
Figure \ref{fig:scaling_values_varying_slit_boxes} shows the change of median offset values for varying aperture sizes for each of the COSMOS mask.  As is evident from the figure, when the slit height increases from $1''.4$ to $2''.8$, most of the light emitted by the galaxies is included within the slit aperture. For any slit height beyond that, there is no significant change to the integrated counts, thus suggesting the addition of noise. Driven by this reason, we choose the $0''.7\times2''.8$ slit size to perform the spectrophotometric calibrations.

\begin{figure*}
\includegraphics[scale=0.9]{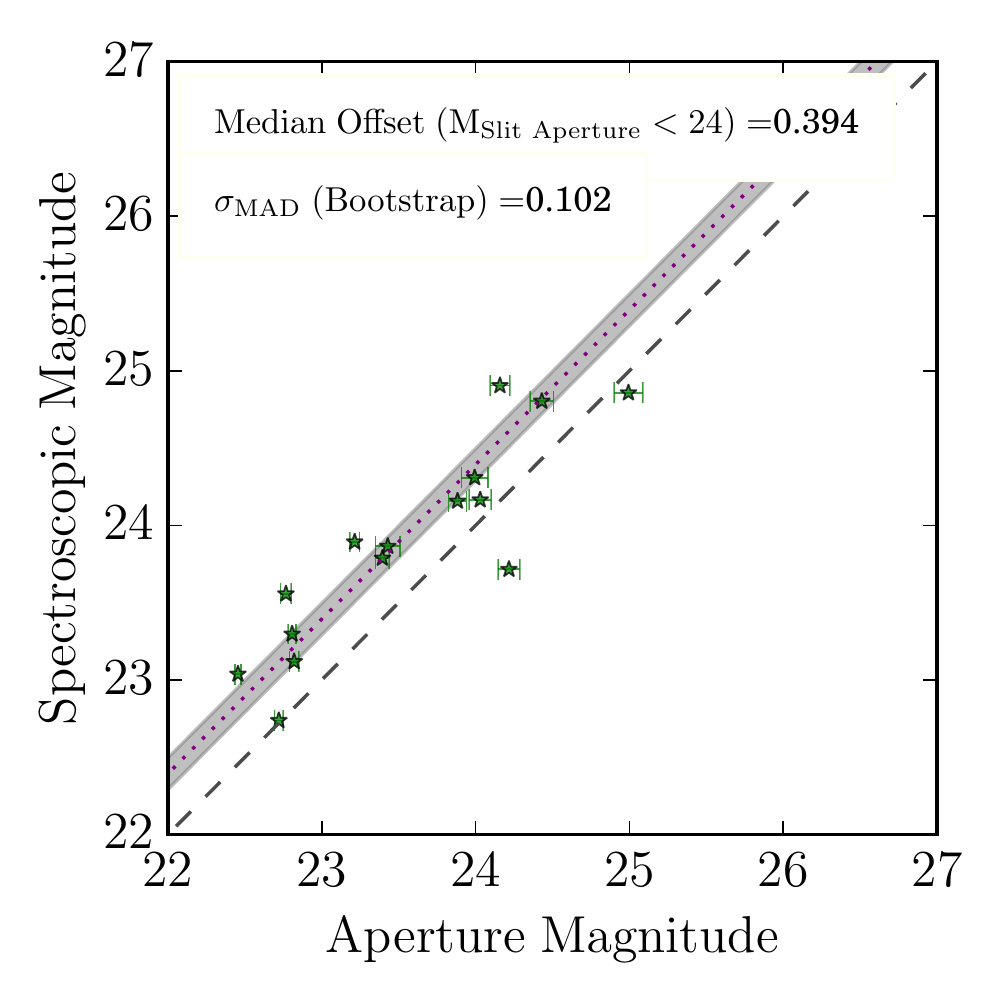}
\includegraphics[scale=0.9]{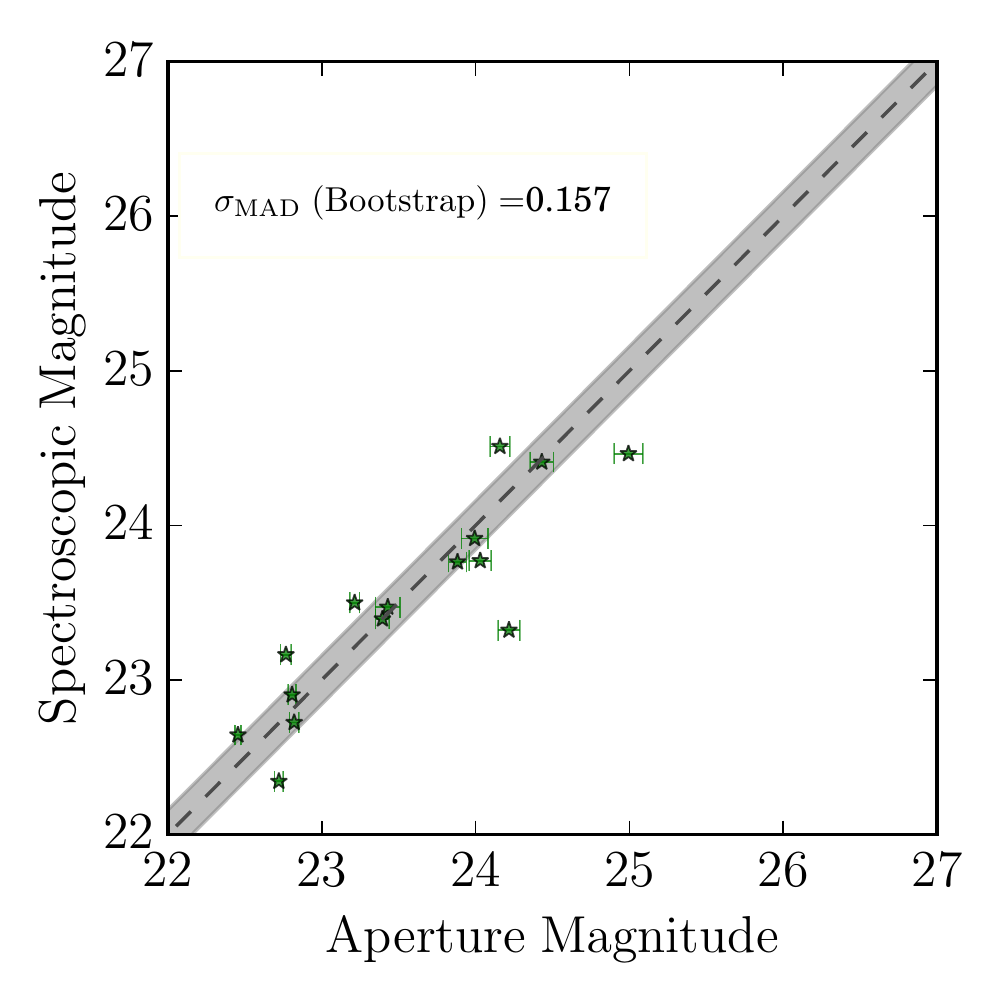}
\includegraphics[scale=0.9]{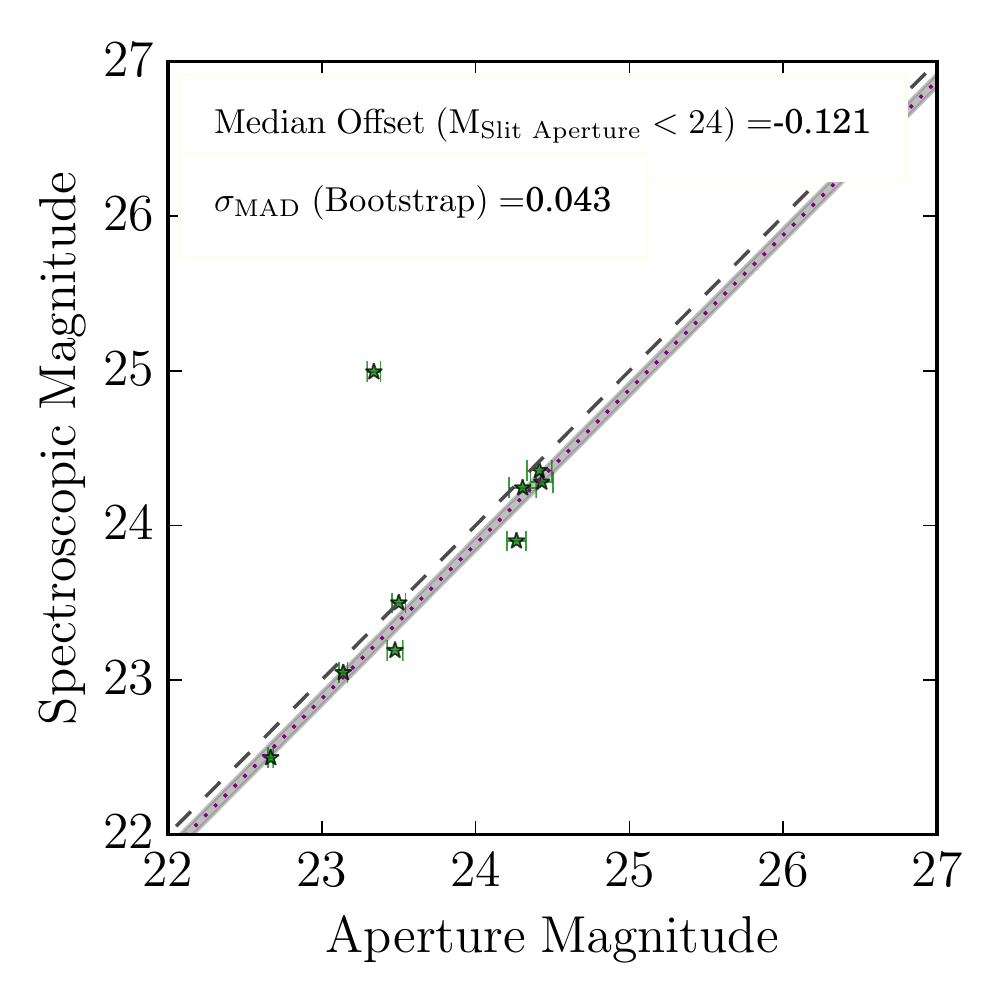}
\includegraphics[scale=0.9]{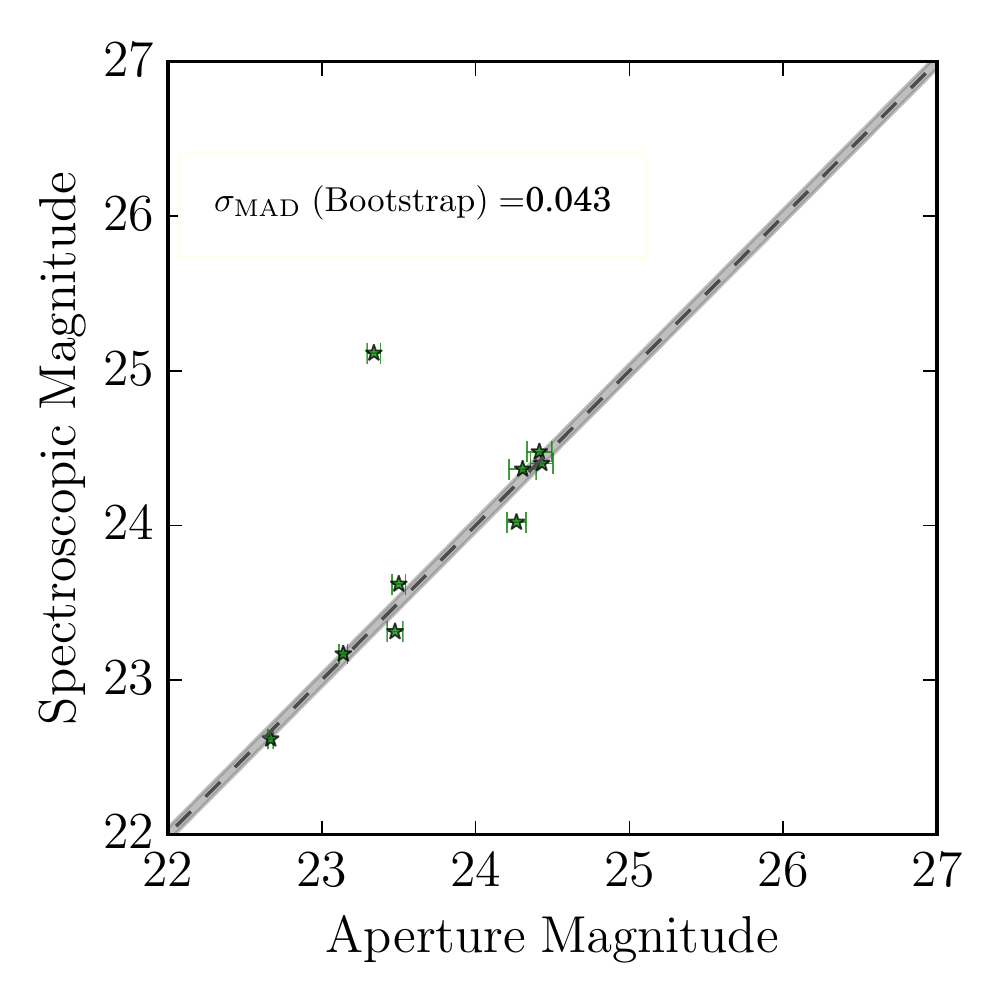}
\caption{Two example masks showing the comparison between spectroscopically derived magnitude to the photometrically derived magnitude using a $0''.7 \times 2''.8$ slit box. 
{\bf Top left:} K-band mask (KL3) before spectrophotometric calibration. The legend shows the median offset of galaxies with slit magnitude $<$24 and the corresponding bootstrap error.
{\bf Top right:} similar to left panel but after the spectrophotometric calibration has been applied. Since the scaling factor is now applied to the data, the median offset for galaxies with slit magnitude $<$24 is now 0. The inset shows the bootstrap error after the scaling is applied. This is considered to be the error of the spectrophotometric calibration process.
{\bf Bottom:} similar to the top panels but for a H-band mask (H1).
The grey shaded area in all the panels is the bootstrap error.
Error bars are from the ZFOURGE photometric catalogue. The flux monitor stars have been removed from the figure to focus the value range occupied by the galaxies. }
\label{fig:mask_scaling_example}
\end{figure*}

We show the magnitude distribution of two example masks in Figure \ref{fig:mask_scaling_example}. Once a uniform scaling is applied to all the objects in a given mask, the agreement between the photometric slit-box magnitude and the spectroscopic magnitude increases.

\clearpage

\section{B: Differences between ZFOURGE versions}
\label{sec:ZFOURGE comparison}

Here we show the effect of minor changes between different versions of ZFOURGE catalogues. 
ZFIRE\ sample selection was performed using an internal data release intended for the ZFOURGE team (v2.1). In this version, detection maps were made from Ks band photometry from FourStar imaging.
The 5$\sigma$ depth for this data release is Ks $\leq$25.3 in AB magnitude (Straatman et al., in press; This is 24.8 in \citet{Spitler2012}). 
All results shown in the paper, except for the photometric redshift analysis, are from v2.1. 

ZFOURGE COSMOS field has now been upgraded by combining the FourStar imaging with VISTA/K from UltraVISTA \citep{McCracken2012} to reach a 5$\sigma$  significance of 25.6 in AB magnitude (v3.1). This has increased the number of detected objects of the total COSMOS field by \around50\%. 

\begin{figure}[b]
\centering
\includegraphics[trim = 10 10 10 5, clip, scale=0.90]{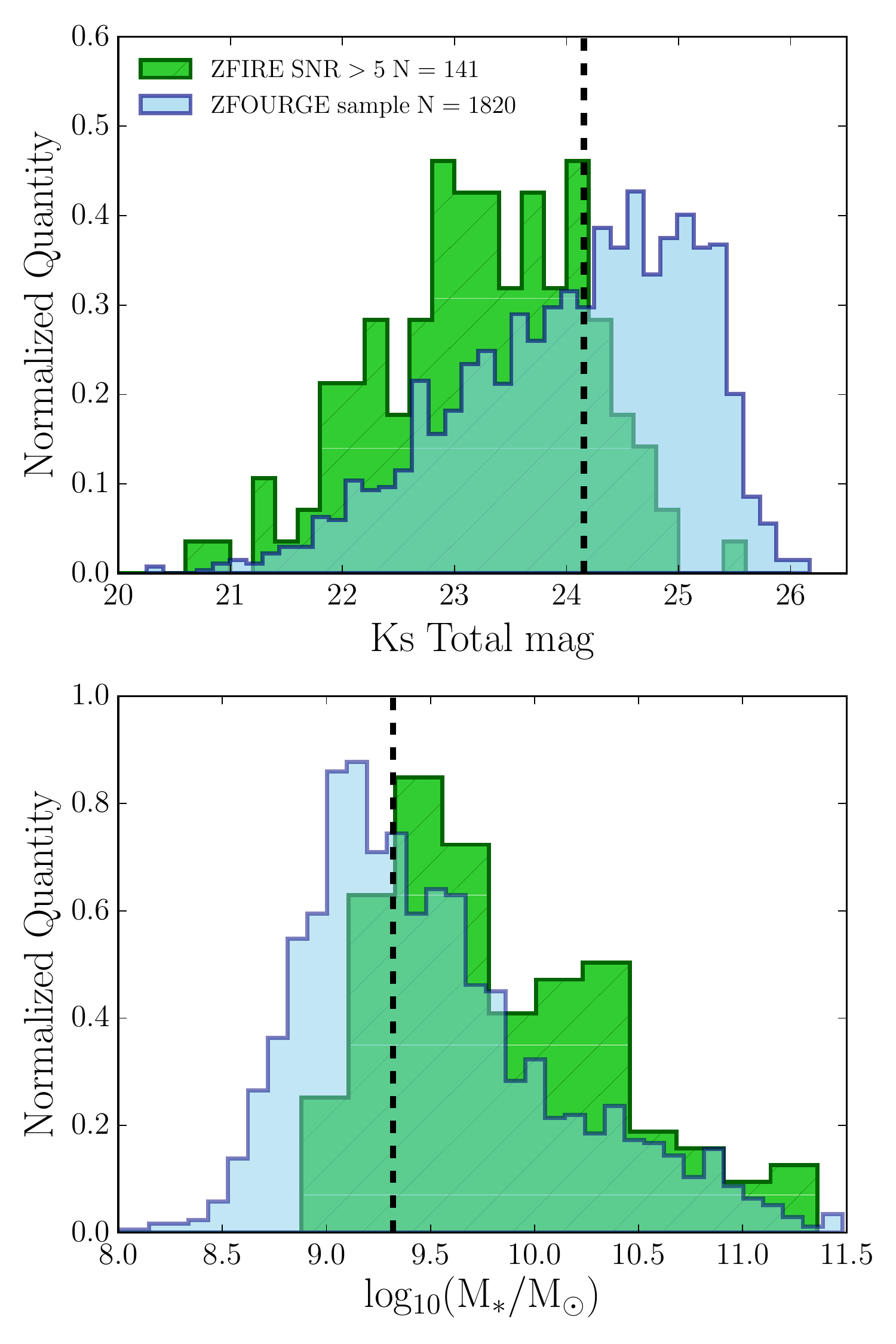}
\caption{The Ks magnitude and mass distribution of the 1.90$<z<$2.66 galaxies from ZFOURGE overlaid on the ZFIRE\ detected sample.
This figure is similar to Figure \ref{fig:detection_limits}, but the ZFOURGE data has been replaced with the updated deeper ZFOURGE catalogue (v3.1) and shows all ZFOURGE and ZFIRE detected galaxies in this redshift bin (In Figure \ref{fig:detection_limits} the quiescent sample is removed to show only the red and blue star-forming galaxies). 
}
\label{fig:detection_limits_newcat}
\end{figure}

All ZFIRE\ galaxies identified by v2.1 of the catalogue are also identified with matching partners by v3.1. 
Figure \ref{fig:detection_limits_newcat} shows the distribution of the Ks magnitude and masses of the updated ZFOURGE catalogue in the redshift bin $1.90<z<2.66$. 
The 80\%-ile limit of ZFOURGE in this redshift bin increases by 0.4 magnitude to to $Ks = 25.0$.
Similarly, the 80\% mass limit is \around$10^9$ \msol\ which is an increase of 0.2 dex in sensitivity.  
It is evident from the histograms that the significant increase of the detectable objects in this redshift bin has largely been driven by faint smaller mass galaxies. 
The 80\% limit for the ZFIRE-COSMOS galaxies is Ks=24.15 with the new catalogue. The change is due to the change of photometry between the two catalogues.

\begin{figure}
\includegraphics[trim = 10 10 10 5, clip, scale=0.925]{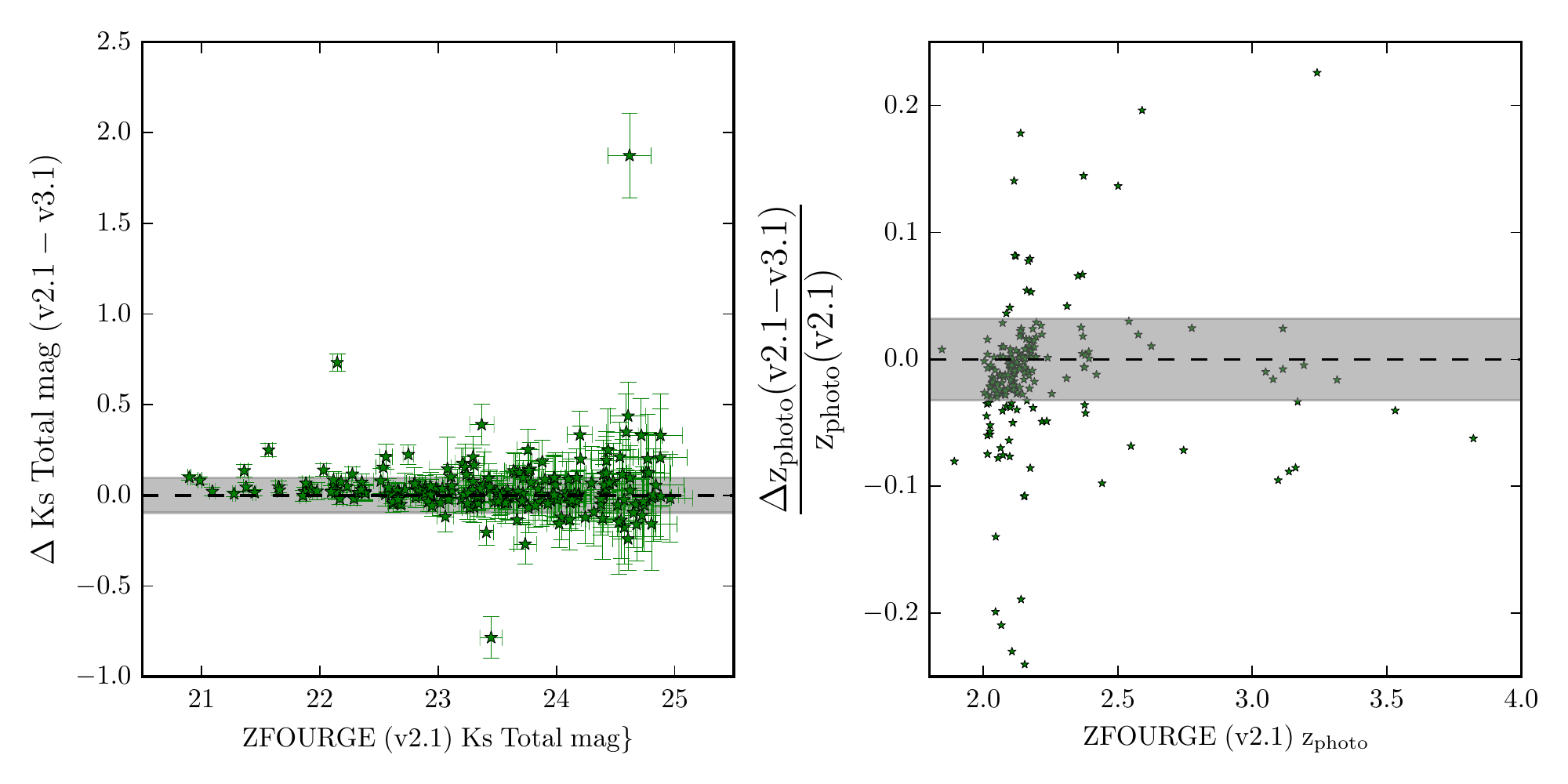}
\caption{Ks magnitude and the photometric redshift differences of ZFOURGE catalogues.
Only galaxies targeted by ZFIRE\ are shown.
{\bf Left:} the Ks band total magnitude difference between v2.1 and v3.1 of the ZFOURGE catalogues. 
{\bf Right:} the photometric redshift difference between v2.1 and v3.1 of the ZFOURGE catalogues. The grey shaded region denotes the \NMAD\ of the distribution. 
In both panels, the grey shaded region denotes the \NMAD\ of the distribution, which are respectively 0.09 magnitude and 0.03. 
}
\label{fig:cat_differences}
\end{figure}

Figure \ref{fig:cat_differences} shows the ZFOURGE catalogue differences between Ks total magnitude and the photometric redshift of the ZFIRE\ targeted galaxies. 
The Ks magnitude values may change due to the following reasons. 
\begin{enumerate}
\item The detection image is deeper and different, which causes subtle changes in the location and the extent of the galaxies. 
\item The zero point corrections applied to the data uses an improved method and therefore the corrections are different between the versions. 
\item The correction for the total flux is applied using the detection image, rather than the Ks image. Due to subtle changes mentioned in 1,  this leads to a different correction factor. 
\end{enumerate}
The \NMAD\ of the scatter for the Ks total magnitude is \around0.1 mag and is shown by the grey shaded region. There are few strong outliers. 
Two of the three catastrophic outliers are classified as dusty galaxies. One of them is close to a bright star and has an SNR of \around5 in v2.1. 
With the updated catalogue, the SNR has increased by \around30\% and therefore the new measurement is expected to be more robust.  
For the remaining galaxy, we see no obvious reason for the difference.

\begin{figure}
\includegraphics[trim = 10 10 10 5, clip, scale=0.65]{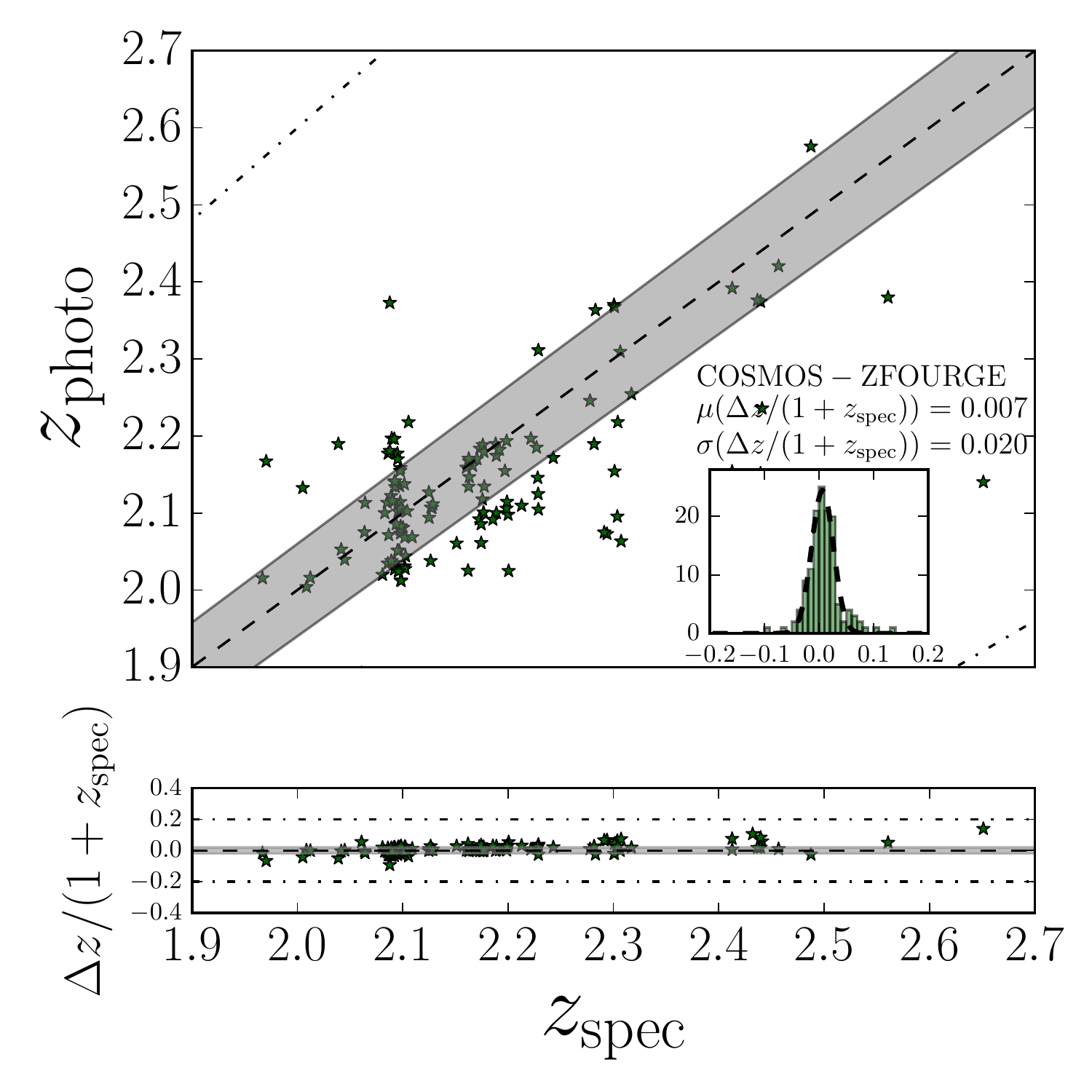}
\caption{ Photometric and spectroscopic redshift comparison between ZFOURGE v2.1 and ZFIRE. 
This Figure is similar to Figure \ref{fig:specz_photoz} with the exception of all photometric redshifts now being from v2.1 of the ZFOURGE catalogue. 
}
\label{fig:specz_photoz_newcat}
\end{figure}

Figure  \ref{fig:specz_photoz_newcat} shows the redshift comparison between ZFIRE spectroscopy and the v2.1 of the ZFOURGE catalogue. In v3.1, the photometric redshifts were updated by the introduction of high \Halpha\ equivalent width templates to EAZY and improved zero-point corrections to the photometric bands.
These changes along with the extra Ks depth have driven the increase in accuracy of the photometric redshifts from \around2.0\% in v2.1 to \around1.6\% in v3.1.

\end{document}